\documentclass[twocolumn]{aastex631}

\DeclareMathSymbol{\ast}{\mathbin}{symbols}{"03}

\begin{document}

\title{What is the nature of Little Red Dots and what is not, MIRI SMILES edition}

\author[0000-0003-4528-5639]{Pablo G. P\'erez-Gonz\'alez}
\affiliation{Centro de Astrobiolog\'{\i}a (CAB), CSIC-INTA, Ctra. de Ajalvir km 4, Torrej\'on de Ardoz, E-28850, Madrid, Spain}

\author[0000-0001-6813-875X]{Guillermo Barro}
\affiliation{Department of Physics, University of the Pacific, Stockton, CA 90340 USA}

\author[0000-0003-2303-6519]{George H. Rieke}
\affiliation{Steward Observatory, University of Arizona, 933 North Cherry Avenue, Tucson, AZ 85721, USA}

\author[0000-0002-6221-1829]{Jianwei Lyu}
\affiliation{Steward Observatory, University of Arizona, 933 North Cherry Avenue, Tucson, AZ 85721, USA}

\author[0000-0002-7893-6170]{Marcia Rieke}
\affiliation{Steward Observatory, University of Arizona, 933 North Cherry Avenue, Tucson, AZ 85721, USA}

\author[0000-0002-8909-8782]{Stacey Alberts}
\affiliation{Steward Observatory, University of Arizona, 933 North Cherry Avenue, Tucson, AZ 85721, USA}

\author[0000-0003-2919-7495]{Christina C. Williams}
\affiliation{NSF’s National Optical-Infrared Astronomy Research Laboratory, 950 North Cherry Avenue, Tucson, AZ 85719, USA}

\author[0000-0003-4565-8239]{Kevin Hainline}
\affiliation{Steward Observatory, University of Arizona, 933 North Cherry Avenue, Tucson, AZ 85721, USA}

\author[0000-0002-4622-6617]{Fengwu Sun}
\affiliation{Steward Observatory, University of Arizona, 933 North Cherry Avenue, Tucson, AZ 85721, USA}

\author[0000-0001-8630-2031]{D\'avid Pusk\'as}
\affiliation{Kavli Institute for Cosmology, University of Cambridge, Madingley Road, Cambridge, CB3 0HA, UK}
\affiliation{Cavendish Laboratory, University of Cambridge, 19 JJ Thomson Avenue, Cambridge, CB3 0HE, UK}


\author[0000-0002-8053-8040]{Marianna Annunziatella}
\affiliation{Centro de Astrobiolog\'{\i}a (CAB), CSIC-INTA, Ctra. de Ajalvir km 4, Torrej\'on de Ardoz, E-28850, Madrid, Spain}

\author[0000-0003-0215-1104]{William M.\ Baker}
\affiliation{Kavli Institute for Cosmology, University of Cambridge, Madingley Road, Cambridge, CB3 0HA, UK}
\affiliation{Cavendish Laboratory, University of Cambridge, 19 JJ Thomson Avenue, Cambridge, CB3 0HE, UK}

\author[0000-0002-8651-9879]{Andrew J.\ Bunker }
\affiliation{Department of Physics, University of Oxford, Denys Wilkinson Building, Keble Road, Oxford OX1 3RH, UK}

\author[0000-0003-1344-9475]{Eiichi Egami}
\affiliation{Steward Observatory, University of Arizona, 933 North Cherry Avenue, Tucson, AZ 85721, USA}

\author[0000-0001-7673-2257]{Zhiyuan Ji}
\affiliation{Steward Observatory, University of Arizona, 933 North Cherry Avenue, Tucson, AZ 85721, USA}

\author[0000-0002-9280-7594]{Benjamin D.\ Johnson}
\affiliation{Center for Astrophysics $|$ Harvard \& Smithsonian, 60 Garden St., Cambridge MA 02138 USA}

\author[0000-0002-4271-0364]{Brant Robertson}
\affiliation{Department of Astronomy and Astrophysics, University of California, Santa Cruz, 1156 High Street, Santa Cruz, CA 95064, USA}

\author[0000-0001-5171-3930]{Bruno Rodríguez Del Pino}
\affiliation{Centro de Astrobiolog\'{\i}a (CAB), CSIC-INTA, Ctra. de Ajalvir km 4, Torrej\'on de Ardoz, E-28850, Madrid, Spain}

\author[0000-0002-0303-499X]{Wiphu Rujopakarn}
\affiliation{National Astronomical Research Institute of Thailand, Don Kaeo, Mae Rim, Chiang Mai 50180, Thailand}
\affiliation{Department of Physics, Faculty of Science, Chulalongkorn University, 254 Phayathai Road, Pathumwan, Bangkok 10330, Thailand}

\author[0000-0003-4702-7561]{Irene Shivaei}
\affiliation{Centro de Astrobiolog\'{\i}a (CAB), CSIC-INTA, Ctra. de Ajalvir km 4, Torrej\'on de Ardoz, E-28850, Madrid, Spain}

\author[0000-0002-8224-4505]{Sandro Tacchella}
\affiliation{Kavli Institute for Cosmology, University of Cambridge, Madingley Road, Cambridge, CB3 0HA, UK}
\affiliation{Cavendish Laboratory, University of Cambridge, 19 JJ Thomson Avenue, Cambridge, CB3 0HE, UK}

\author[0000-0001-9262-9997]{Christopher N.\ A.\ Willmer}
\affiliation{Steward Observatory, University of Arizona, 933 North Cherry Avenue, Tucson, AZ 85721, USA}

\author[0000-0002-4201-7367]{Chris Willott}
\affiliation{NRC Herzberg, 5071 West Saanich Rd, Victoria, BC V9E 2E7, Canada}

\begin{abstract}
We study 31 little red dots (LRD) detected by JADES/NIRCam and covered by the SMILES/MIRI survey, of which  $\sim70$\% are detected in the two bluest MIRI bands and 40\% in redder MIRI filters. The median/quartiles redshifts are $z=6.9_{5.9}^{7.7}$ (55\% spectroscopic). 
The spectral slopes flatten in the rest-frame near-infrared, consistent with a 1.6~$\mu$m stellar bump but bluer than direct pure emission from active galactic nuclei (AGN) tori. The apparent dominance of stellar emission  at  these wavelengths for many LRDs expedites  stellar mass estimation: the median/quartiles are $\log \mathrm{M_\star/M_\odot}=9.4_{9.1}^{9.7}$. The number density of LRDs is $10^{-4.0\pm0.1}$~Mpc$^{-3}$, accounting for $14\pm3$\% of the global population of galaxies with similar redshifts and masses. The rest-frame near/mid-infrared (2--4~$\mu$m) spectral slope reveals significant amounts of warm dust (bolometric  attenuation $\sim3-4$~mag). Our spectral energy distribution modeling implies the presence of $<0.4$~kpc diameter knots, heated by either dust-enshrouded OB stars or an AGN producing a similar radiation field, obscured by $\mathrm{A(V)}>10$~mag.  We find a wide variety in the nature of LRDs. However, the best-fitting models for many of them  correspond to  extremely intense and compact starburst galaxies with mass-weighted ages 5--10~Myr, very efficient in producing dust, with their global energy output dominated by the direct (in the flat rest-frame ultraviolet and optical spectral range) and dust-recycled emission from OB stars with some contribution from an obscured AGN (in the infrared).
\end{abstract}



\section{Introduction}
\label{sec:intro}

In the very first month of JWST science operations, \citet{2023Natur.616..266L} identified a sample of compact sources with distinct spectral energy distributions (SEDs) presenting two clear breaks (Lyman and Balmer) in the NIRCam$+$HST data. The SEDs were also characterized by a change of slope: they had red colors at observed wavelengths between $\sim$2 and (at least) $\sim5\,\mu$m, the range covered by the NIRCam long-wavelength (LW) channels, and a flat SED in the short wavelength (SW) bands. \cite{2023Natur.616..266L} identified these SW-blue$+$LW-red sources as very massive, $M_\ast>10^{10}$~M$_\odot$, maybe significantly-obscured, $\mathrm{A(V)}>1.5$~mag, galaxies at $z=7-9$. They have been claimed to represent a new galaxy population revealed through the new capabilities provided by JWST,  easily reaching magnitudes down to 28-29 mag from 1 to 5\,$\mu$m. 


The existence of such early massive galaxies was quickly identified as a possible severe problem in our understanding of how galaxies form and evolve. Indeed, the large masses would be very difficult to explain with the current $\Lambda$CDM  paradigm, since the amount of baryons already transformed into stars could exceed their abundance in early halos \citep{2022ApJ...939L..31H,2023NatAs...7..731B,2023MNRAS.523.3201D}.

This high-mass interpretation was challenged by adopting models with prominent nebular emission, which could account for their red nature due to the non-negligible contribution of emission lines (and nebular continuum) to the broad-band fluxes as they enter the NIRCam passbands for different redshifts, implying significantly smaller (by a factor of 10 or more) stellar masses \citep[see][]{2023MNRAS.524.2312E,2023arXiv230605295E,2023ApJ...946L..16P,2023arXiv231003063D}. The estimated masses of early galaxies would also be reduced significantly if they had top-heavy stellar initial mass functions \citep{2023arXiv231018464W,2023arXiv231006781W}.

A simpler explanation in some cases is redshift errors or mis-identifications. Indeed, \citet{2023ApJ...954L...4K} presented spectroscopy for one of the claimed massive galaxies in \cite{2023Natur.616..266L}, showing an overestimation of the photometric redshift which could partly explain the high mass value for this and some other sources, as opposed to the more accurate photometric redshift and (10 times) smaller mass found in \citet{2023ApJ...946L..16P} for this same source. However, the general agreement among different authors in the determination of photometric redshifts for this type of source implies that redshift errors are not worrisome in a statistical sense (estimations agree well between papers for $\sim75$\% of the samples).
In addition, some purported LRDs (at the 10--20\% level) have been found to be brown dwarfs based on NIRCam colors, proper motions, and spectroscopy \citep{2023arXiv230903250H,2023ApJ...957L..27L,2023arXiv230812107B}.

Jointly with their colors, the extreme compactness of the {\it bona fide} high-redshift  objects led to the term Little Red Dots (LRDs, as first suggested by \citealt{2023arXiv230605448M}). Understanding LRDs has become more complex given that there is a variety of selection techniques that arrive at very similar types of objects in terms of compact morphology and red LW-to-SW colors, with high levels of overlap but also  ``contamination" from other types of sources (e.g., not so compact red galaxies or little not-so-red dots). \citet{2023Natur.616..266L} originally looked for sources with SED breaks, which resulted in a flat or blue SED at wavelengths shorter than $\sim2$~$\mu$m and a very red one at longer wavelengths to $\ge$ 5 $\mu$m (SW-blue$+$LW-red SEDs). Eventually arriving at similar objects, \citet{2023arXiv230514418B}, \citet{2023ApJ...956...61A}, or \citet{2023arXiv230607320L} used colors such as F277W-F444W in their selection, adding additional constraints in the SW channels. Some LRDs also entered in the selection carried out by \citet{2023ApJ...946L..16P}, \citet{2023MNRAS.522..449B} and \citet{2023arXiv231107483W} based on F150W-F356W and F150W-F444W (see also red sources selected with F200W-F444W in \citealt{2023MNRAS.518L..19R} and \citealt{2023A&A...676A..76B}), which were in some cases included in  larger investigations about the nature and cosmic relevance of ``HST-dark'' sources. Finally, complementary to purely photometric techniques, spectroscopic data with less comprehensive photometry  have also been able to identify LRDs, mainly based on the selection of strong H$\alpha$ or [OIII] emitters in blind NIRCam/grism spectroscopic surveys \citep{2023ApJ...950...67M,2023arXiv230605448M}.


The question about the general nature of LRDs
was further tangled by finding evidence for the presence of Active Galactic Nuclei (AGN). \citet{2023ApJ...954L...4K} identified a broad (1000~km~s$^{-1}$) component in the H$\alpha$ emission of one galaxy in the \citet{2023Natur.616..266L} sample, which implied the presence of a 10$^7$~M$_\odot$ super-massive black hole (SMBH). This large SMBH mass is consistent with the discoveries of high-redshift AGN in other studies, all indicating that black holes grew rapidly in the early Universe \citep{2021A&A...654A..37D,2023ApJ...953L..29L,2023arXiv230512492M,2023arXiv230605448M,2023A&A...677A.145U,2023arXiv230311946H,2023arXiv230802654N,2023NatAs.tmp..223B}, many lying above the \citet{1998AJ....115.2285M} relationship \citep{2023arXiv230812331P, 2023arXiv231018395S}.
 Further spectroscopic analyses identified AGN in some additional LRDs and stellar emission in others, even both at different wavelengths in the same LRD \citep{2023arXiv230605448M,2023arXiv230905714G,2023arXiv231203065K}. ALMA non-detections have also been used to defend the AGN nature of the bulk of the LRD emission  \citep{2023arXiv230607320L}.

Solving the challenges outlined above requires going further  into the mid-infrared where stellar emission and AGN can be distinguished. MIRI data at observed wavelengths shorter than 8~$\mu$m for a handful of LRDs 
was presented in \citet{2023ApJ...956...61A}, \citet[][also with a F1000W detection]{2023arXiv230514418B}, indicating that they could be either dusty starbursts (see also \citealt{2023MNRAS.518L..19R}) or obscured AGN, but no definitive conclusions could be reached  with just the bluest MIRI bands. \citet{2023arXiv231107483W} used longer wavelength MIRI data from the SMILES program and found that the averaged SEDs (stacked in observed bands) of the LRDs flattened beyond 5 $\mu$m, interpreting this behavior as the expected turnover of a normal stellar SED at rest $\sim1.6$~$\mu$m. In addition, and at odds with the AGN interpretation of the ALMA data of LRDs in the A2744 field presented by  \citealt{2023arXiv230607320L},  \citet{2023arXiv231107483W} found that even deeper ALMA observations of LRDs in GOODS-S could agree with SED models mostly dominated by stars (with some implications for the global dust emission SED shape). 

As is clear from the preceding discussion, the exact nature of the LRDs is still unclear, and there is no definitive proof that they are a single type of phenomenon, i.e., they could be a mixture of the various hypotheses advanced, their nature being also affected by selection biases. 


In this paper, we investigate the nature of LRDs using the best multi-wavelength, large area MIRI data available to date, those coming from the SMILES survey \citep{2023arXiv231012330L}. MIRI, in fact, is essential to disentangle the nature of LRDs. If dominated by star formation, the SEDs of LRDs should show a 1.6\,$\mu$m bump, redshifted to $\sim10-15$~$\mu$m at $z=5-8$ \citep[e.g.][]{2023arXiv231107483W}. On the other hand, an obscured AGN will yield a steep SED. The significant uncertainty in the relative importance of star formation and nuclear activity needs both spectroscopic \citep[e.g.,][]{2023ApJ...954L...4K,2023arXiv231203065K} and photometric data to cover the widest spectral range possible, as well as a careful SED analysis of a statistically representative sample. In modeling these sources, we will also explore the uncertainties by using four different sets of modeling software with a variety of approaches and assumptions. 

The main goals of this paper are to select and study LRDs in a comprehensive way, including examples representing a range of properties, and to see how our conclusions about them are influenced by different elaborated modeling approaches. To achieve these ends, the paper is organized as follows. Section~\ref{sec:sample} presents the NIRCam and MIRI data used to select and characterize a sample of LRDs in the GOODS-S field. Section~\ref{sec:avesed} describes the average SEDs of LRDs and  Section~\ref{sec:modeling} presents stellar and AGN emission models that will be used to characterize the physical properties of LRDs on galaxy-by-galaxy and statistical bases in Section~\ref{sec:properties}. Finally, Section~\ref{sec:conclusions} discusses and summarizes our conclusions. We also include in this paper three appendices that describe in detail the MIRI data reduction, the dust emission models that are the most important ingredients of the SED modeling, and provide SED fits for the whole sample. 

Throughout the paper, we assume a flat cosmology with $\mathrm{\Omega_M\, =\, 0.3,\, \Omega_{\Lambda}\, =\, 0.7}$, and a Hubble constant $\mathrm{H_0\, =\, 70\, km\,s^{-1} Mpc^{-1}}$. We use AB magnitudes \citep{1983ApJ...266..713O}. All stellar mass and SFR estimations assume a universal \citet{2003PASP..115..763C} IMF, which is a very relevant assumption for these parameters (but might not be accurate given the extreme conditions).

\section{Data and sample selection}
\label{sec:sample}

\begin{figure*}[htp!]
\centering
\includegraphics[width=18cm,angle=0]{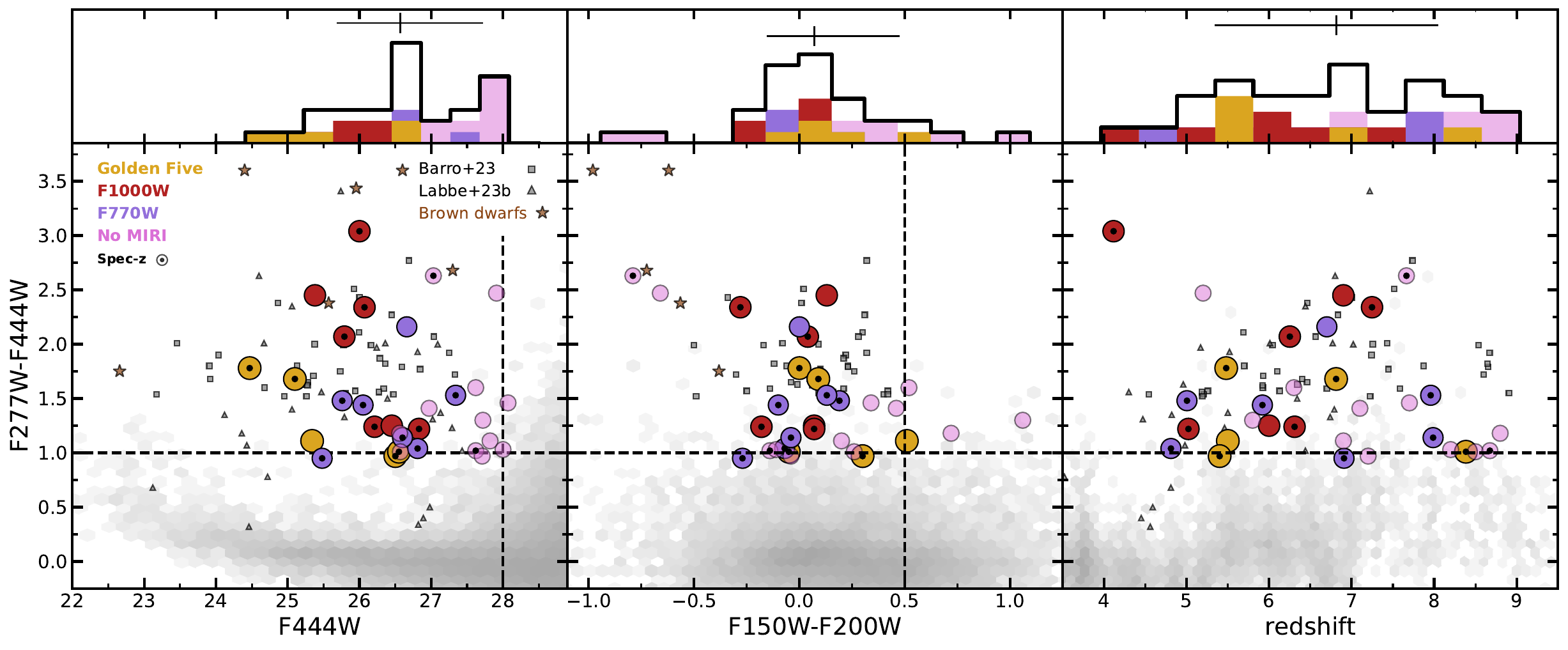}
\caption{\label{fig:selection1}The left and central panels show the $F277W-F444W$ vs$.$ $F444W$ color-magnitude and 
$F277W-F444W$ vs. $F150W-F200W$ color-color diagrams, as well as histograms of NIRCam colors, indicating the selection thresholds for LRDs ($F277W-F444W>1$~mag, $F150W-F200W<0.5$~mag and $F444W<28$~mag; dashed lines) relative to the bulk of the JADES DR2 galaxy catalog. The different colors indicate the subsets of LRDs detected in different MIRI bands: up to F1280W and beyond (Golden Five galaxies), up to F1000W, up to F770W, or not detected in MIRI (at the SMILES depth). Comparison LRD samples from \citet{2023arXiv230514418B} and \citet{2023arXiv230607320L} in the CEERS and UNCOVER fields  are shown with squares and triangles, respectively. The right panel shows the color $vs.$ redshift diagram for the LRDs and JADES galaxies and the redshift distribution of LRDs (including median and quartiles). The 55$\%$ of the LRDs in our paper that have secure NIRSpec- and NIRCam-based spectroscopic redshifts are marked with a black dot.} 
\end{figure*}

\subsection{NIRCam-based selection}

The sample of LRDs analyzed in this paper has been gathered from the JWST Advanced Deep Extragalactic Survey, JADES \citep{2023arXiv230602465E} Data Release 2 catalogs (\citealt{2023arXiv231012340E}), which also gathered data from the First Reionization Epoch Spectroscopically Complete Observations (FRESCO, \citealt{2023MNRAS.525.2864O}) and the JWST Extragalactic Medium-band Survey (JEMS, \citealt{2023ApJS..268...64W}). We also used the \citet{2023arXiv230602466R} catalog for spectroscopic redshifts, which included values measured with data taken by NIRCam \citep{2023MNRAS.525.2864O} and NIRSpec \citep{2023arXiv230602467B}. Given the small (nearly point-like) nature of our sources of interest and the fact that eventually we wanted to analyze SEDs including MIRI data, we used 0.25\arcsec\, radius PSF-matched photometry as our fiducial aperture for selection, but also checked the results using smaller radii, more suitable for NIRCam only. We searched for NIRCam SW-blue$+$LW-red sources defined by F277W-F444W$>$1~mag and F150W-F200W$<$0.5~mag colors, and magnitudes F444W$\leq$28~mag, following the strategies presented in \citet{2023arXiv230514418B}, \citet{2023arXiv230607320L} and \citet{2023arXiv230905714G}. The two latter references further use a concentration criterion to select compact sources, but we found no need for it after applying the F150W-F200W$<$0.5~mag cut. In any case, the typical concentration of our original sample is indicated by the flux ratio (after applying the appropriate aperture corrections) of  F(F444W)$_{r=0.25\arcsec}$/F(F444W)$_{r=0.10\arcsec}$$=1.04_{0.98}^{1.10}$ (median and quartiles). That is, 
most of the sources are nearly point-like.


The selection procedure is summarized in Figure~\ref{fig:selection1}, where we compare our sample with others found in the literature. Overall, our color selection criterion is bluer than that used in \citet{2023arXiv230514418B}, who imposed a redder color cut (F277W-F444W$>$1.5~mag instead of 1.0~mag), and very similar to the one used for the sample presented in \citet{2023arXiv230607320L}, although we impose a magnitude cut $\sim0.5$~mag deeper in F444W.

\begin{figure*}[htp!]
\centering
\epsscale{1.0}
\plotone{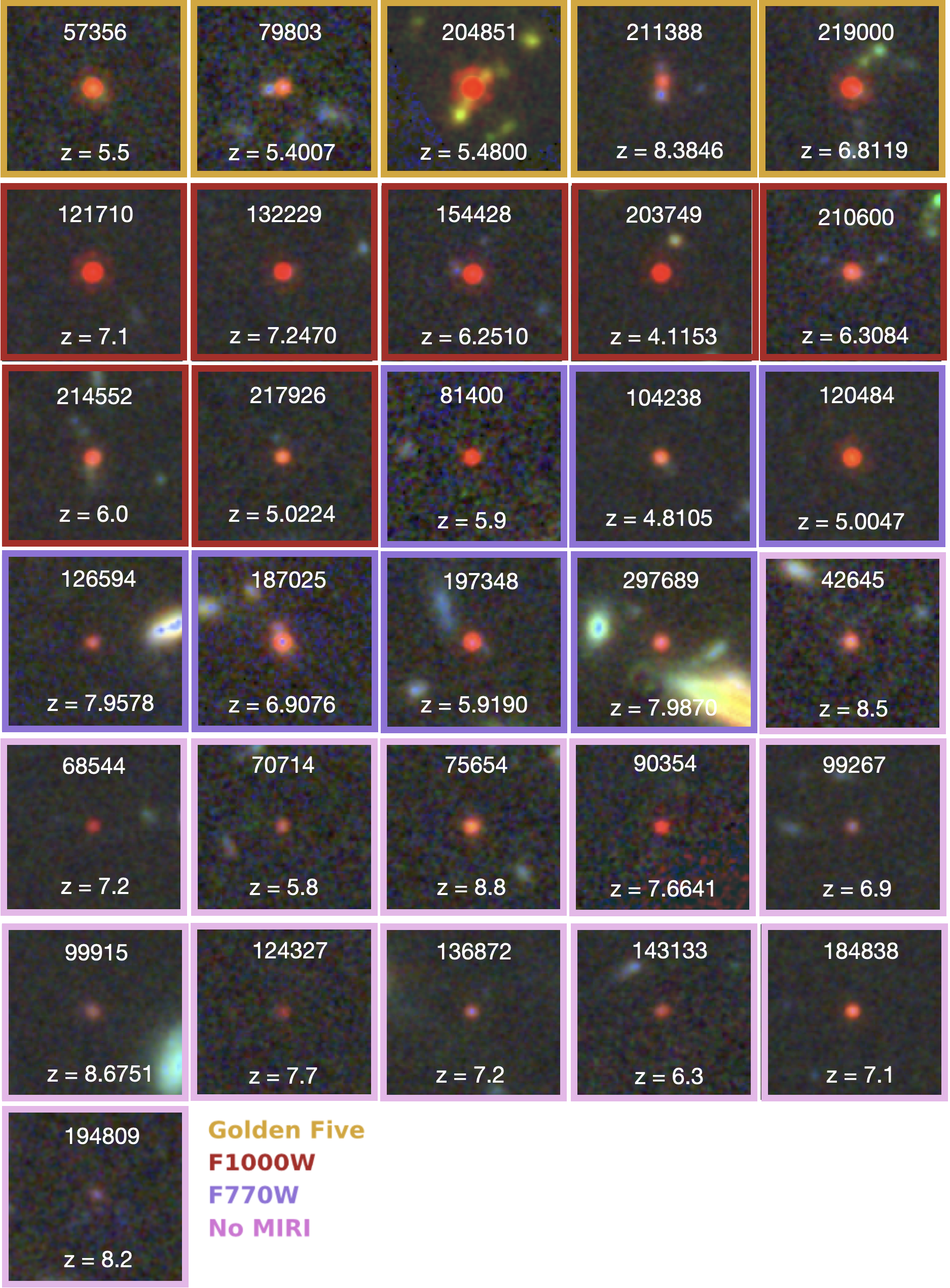}
\caption{\label{fig:selection2}NIRCam F150W+F277W+F444W color composite postages of the sample of 31 LRDs in this paper. North is up, East is left, sizes are 2.5$\times$2.5 arcsec$^2$. Galaxies detected by MIRI at wavelengths longer than F1280W (Golden Five galaxies), F1000W, and F770W are marked in gold, red, and purple (and the sources are ordered according to this MIRI detections). Those not detected in MIRI are marked in magenta. We display redshifts, with spectroscopic values written with 4 decimals.}
\end{figure*}

We further restrict the sample to the area covered by MIRI within the Systematic Mid-infrared Instrument Legacy Extragalactic Survey (SMILES, \citealt{2023arXiv231012330L}), arriving at a sample of 37 galaxies. As shown by \citet{2023arXiv230903250H}, this type of color selection of high redshift galaxies might be contaminated by brown dwarfs. To account for that, and following their analysis, we removed from the original sample the 4 sources with color F115W-F150W$<$--0.5~mag. Two other LRDs were removed since they were identified with brown dwarfs with proper motion \citep{2023arXiv230903250H}. Our final sample contains 31 galaxies detected in the SMILES 34.9~arcmin$^2$ area, i.e., the surface density is 0.9~LRD\,arcmin$^{-2}$. Postage stamps in RGB format constructed with NIRCam data are shown in Figure~\ref{fig:selection2}. Table~\ref{tab:sample} provides all relevant information about the selection of our sample of LRDs.

Our final sample includes 14 sources studied in \citet{2023arXiv231107483W}, as well as 2 sources included in the sample of high redshift obscured AGN candidates presented in \citet{2023arXiv231012330L}, one source identified as an AGN in \citet{2023arXiv230605448M}, and 5 galaxies selected as high-redshift candidates in \citet{2023arXiv230602468H}. Compared to the galaxies in common with \citet{2023arXiv231107483W}, the sample in this paper is 0.5~mag fainter in F444W (medians and quartiles $26.5_{26.0}^{27.5}$~mag $vs.$ $26.0_{25.4}^{26.9}$~mag), and  bluer in F277W-F444W ($+1.3_{+1.1}^{+1.7}$~mag  $vs.$ $+1.6_{+1.5}^{+2.5}$~mag) and F150W-F200W ($+0.0_{-0.1}^{+0.2}$~mag  $vs.$ $+0.2_{+0.0}^{+0.3}$~mag). The sources in common with \citet{2023arXiv230602468H} are among the faintest (median F444W 27.1~mag), slightly bluer (median F277W-F444W and F150W-F200W 1.1~mag and 0.0~mag, respectively), and at the highest redshifts in our sample (all with $z>8.2$, median 8.4).

\subsection{Mid-infrared properties}
\label{sec:sample0}

In this section, we discuss the detections of our sample of LRDs in the MIRI data gathered by  SMILES \citep{2023arXiv231012330L}. We first describe briefly this dataset, and then discuss the detection fractions in MIRI as a function of wavelength.  


\subsubsection{MIRI data description}

The MIRI data gathered by the SMILES survey (program ID 1207, PI: G. Rieke, \citealt{2023arXiv231012330L}) were reduced with {\it JWST} pipeline version 1.12.3, reference files in pmap version 1138, which includes improved absolute photometric calibration and aperture corrections based on a better characterization of the PSFs, especially at the shortest wavelengths. 

Apart from the official pipeline steps, our reduction includes a special treatment of the background to deal with artifacts seen in the MIRI data (e.g., striping, tree rings, and inhomogeneities in rows and columns). This is achieved with a super-background strategy that was explained briefly in \citet{2023A&A...671A.105A} and \citet{2023arXiv231012330L}. Given the importance of reaching the deepest fluxes possible in the analysis of high redshift sources presented in this paper, we explain and characterize in detail the bespoke MIRI reduction procedures in Appendix~\ref{app:mirireduction}.

Photometry in the MIRI bands was performed in circular apertures with radii of 0.2\arcsec, 0.3\arcsec, 0.4\arcsec, and 0.5\arcsec, applying the corresponding aperture corrections to each measurement. Uncertainties were calculated following the procedure explained in \citet{2023ApJ...951L...1P}, which gathers non-contiguous (i.e., independent) pixels to avoid the effect of correlated noise. Different measurements for our sample of LRDs agreed within 0.1~mag and we eventually chose for each band the weighted mean of all fluxes, typically the one corresponding to $\sim$70\% encircled energy for each band. The photometry was revised visually to avoid contamination from cosmic ray showers, which we found to affect one source. We only kept fluxes with S/N$>$5 and considered 5$\sigma$ upper limits otherwise.

\subsubsection{MIRI detections}

Out of the 31 LRDs selected with the JADES NIRCam data down to F444W$=$28~mag, we detect [19,22,12,7,4,2,1,0] galaxies for MIRI bands  [F560W,F770W,F1000W,F1280W,F1500W,F1800W, F2100W,F2550W] respectively. The corresponding detection fractions (in \%) are [61,71,39,23,13,7,3,0] at $>5\sigma$ levels of [26.1,25.8,24.7,24.3,24.2,23.0,22.6,20.8]~mag (these magnitudes are measured in circular apertures of radius equal to the PSF FWHM, corrected to infinite extent).
The detection fractions depend strongly on F444W magnitude. For the brightest galaxies, F444W$<$26.5~mag,  the detection fractions are (in \%) [93,100,73,47,20,13,7,0] in [F560W,F770W,F1000W,F1280W,F1500W,F1800W, F2100W,F2550W], respectively. 

We divided the sample of LRDs into 4 subsets detected up to a given MIRI band: (1) 5 galaxies detected in F1280W and longer wavelengths, hereafter the Golden Five; (2) 7 galaxies detected only up to F1000W (i.e., excluding the previous type); (3) 7 sources detected only up to F770W (excluding the previous types); (4) and, lastly, the 12 galaxies not detected in any MIRI band. The panels and histograms in Figure~\ref{fig:selection1}, as well as the postage stamps in Figure~\ref{fig:selection2}, show these subsets with different colors. The first figure illustrates that the MIRI detection fraction depends strongly on the F444W magnitude and it is nearly independent of the F150W-F200W color. Interestingly, the F277W-F444W color does exhibit some differences between subsets, with the Golden Five being among the bluest, and the F1000W sample among the reddest (we will comment on this in the following sections). It is worth mentioning that only two of the Golden Five galaxies would be identified as LRDs under a more restrictive color selection of F277W-F444W$>$1.5~mag.

\subsection{Redshifts}

A total of 18 galaxies out of the 31 in the whole sample (55\%) have  spectroscopic redshifts provided by CANDELS \citep{2013ApJS..207...24G}, FRESCO \citep{2023MNRAS.525.2864O} data (measured by the JADES team) and JADES NIRSpec data \citep{2023arXiv230602467B,2023arXiv231012340E}, all of them flagged as highly secure. 

Photometric redshifts for all galaxies were estimated with {\sc eazy} \citep{2008ApJ...686.1503B}, using either direct flux measurements for all bands or with a modified version of the program that penalizes solutions implying fluxes above upper limits. We used v1.3 templates, which include a dusty galaxy and a high-EW emission-line galaxy spectrum, and we added a new extreme emission-line galaxy template as well as an AGN$+$torus model (based on the spectrum in \citealt{2023arXiv231203065K}). Based on the comparison with spectroscopic values, we chose the maximum $\chi^2$ photometric redshift as our main solution. We estimated redshifts both using only NIRCam fluxes and also adding MIRI measurements, obtaining better results when using NIRCam only (a conclusion based on the comparison with spectroscopic values). 

The quality of the photometric redshifts, when compared to spectroscopic values, is remarkably high, partly as a consequence of the strong emission lines present in many of the LRDs in our sample. The average absolute difference between photometric and spectroscopic values is 0.01, and we do not have any outliers (defined as galaxies with $\Delta(z)/(1+z)>0.1$).

A histogram of the redshift distribution of our sample is given in Figure~\ref{fig:selection1}. The distribution is relatively flat between $z\sim5$ and $z\sim9$, peaking at $z\sim7$. This behavior is probably related to the fact that strong emission lines entering the F444W filter affect the selection significantly (as also seen in extreme emission line galaxies included in \citealt{2023ApJ...946L..16P}). These lines are mainly [OIII]+H$\beta$ and H$\alpha$+[NII]+[SII], which translate to $z\sim8$ and $z\sim6$ for the lines lying within the F444W passband. Summarizing, the typical redshift (median and quartiles) of LRDs selected down to F444W$\sim$28~mag is $z={6.9}_{5.9}^{7.7}$.

\subsection{Additional SED constraints}

Apart from the NIRCam and MIRI fluxes, we used {\it Hubble} ACS fluxes measured in Hubble Legacy Field images \citep{2016arXiv160600841I,2019ApJS..244...16W}. Except for one source, not covered by the NIRCam SW channels, we did not use WFC3 data. We checked that none of the LRDs in this paper are detected by {\it Spitzer} with the MIPS instrument or by Herschel with PACS and SPIRE. Taking into consideration the catalogs used in \citet{2019ApJS..243...22B}, we imposed in our SED analysis 5$\sigma$ upper limits of 30~$\mu$Jy for the MIPS 24~$\mu$m band, and  4, 5, 9, 11, and 11 mJy for Herschel bands at 100, 160, 250, 350, and 500~$\mu$m, respectively. 


\section{Spectral energy distributions of LRDs}
\label{sec:avesed}

\begin{figure*}[htp!]
\centering
\hspace{-0.5cm}
\includegraphics[clip, trim=0.1cm 0.0cm 1.6cm 1.0cm,width=8.8cm,angle=0]{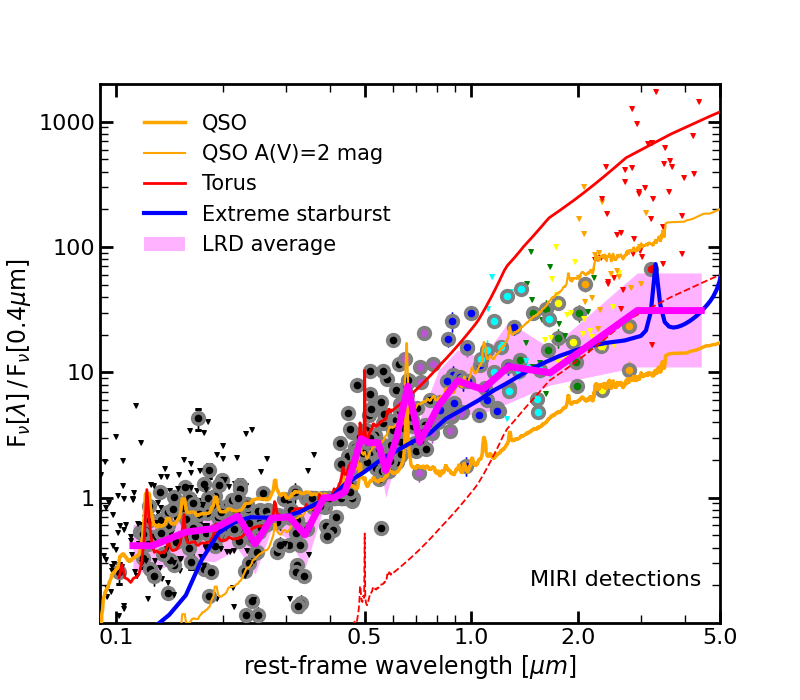}
\includegraphics[clip, trim=0.1cm 0.0cm 1.6cm 1.0cm,width=8.8cm,angle=0]{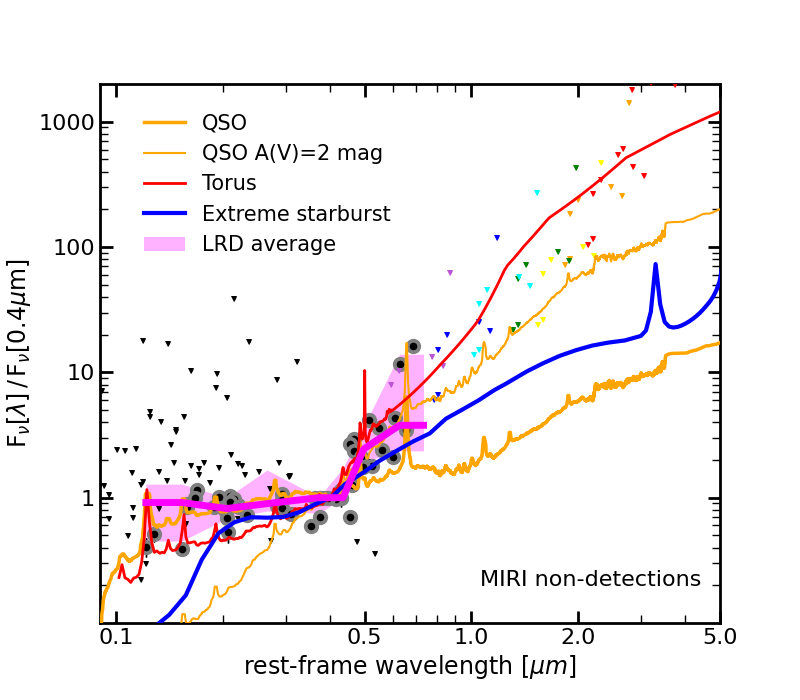}
\caption{Stacked LRD SEDs for sources detected by MIRI (left panel) and not (right panel), normalized at rest-frame wavelength 0.4~$\mu$m. Black points show NIRCam fluxes for individual sources, while rainbow color points show MIRI fluxes. Arrows depict 5$\sigma$ upper limits. The average SED is shown with a magenta line (10 point averages) and its rms with a magenta shaded region. The observed average SED for LRDs is compared to 5 different templates. The orange lines show an average QSO spectrum (see text for details), and the same template extincted by $\mathrm{A(V)}=2$~mag using a \citet{2000ApJ...533..682C} attenuation law. The red lines stand for the torus template in \citet{2007ApJ...663...81P}, normalized to the same wavelength as the observations (continuous line) as well as normalized at the 2~$\mu$m average SED level (dashed line) in the case of the MIRI detections plot. The blue line shows the model for an intense starburst presented in \citet{2007A&A...461..445S}, more specifically, the sub-LIRG model with a size of 350~pc, 90\% of the total luminosity coming from OB stars, optical attenuation of $\mathrm{A(V)}=36$~mag, and $10^3$~cm$^{3}$ density of dust in hot spots (gas clouds surrounding and directly heated by OB stars).}
\label{fig:average_sed}
\end{figure*}

\subsection{Average spectral energy distributions}
\label{sec:sample3}

Figure~\ref{fig:average_sed} shows the rest-frame SEDs, normalized at 0.4~$\mu$m, and their average (more specifically, the weighted mean every 10~points calculated on a wavelength-ordered flux list) for the LRDs detected by MIRI and those with just upper limits. The data are compared qualitatively with some relevant templates to understand what the MIRI observations are telling us about the general nature of LRDs. 

Overall, the plots highlight very clearly the characteristic flat, blue UV continuum plus steep, red optical continuum of LRDs. Such peculiar SEDs are difficult to model with traditional stellar population models under conventional assumptions and therefore require more complex or unusual treatment (to avoid biases in the determination of stellar masses, for example).  Recent works have put forward different solutions to try to explain the emission of the LRDs using: strong, high EW ($>$1000\AA) emission lines that boost even the broad-band photometry \citep{2023MNRAS.524.2312E,2023arXiv230605295E,2023arXiv230605448M,2023MNRAS.519.3064F,2023ApJ...946L..16P,2023arXiv231003063D}, complex stellar models with a flexible treatment of the dust attenuation \citep{2023Natur.616..266L,2023arXiv230514418B,2023ApJ...956...61A,2023arXiv231107483W}, or hybrid models that combine stellar and AGN-driven emission with either component dominating different spectral regions or the full SED \citep{2023ApJ...954L...4K,2023arXiv230514418B,2023arXiv230607320L,2023arXiv230905714G}. 

The fully AGN-dominated models have received more attention lately based on the spectroscopic observations of LRDs showing strong, and sometimes broad emission lines (FWHM$>$2000 km/s; \citealt{2023ApJ...954L...4K,2023arXiv230311946H,2023arXiv230605448M,2023arXiv230905714G,2023arXiv230801230M}). Briefly, these models combine AGN continuum emission from 1) a highly obscured accretion disk, 2) a small fraction ($\sim3\%$) of unobscured, scattered light, and 3) the dust emission of the surrounding torus (overall similar to the \citealt{2007ApJ...663...81P} torus template shown in Figure~\ref{fig:average_sed}). This model also explains the presence of both narrow and broad emission lines because there is a direct sight line (albeit obscured) to the Broad Line Region, and it predicts that the red optical colors extend toward the NIR, which is roughly consistent with results showing that the handful of LRDs observed in the short-wavelength MIRI bands (F560W and F770W) continue the steep SED trend \citep{2023arXiv230514418B}.

Figure~\ref{fig:average_sed} indicates that the stacked SED of the SMILES LRDs agrees well with the results from previous works based on NIRCam data: the rest-frame UV up to 0.4~$\mu$m nicely matches the emission of an unobscured, or slightly reddened, QSO, which is consistent with the low-luminosity scattered light component of the accretion disk. Remarkably, the average SED presents a bump around the MgII emission feature, indicating that the AGN makes a non-negligible  contribution to the rest-frame UV spectral range. However, the scatter around the median suggests that there might be sources other than an AGN contributing to the UV, e.g., emission from stars in the host galaxy. 

The rest-frame optical presents a very steep slope from 0.4 to 1 $\mu$m consistent with the emission of a heavily obscured supermassive black hole, here indicated with the \citet{2007ApJ...663...81P} circumnuclear torus template. In addition, aided by the inclusion of the JEMS medium-band photometry, the stack exhibits statistically significant peaks at the wavelengths where we would expect strong H$\alpha$+[NII]+[SII] and [OIII]+H$\beta$ emission. The average LRD SED might be identified with a torus up to $\sim0.7\,\mu$m; although the torus does not have optical emission lines, the lines from the central engine can  masquerade as a steeper (dustier) continuum. In summary, up to the reddest bands covered by NIRCam (F444W, F460M, and F480M filters, typically probing up to H$\alpha$), the average UV-optical SED presented in this paper roughly matches an obscured  QSO spectrum plus a torus.


However, the interpretation of LRD SEDs being dominated by a pure and/or complex (with a gray attenuation law) AGN model is not supported when we consider the rest-frame near-infrared (NIR) spectral region probed by MIRI, the key addition in this paper, as first presented in \citet{2023arXiv231107483W}. The MIRI data (colored circles in Figure~\ref{fig:average_sed}) and average SED are consistently lower than the predictions of the \citet{2007ApJ...663...81P} torus template. A more general evaluation can be made by applying different reddening levels to a standard unreddened AGN template. In the $\sim$0.5--3~$\mu$m  region where the MIRI measurements fall, there is general agreement on the shape of the intrinsic template, as determined by averaging the behavior of large samples \citep[e.g.,][]{elvis1994,richards2006,2006ApJ...640..579G,Lyu2017}. \citet{lyu2022c} show that this is an appropriate average behavior including the variations such as warm dust deficient and hot dust deficient cases \citep{Lyu2017}. It is therefore the appropriate foundation for fitting the LRD photometry, but with the large attenuations required to match the SED slope. To be specific, we adopt a SED from \citet{2006ApJ...640..579G}, extended to the mid-infrared (MIR) with models from \citet{2015A&A...583A.120S}. Figure~\ref{fig:average_sed} shows that there is no suitable solution; if reddening is selected to match the behavior between 0.5 and 1 $\mu$m, the SED falls substantially higher than the MIRI photometry near 2 $\mu$m (rest). Reducing the reddening to solve this problem yields a SED too low near 1 $\mu$m. Overall, the MIRI photometry indicates a change to a significantly bluer slope at $\sim$ 1 $\mu$m and even stabilizes at a roughly flat value at 1-2~$\mu$m, where stellar emission is expected to peak. This is a fundamentally different shape from AGN SEDs. There are also hints of another steepening at $>3\mu$m, but this behavior is traced only by a small subset of LRDs detected beyond 1.6~$\mu$m rest-frame (see discussion in \S~\ref{sec:dust_emission}); deeper MIRI data at the longest wavelengths would be necessary to probe this region.


The comparison of photometry and templates in Figure~\ref{fig:average_sed} strongly suggests that the rest-NIR emission of the LRDs (around 1-2~$\mu$m) is very different from the hot-dust (T$\sim$1500~K) dominated emission of the typical QSO templates used in recent LRD papers dominated by the fit of NIRCam data alone (e.g., \citealt{2023arXiv230514418B}, \citealt{2023arXiv230607320L}, or \citealt{2023arXiv230905714G}). Reproducing the observed MIRI colors would require a more flexible modeling of the torus emission that can vary its IR luminosity (i.e., the relative flux level with respect to other components such as stars or the accretion disk) and the wavelength at which the emission peaks. This can be accomplished using either radiative transfer models
\citep[e.g.,][]{2008ApJ...685..160N,2012MNRAS.420.2756S,2016MNRAS.458.2288S,2015A&A...583A.120S}, 
modified blackbody templates (\citealt{kim15}, \citealt{2016MNRAS.463.2064H} or \citealt{2023arXiv231203065K}), or empirical templates that can account for separate contributions of
the torus dust emission at different temperatures (e.g., warm and hot, as in \citealt{Lyu2017}). However, such models require significant departures from typical AGN behavior in the red and near infrared. Alternatively, the result could indicate that the SED at 0.5--2.0~$\mu$m is not AGN-dominated but rather stellar-dominated, as recently discussed in \citet{2023arXiv231107483W}. We demonstrate this point by plotting in Figure~\ref{fig:average_sed} the torus template normalized to the average SED level of LRDs at 2~$\mu$m (dashed red line). If the emission at rest-frame wavelengths longer than 2~$\mu$m is dominated by a torus, then the flux at around 1~$\mu$m must be dominated by a different component, such as the accretion disk or, more likely, stars, whose emission peaks exactly in that spectral zone (except for very young ages).


The right panel of Figure~\ref{fig:average_sed} shows the average SED of LRDs that are not detected by MIRI. The UV range is bluer than in the MIRI-detected case, more in line with the slope of the average QSO spectrum. The change in slope at 0.4~$\mu$m is also clear (this being one of the main selection criteria of our sample). Strong emission lines might also be present; indeed we observe a similar SED rise in the [OIII] region, and two NIRCam points may be revealing  strong H$\alpha$ emission. However, the small number of sources does not support detecting emission lines in the average SED as well as in the case for MIRI detections. For these sources, the MIRI upper limits also imply that the SED flattens at $\sim1$~$\mu$m, i.e., stellar emission could dominate this spectral range. However, the SED possibly has a steepening compatible with the presence of strong hot dust emission at wavelengths around 2~$\mu$m and redward; the currently available MIRI data cannot constrain this spectral region appropriately.

\subsection{MIRI colors}
\label{sec:sample6}

\begin{figure*}[htp!]
\centering
\hspace{-0.8cm}
\includegraphics[clip, trim=4.0cm 2.0cm 1.cm 1.0cm,width=19.5cm,angle=0]{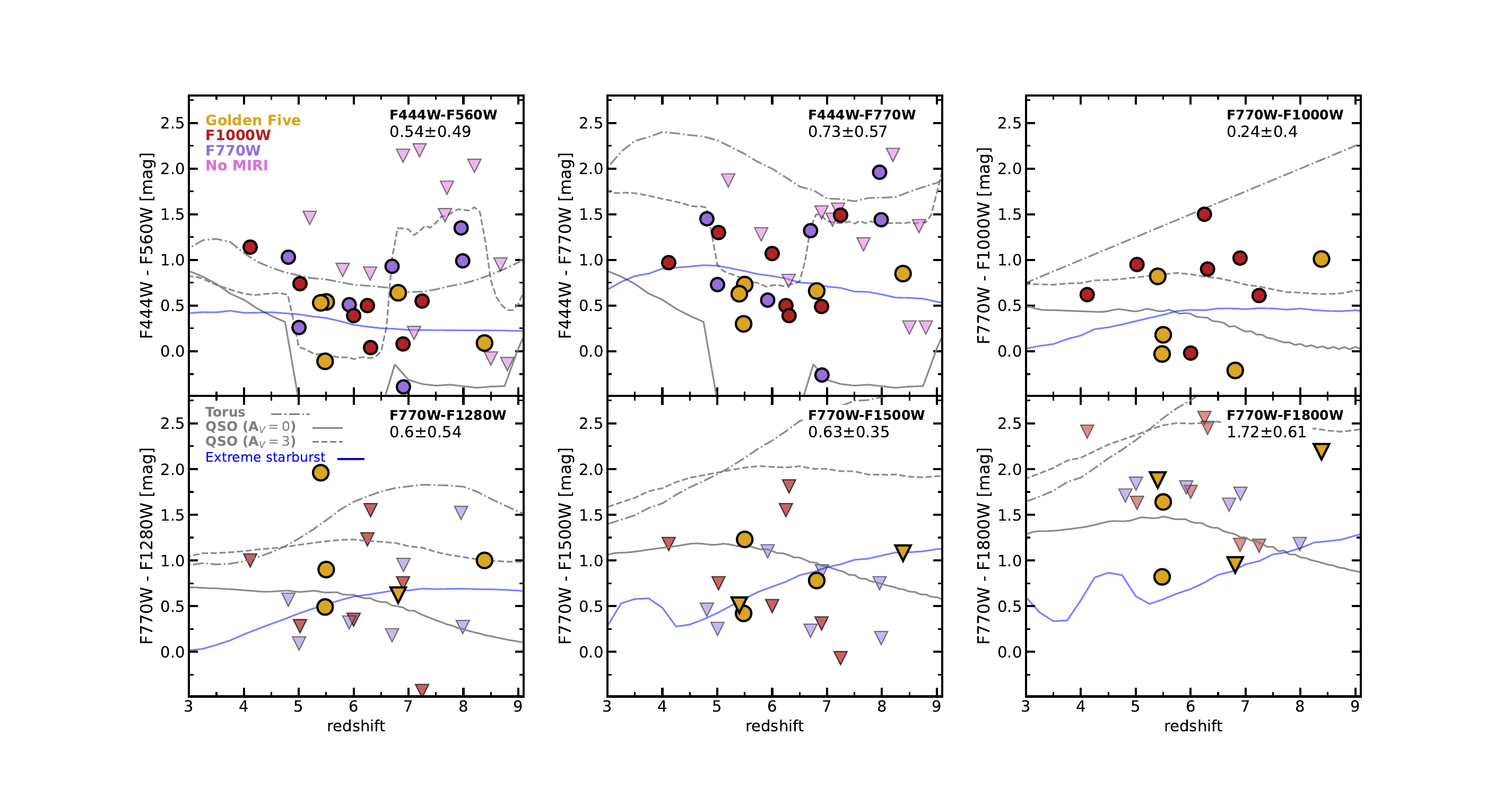}
\caption{NIRCam and MIRI colors $vs.$ redshift for all LRDs detected at least in F770W. The lines illustrate the color-redshift tracks for the same templates shown in Figure~\ref{fig:average_sed}. Galaxies from different samples are marked with different colors. The first two panels in the top row show the F444W-F560W(F770W) colors which are available for the majority of the sources. The last panel at the top and the bottom panels show the MIRI-MIRI colors relative to F770W. The median and scatter of the colors are indicated in the top left corner.}
\label{fig:colors}
\end{figure*}




Figure~\ref{fig:colors} illustrates further what the MIRI observations reveal about the nature of LRDs, probing further into the redshift effects. Two of the top panels show the expected NIRCam-MIRI colors for different types of templates as a function of redshift. The nature of the LRD emission in the spectral range probed by the F444W-F560W or F444W-F770W colors cannot be disentangled: both AGN and stellar dominated emission would present very similar color differences of about 0.5~mag, and the influence of emission lines in one and/or the other filter at specific redshifts complicates the problem. Only when observing at longer wavelengths with MIRI, 10--15~$\mu$m, would the data start to differentiate between the models. For our detections at these wavelengths, the MIRI data reveal flat SEDs, F770W-F1000W, and F770W-F1280W values, bluer than 0.5~mag for 60\% of the sample and consistent with stellar dominated emission presenting a 1.6~$\mu$m bump. We note however that stellar emission alone leads to flat or even negative MIRI colors, whereas the median colors of the LRDs, including the upper limits, show small, but increasingly redder colors with wavelength which suggest that there is emission from another source, such as nebular continuum or dust, but not as dominant as in the QSO templates (see also discussion in \S~\ref{sec:dust_emission}).

In summary, as found in the preceding section, this more detailed analysis shows how SEDs constructed up to the reddest NIRCam bands are compatible with a dominant AGN with composite emission: blue in the UV, with emission lines coming from the broad-line region, and starting to rise as expected for dust torus emission up to 5~$\mu$m. MIRI data at 5.6~$\mu$m deviates very little from this scenario, but at 7.7~$\mu$m starts to tell a different story. Even longer wavelength data clearly point to a restricted contribution of a dust torus at rest-frame wavelengths around 1--2~$\mu$m, where the stellar emission is expected to peak.

Both in Figures~\ref{fig:average_sed} and \ref{fig:colors}, we show one of the radiative transfer models of dust-rich compact nuclear starbursts and luminous infrared galaxies (LIRGs) presented in \citet{2007A&A...461..445S}. In particular, we plot (in blue) the template for an extreme starburst with a 350~pc star-forming region with a sub-LIRG-like luminosity ($10^{10.1}$~L$_\odot$; but LIRG-like luminosity density), with 90\% of its luminosity arising from OB stars embedded in very thick [$\mathrm{A(V)}=36$~mag] and dense ($10^3$~cm$^{-3}$ hydrogen density) birth clouds. Remarkably, this stellar-only model, mainly based on the presence of highly embedded OB stars within a more general stellar population, nicely reproduces the main characteristics of LRDs across the whole spectrum: the change in spectral slope at rest-frame $\sim0.4$~$\mu$m and the flattening of the SED probed by MIRI. It, however, presents a steeper slope in the FUV, but we remark the stellar emission is constant across all these models and does not take into account nebular emission. The redshift dependence of the color of this template, shown in Figure~\ref{fig:colors}, is also completely consistent with the measurements.

With the general trends discussed with  Figures~\ref{fig:average_sed} and \ref{fig:colors} in mind, in the following section we investigate several different new scenarios to model the NIRCam+MIRI SEDs of each LRD in our sample.



\section{SED modeling codes for LRDs}
\label{sec:modeling}

In this section, we present the detailed SED fitting methods that will be applied to the analysis of individual objects in our sample to infer their physical properties. The results will be presented in the next section, concentrating on those galaxies that have spectroscopic redshifts and are detected in several MIRI bands. The SEDs of all our LRDs were fitted with 4 codes, each one with different assumptions and approaches: \textsc{synthesizer-AGN}, \textsc{prospector-SF}, \textsc{prospector-AGN}, and \textsc{prospector-AGN$+$}. 

\subsection{The \textsc{synthesizer-AGN} code}

For the \textsc{synthesizer-AGN} code \citep{2003MNRAS.338..508P,2008ApJ...675..234P} we assumed a composite stellar population with both a young and a more evolved starburst. The stellar emission is described by the \citet{2003MNRAS.344.1000B} models, assuming a \citet{2003PASP..115..763C} IMF with stellar mass limits between 0.1 and 100~$\mathrm{M}_\odot$. Both star-forming events are characterized by a Star Formation History (SFH) described by delayed exponential function with timescales between 1~Myr and 1~Gyr, and with ages from 1~Myr up to the age of the Universe at the redshift of the source. Nebular emission is also considered (see \citealt{2003MNRAS.338..508P} for details). The attenuations of the stellar and nebular emission are described by the \citet{2000ApJ...533..682C} law, with A(V) values ranging from 0 to 4~mag for each population, considered to have completely independent attenuations.    The dust emission is also modeled (beyond what the AGN might contribute) with the radiative transfer templates of nuclear starbursts  presented in \citet{2007A&A...461..445S}. We use the templates for 350~pc diameter star-forming regions, with different SED shapes parametrized by the total IR luminosity (from sub-LIRG to Hyper-LIRG ranges), visual extinction of the starburst (from 2 to 144~mag), ratio to the total luminosity (from 40-90\%) of the emission arising from OB stars embedded in dense molecular clouds of a variety of hydrogen number densities (from $10^2$ to $10^4$ cm$^3$, assuming a gas-to-dust ratio of 150).

Apart from the stellar population, we include an AGN component described by a QSO template constructed with the average spectrum of QSOs presented in \citet{2001AJ....122..549V} and \citet{2006ApJ...640..579G}, extended to the far-IR with the dust torus template in \citet{2015A&A...583A.120S}. This model was shown in Figure~\ref{fig:average_sed}. The attenuation of the AGN model is also assumed to follow the \citet{2000ApJ...533..682C} law, and we impose an energy balance criterion so the energy absorbed by dust is reradiated in the mid- and far-IR, scaling up the dust torus emission. 

There are 9 free parameters to describe the stellar population. Apart from the 4 parameters for each stellar population, i.e., age, timescale, attenuation and metallicity, we fit the stellar mass ratio between the youngest stars and the total mass, the burst strength.  We add to the 9 stellar parameters, 2 more that describe the scaling of the AGN template with respect to the stars and its attenuation.

We remark that \textsc{synthesizer-AGN} assumes independent extinctions for the 2 stellar populations in the host (as well as for the AGN). This is different from what is done by the next codes, where the attenuation for the stellar emission is governed by common parameters (visual attenuation and attenuation law slope), although internally young stars are treated differently from old stars.

\subsection{The \textsc{prospector-SF} code}

For \textsc{prospector-SF} (\citealt{leja17}, \citealt{2021ApJS..254...22J}), we use the MIST stellar evolutionary tracks and isochrones \citep{choi16}, a \citet{2003PASP..115..763C} IMF, a range in stellar metallicity between -1.0 and 0.19, and gas phase metallicity $\log (\mathrm{Z}/\mathrm{Z}_\odot)$ between -2.0 and 0.5. For the SFH, we use a non-parametric model, following the flexible SFH prescription \citep{leja19} with 6 time bins and the bursty-continuity prior \citep{tacchella22a}. The ionization parameter for the nebular emission ranges from $\log \mathrm{U}$ =~ -4 to -1. The nebular line and continuum emission are  generated using \texttt{CLOUDY} \citep{cloudy}. We base the attenuation law on a dust model that combines: (1) the \citet{cf00} two-component approach, birth-cloud vs. diffuse dust screens; and (2) the \citet{kriek13} method that parametrizes the diffuse component as a combination of a Calzetti attenuation plus a Lorentzian Drude to model the strength of the UV bump. Both components of the diffuse dust are then modulated by a power-law factor $n$ that varies the slope of the attenuation between -1 to 0.4 relative to Calzetti ($n=0$). 

 This model includes 5 free parameters that control the ratio of SFR in six adjacent time bins; the first two bins are spaced at $0 - 5 $ Myr and $5 - 10$ Myr of lookback time, and the remaining four bins are log-spaced to a maximum age of 100 Myr. In addition, it fits (1) for the ratio of the nebular to diffuse attenuation, which ranges between 0 and 2, but follows a clipped normal prior centered on 1, and also (2) for the dust index $n$. 

\subsection{The \textsc{prospector-AGN} code}

A third type of fit, \textsc{prospector-AGN} hereafter, makes use of a hybrid galaxy+AGN model similar to the one described in \citet{2023arXiv230514418B}. Briefly, we use a combination of a stellar emission component that dominates in the rest-frame UV and an AGN one that dominates in the optical to IR wavelengths. By construction, this is the only model where the AGN makes up the bulk of the luminosity and, as such, it serves as a feasibility test for AGN-dominated SEDs (forced to be dominant, not free as in the case of the \textsc{prospector-AGN$+$} code presented below). The stellar emission is modeled with Prospector using a parametric delayed-$\tau$ SFH with no attenuation. The AGN has two distinct components that account for the emission of the accretion disk and the torus, respectively.  The accretion disk is modeled after the empirical QSO templates mentioned above from 0.1~$\mu$m up to 0.6~$\mu$m rest-frame, followed by a power-law (f$_{\nu}=\nu^{\alpha}$, see for example \citealt{2016MNRAS.463.2064H} or more recently in \citealt{2023arXiv230714414B} and \citealt{2023arXiv231203065K}) with variable spectral index ranging between $\alpha=$0 to 0.5, with default value of $\alpha=1/3$, and attenuated with a \citet{2000ApJ...533..682C} law. The torus emission is modeled using the clumpy torus models from \citet{2008ApJ...685..160N} included in \textsc{prospector} \citep{leja18}. These have two free parameters: the optical depth, which ranges between $\tau_{V}=1$ and 150, and the total IR luminosity. Since these parameters are independent of each other, the models allow for AGN-dominated SEDs where the emission from the torus dominates at longer wavelengths ($\lambda\gtrsim2$~$\mu$m), as opposed to the typical QSO templates where the torus emission dominates at $\lambda\sim1$~$\mu$m or blueward. 

\subsection{The \textsc{prospector-AGN+} code}

Last but not least, \textsc{prospector-AGN$+$} is a modified version of the original \textsc{prospector} code with the Flexible Stellar Population Synthesis (FSPS; \citealt{Conroy2009, Conroy2010}) for the stellar component. We assumed a \citet{2001MNRAS.322..231K} initial mass function and delayed-tau star-formation history. The stellar nebular line and continuum emission are also turned-on, as pre-configured in FSPS \citep{Byler2017}. We adopted the Calzetti attenuation curve with a flexible slope as introduced in \citet{Kriek2013}. For the galaxy dust emission, since the objects of this work have $z\gtrsim5.0$, we adopted the empirical IR SED model of Haro~11, a low-metallicity star-bursting dwarf galaxy that is believed to share typical features of first-generation galaxies in the early Universe \citep{Lyu2016, DeRossi2018}. 

This code uses a set of semi-empirical AGN SED models,  which have been optimized for AGN identification and characterization \citep{Lyu2022, 2023arXiv231012330L}. For this work, we adopted a similar model configuration as in \citet{2023arXiv231012330L} for the SMILES+JADES AGN identification: the AGN component includes both the AGN-powered continuum from the UV to the far-IR and the narrow and broad emission lines from the UV to the NIR derived based on empirical observations. The continuum shape and line strengths of the AGN SED can be modified by a hybrid extinction configuration featuring the SMC-like curve for the commonly seen UV-optical extinction in Type-1 AGNs and an empirical attenuation law for the IR obscuration. Further details can be found in \citet{2023arXiv231012330L} and references therein.

In total, this code uses 7 free parameters to describe the stellar component and 2 for the AGN.

\subsection{Summary}

These four codes let us compare models of the LRDs with a wide variety of input assumptions. This is important to let us deduce which properties are likely to be intrinsic rather than due to modeling issues. \textsc{synthesizer-AGN}, \textsc{prospector-SF}, and \textsc{prospector-AGN+} use three different prescriptions for the stellar populations, with three varieties for the interstellar extinction law.  \textsc{synthesizer-AGN} and \textsc{prospector-AGN+} treat the stellar-heated dust emission in two different ways. \textsc{synthesizer-AGN}, \textsc{prospector-AGN}, and \textsc{prospector-AGN+} use three different models for the AGN emission. This latter variety is particularly important given the diversity of possibilities for AGN SEDs and the importance of capturing this variety in modeling LRDs. The first case, \textsc{synthesizer-AGN}, uses an empirical fixed model SED.  \textsc{prospector-AGN$+$} fits  empirical templates obscured by an attenuation law derived specifically for AGNs (see e.g. \citealt{Lyu2017}), while \textsc{prospector-AGN} uses a declining power-law with $\alpha=1/3$ in place of the empirical templates.  

\section{Physical properties of LRDs}
\label{sec:properties}

\subsection{Galaxy by galaxy analysis for the Golden Five LRDs}

In this section, we present our analysis of the SEDs of the LRDs detected in several MIRI bands. We concentrate our discussion of properties on the five galaxies detected in  F1280W and redder bands, which we call the Golden Five sample. We compare the global results of the SED fitting provided by each one of our 4 different codes in different subsections, and comparatively discuss how they fit the UV, optical and near-infrared spectral regions. At the end of the section, we describe the properties for the galaxies detected up to F1000W and we briefly discuss the rest of the sample. A more detailed discussion of all the galaxies in our our sample is presented in Appendix~\ref{app:allSEDs}. The main physical properties of all the LRDs are given in Table~\ref{tab:props}.

\begin{figure*}
\centering
\includegraphics[clip,trim=0.0cm 0.0cm 0.0cm 0.0cm,width=8.cm,angle=0]{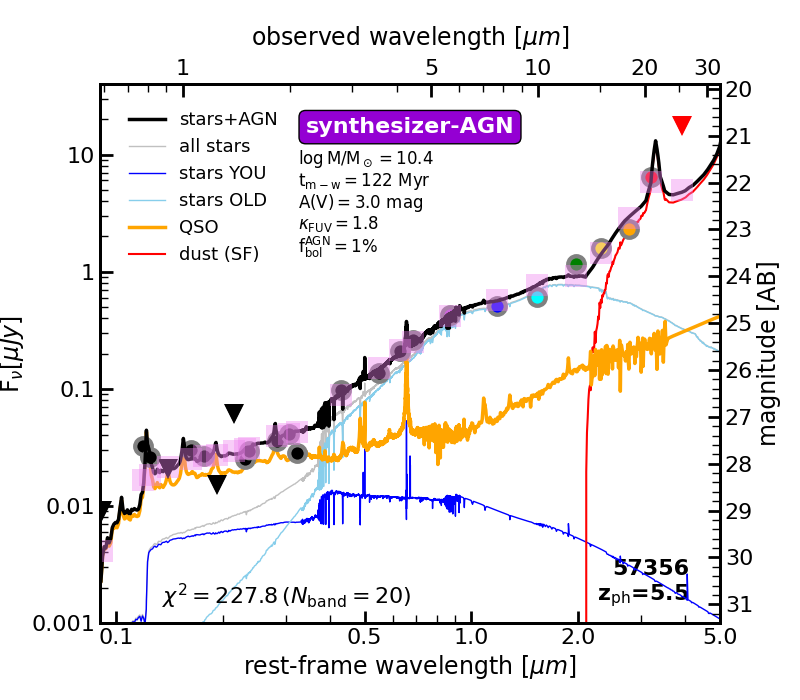}
\includegraphics[clip,trim=0.0cm 0.0cm 0.0cm 0.0cm,width=8.cm,angle=0]{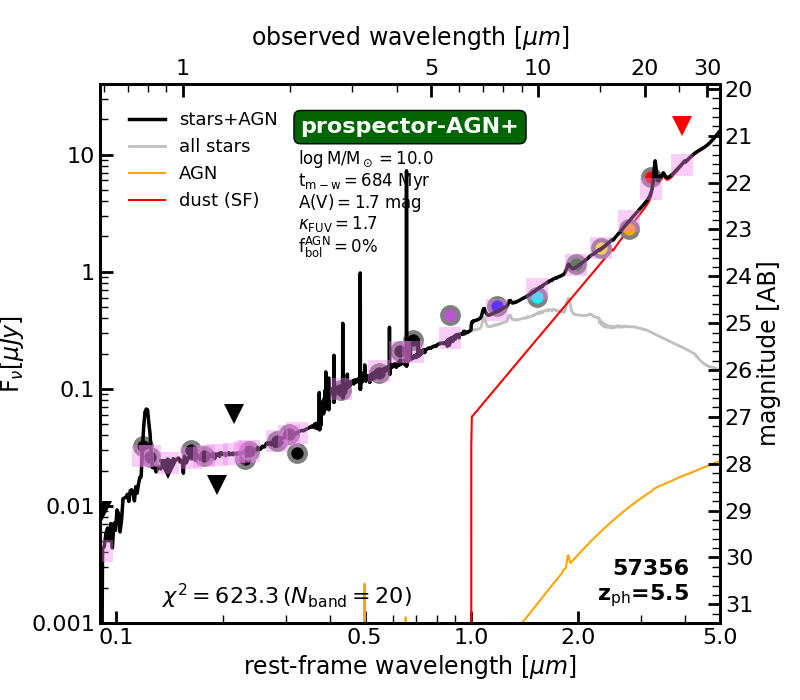}
\includegraphics[clip,trim=0.0cm 0.0cm 0.0cm 0.0cm,width=8.cm,angle=0]{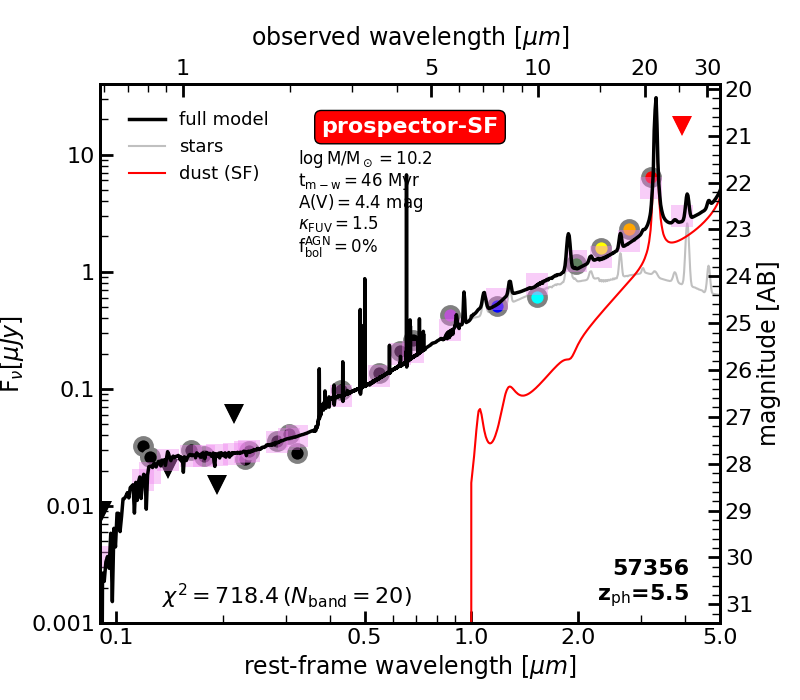}
\includegraphics[clip,trim=0.0cm 0.0cm 0.0cm 0.0cm,width=8cm,angle=0]{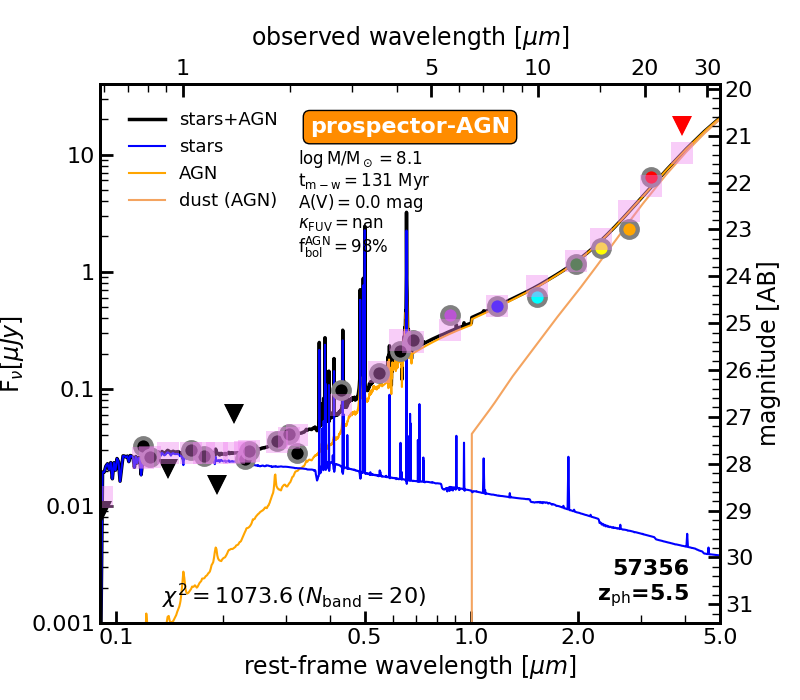}
\includegraphics[clip,trim=0.0cm 1.6cm 0.0cm 1.0cm,width=18cm,angle=0]{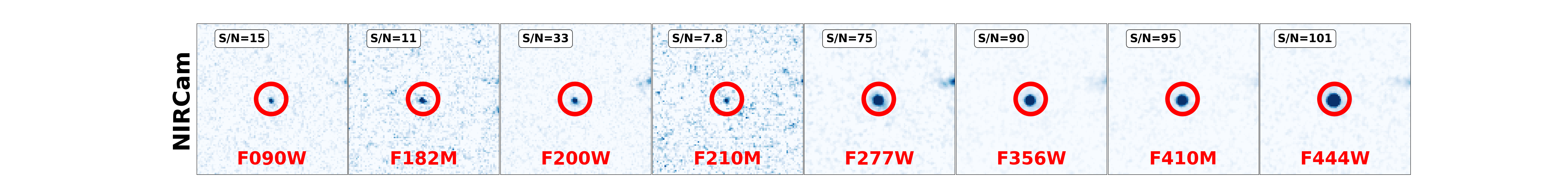}
\includegraphics[clip,trim=0.0cm 1.6cm 0.0cm 1.0cm,width=18cm,angle=0]{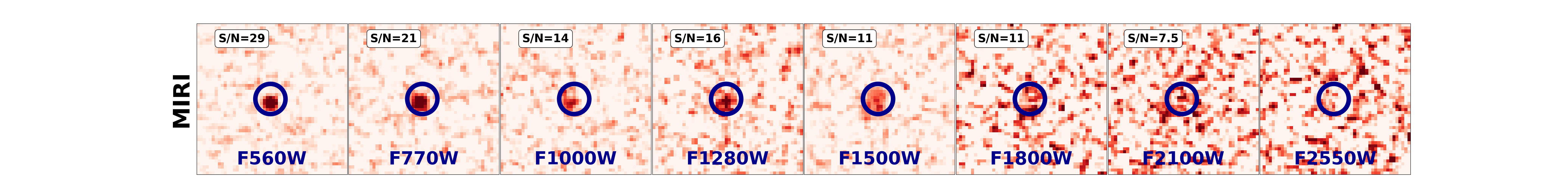}
\includegraphics[clip,trim=0.0cm 1.6cm 0.0cm 1.0cm,width=18cm,angle=0]{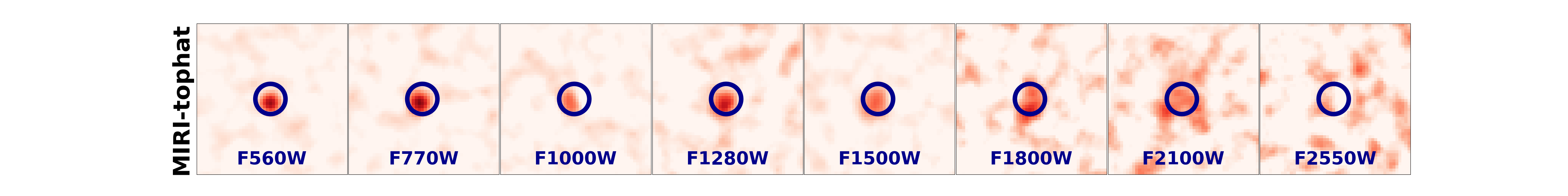}
\caption{SED fitting results for source JADES$-$57356, the  LRD in our sample detected up to F2100W. The four upper panels show: (1) the fits for \textsc{synthesizer-AGN$+$} on the top left panel, including the individual components of the model, i.e., young, old, and all stars in blue, cyan, and gray, (un)obscured AGN in orange, and regular dust emission (in principle, linked to star formation) in red; (2) results for \textsc{Prospector-AGN$+$} on the top right panel, showing the three components, stars in gray, AGN in orange, and star-formation heated dust in red; (3) \textsc{Prospector-SF} on the bottom left, showing stars in gray and star-formation heated dust in red; and (4) \textsc{Prospector-AGN} on the bottom right, including young stars in blue, total AGN emission in orange, with the torus emission shown with a dashed line. Number of bands fitted and direct (i.e., not reduced) $\chi^2$ values are provided, as well a stellar masses, stellar mass-weighted ages, $V$-band stellar attenuation, ratio between the FUV and optical stellar attenuation, and fraction of bolometric luminosity coming from the AGN. The fits include NIRCam bands, shown in black, and MIRI fluxes, in color; upper limits, depicted with triangles, are also used. Below the SEDs, we give 10\arcsec$\times$10\arcsec\, postage stamps in NIRCam (upper row), MIRI (middle row), and MIRI convolved with a 5-pixels wide tophat filter, the LRD being marked with a 0.3\arcsec\, radius circle, and the $S/N$ provided (when it is above 5).}
\label{fig:fits1}
\end{figure*}

\begin{figure*}
\centering
\includegraphics[clip,trim=0.0cm 0.0cm 0.0cm 0.0cm,width=8.cm,angle=0]{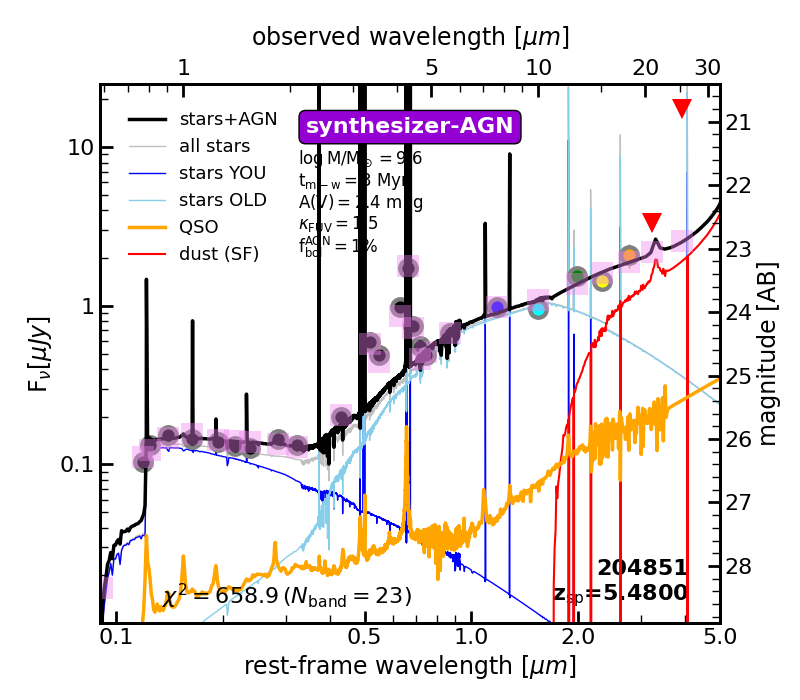}
\includegraphics[clip,trim=0.0cm 0.0cm 0.0cm 0.0cm,width=8.cm,angle=0]{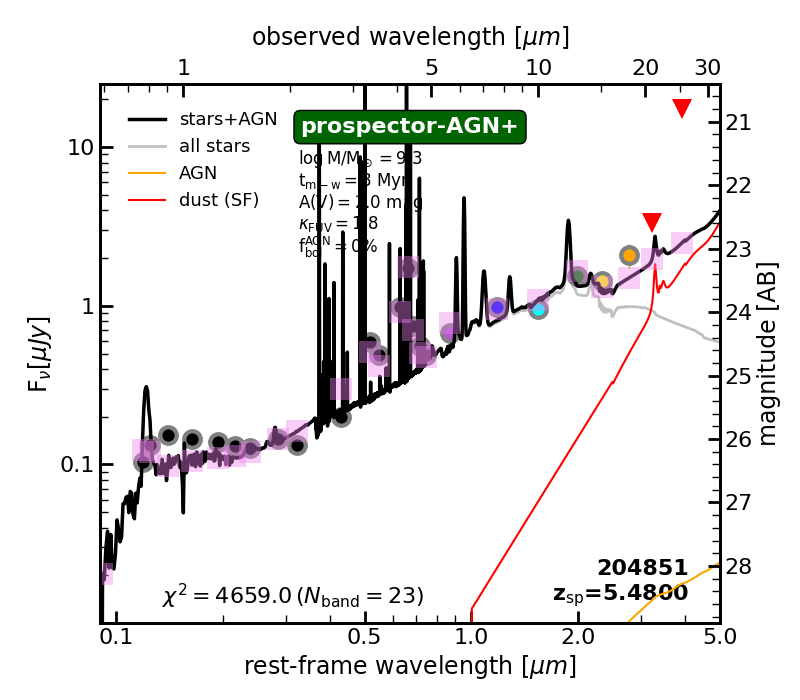}
\includegraphics[clip,trim=0.0cm 0.0cm 0.0cm 0.0cm,width=8.cm,angle=0]{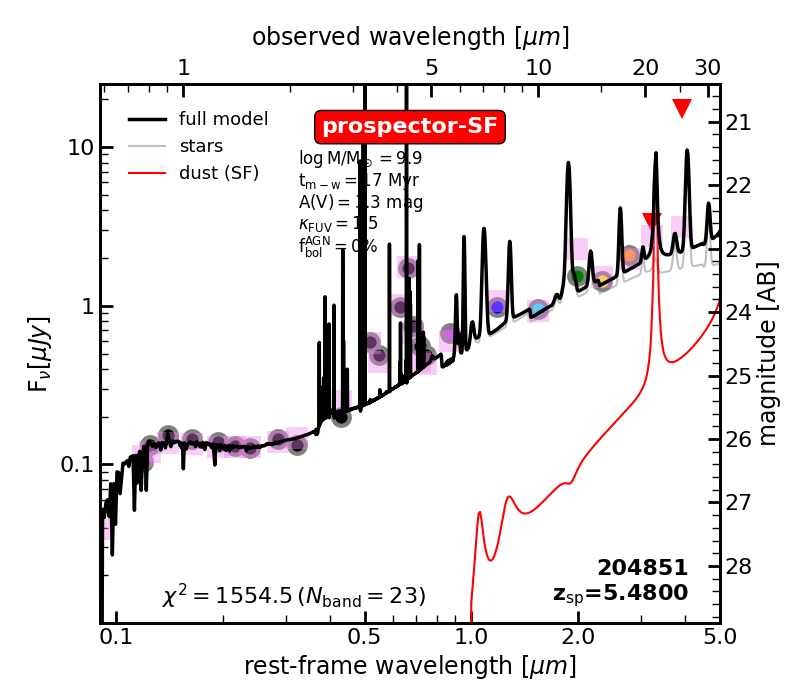}
\includegraphics[clip,trim=0.0cm 0.0cm 0.0cm 0.0cm,width=8cm,angle=0]{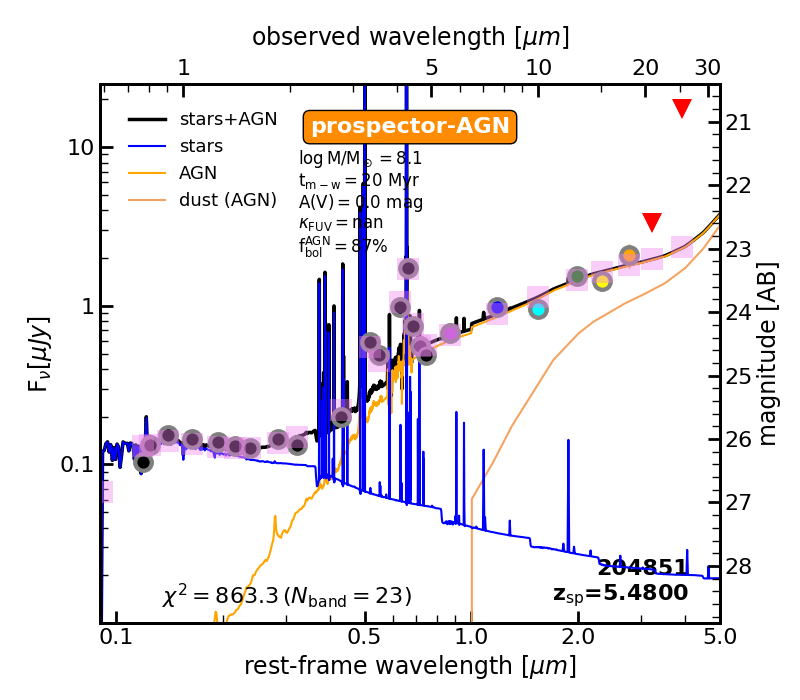}
\includegraphics[clip,trim=0.0cm 1.6cm 0.0cm 1.0cm,width=18cm,angle=0]{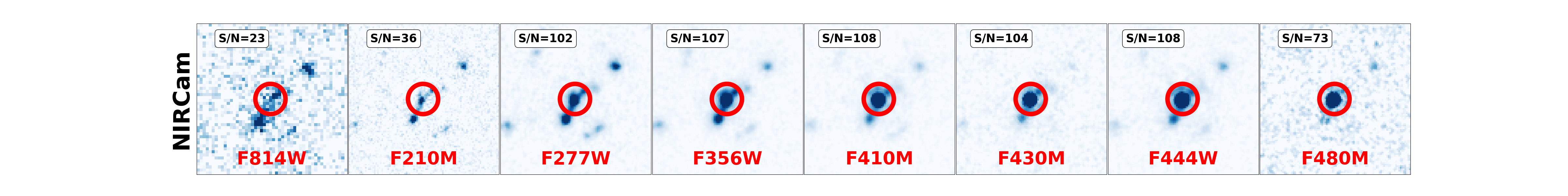}
\includegraphics[clip,trim=0.0cm 1.6cm 0.0cm 1.0cm,width=18cm,angle=0]{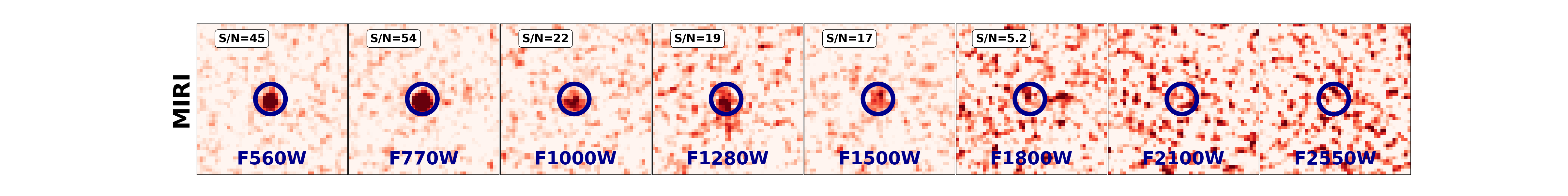}
\includegraphics[clip,trim=0.0cm 1.6cm 0.0cm 1.0cm,width=18cm,angle=0]{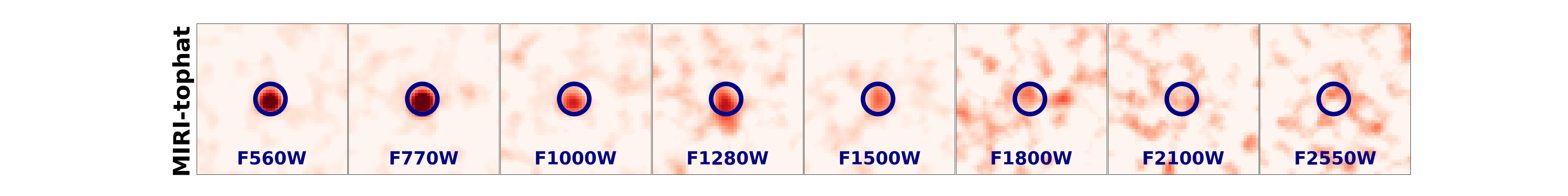}
\caption{Same as Figure~\ref{fig:fits1}, but for source JADES$-$204851, a source detected up to F1800W.}
\label{fig:fits2}
\end{figure*}

\begin{figure*}
\centering
\includegraphics[clip,trim=0.0cm 0.0cm 0.0cm 0.0cm,width=8.cm,angle=0]{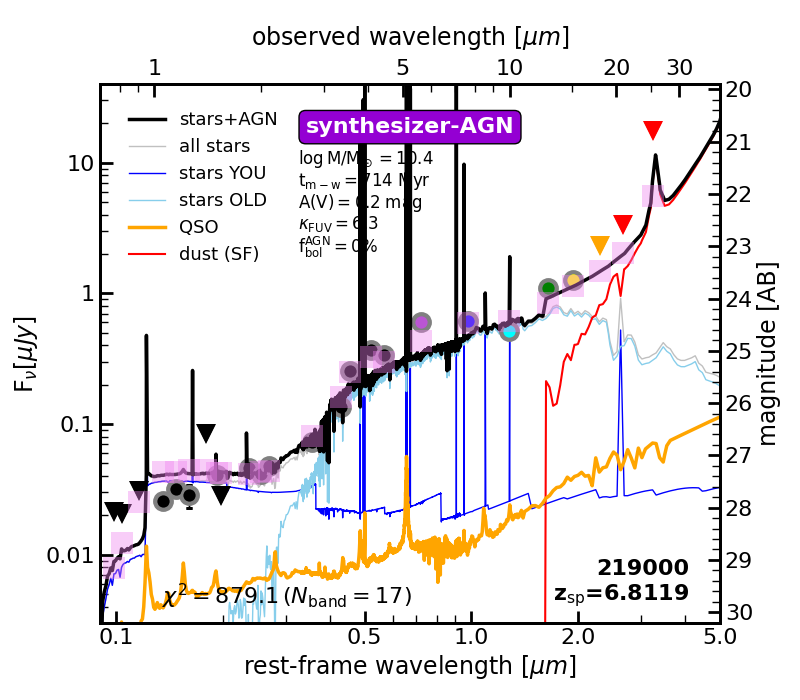}
\includegraphics[clip,trim=0.0cm 0.0cm 0.0cm 0.0cm,width=8.cm,angle=0]{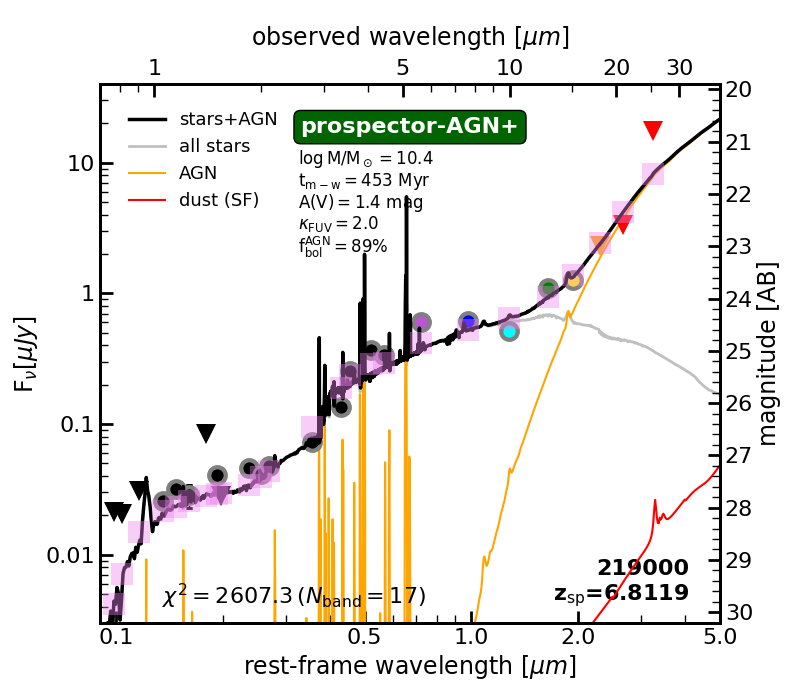}
\includegraphics[clip,trim=0.0cm 0.0cm 0.0cm 0.0cm,width=8.cm,angle=0]{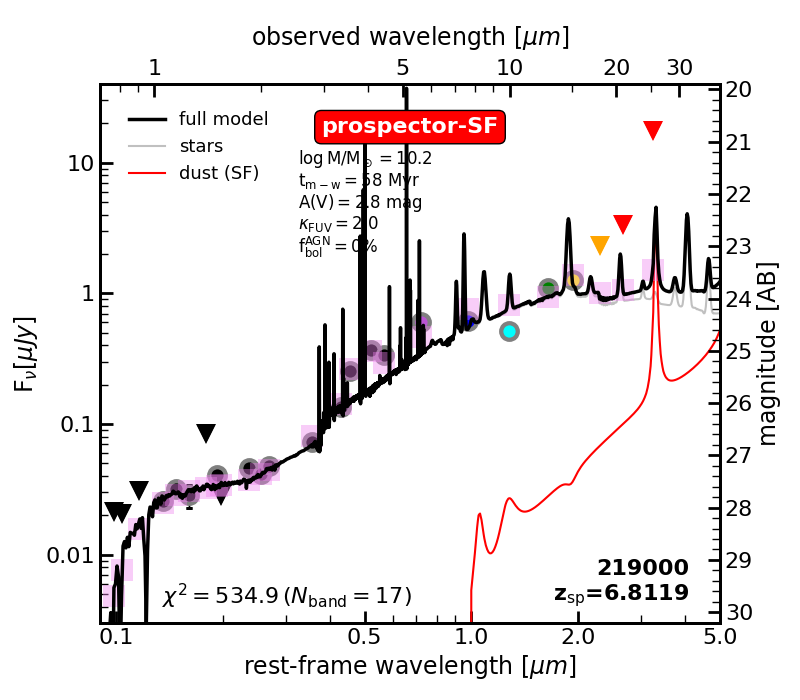}
\includegraphics[clip,trim=0.0cm 0.0cm 0.0cm 0.0cm,width=8cm,angle=0]{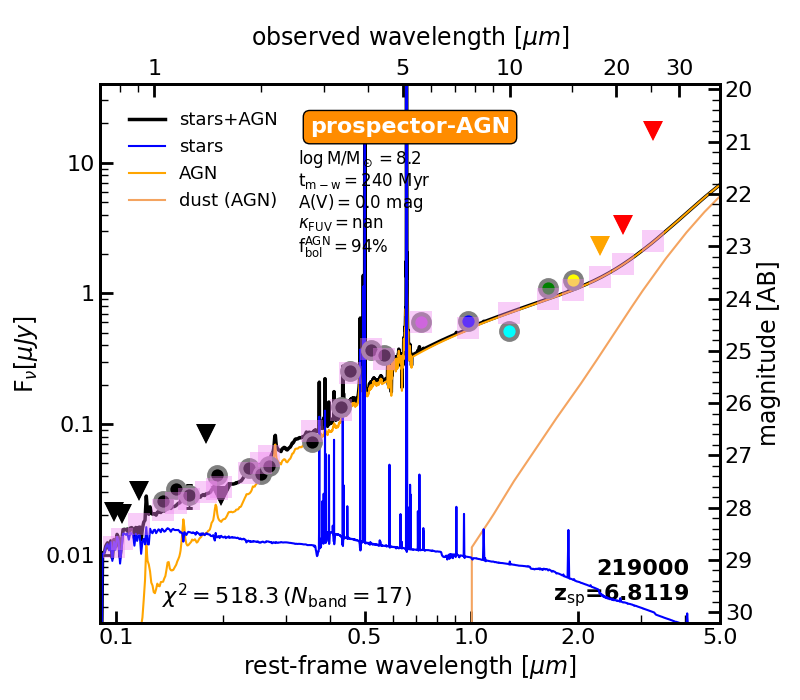}
\includegraphics[clip,trim=0.0cm 1.6cm 0.0cm 1.0cm,width=18cm,angle=0]{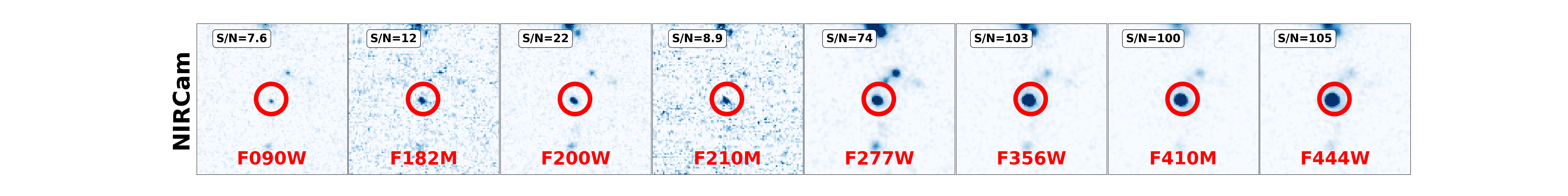}
\includegraphics[clip,trim=0.0cm 1.6cm 0.0cm 1.0cm,width=18cm,angle=0]{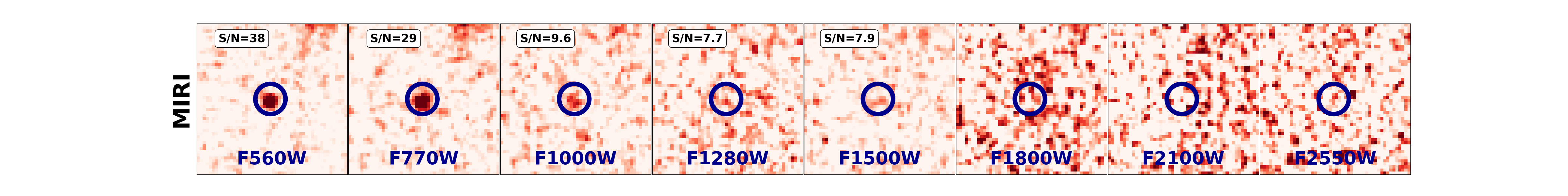}
\includegraphics[clip,trim=0.0cm 1.6cm 0.0cm 1.0cm,width=18cm,angle=0]{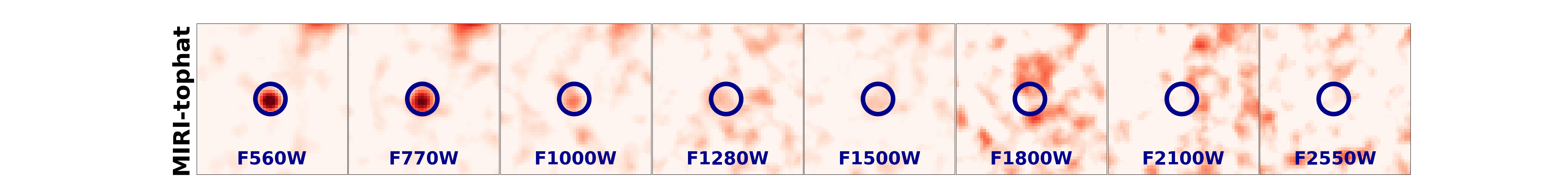}
\caption{Same as Figure~\ref{fig:fits1}, but for source JADES$-$219000, a source detected up to F1500W.}
\label{fig:fits3}
\end{figure*}

\begin{figure*}
\centering
\includegraphics[clip,trim=0.0cm 0.0cm 0.0cm 0.0cm,width=8.cm,angle=0]{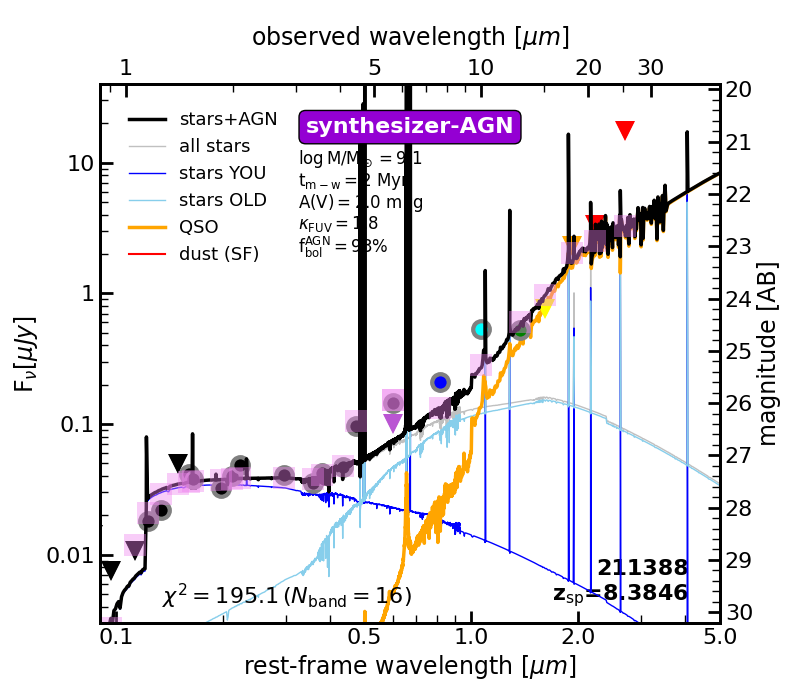}
\includegraphics[clip,trim=0.0cm 0.0cm 0.0cm 0.0cm,width=8.cm,angle=0]{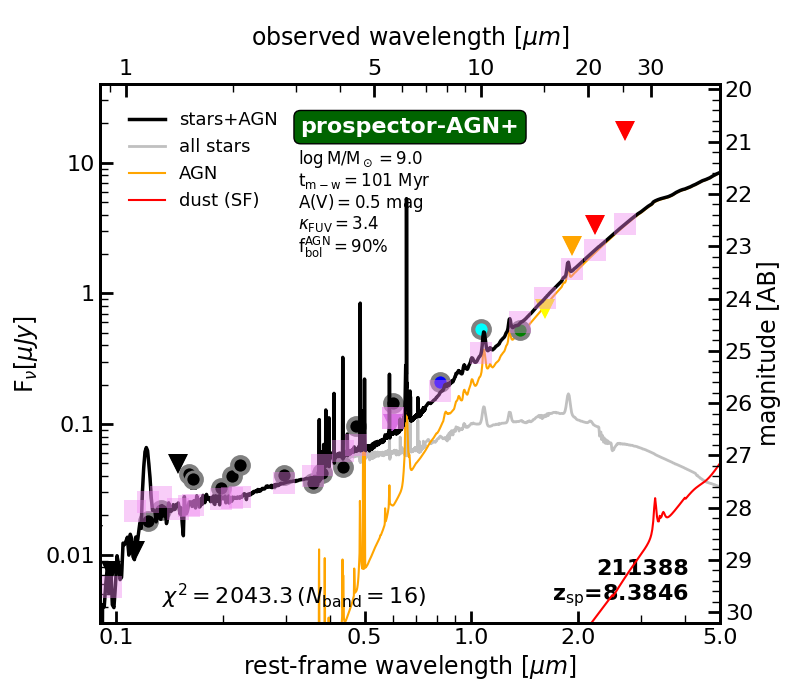}
\includegraphics[clip,trim=0.0cm 0.0cm 0.0cm 0.0cm,width=8.cm,angle=0]{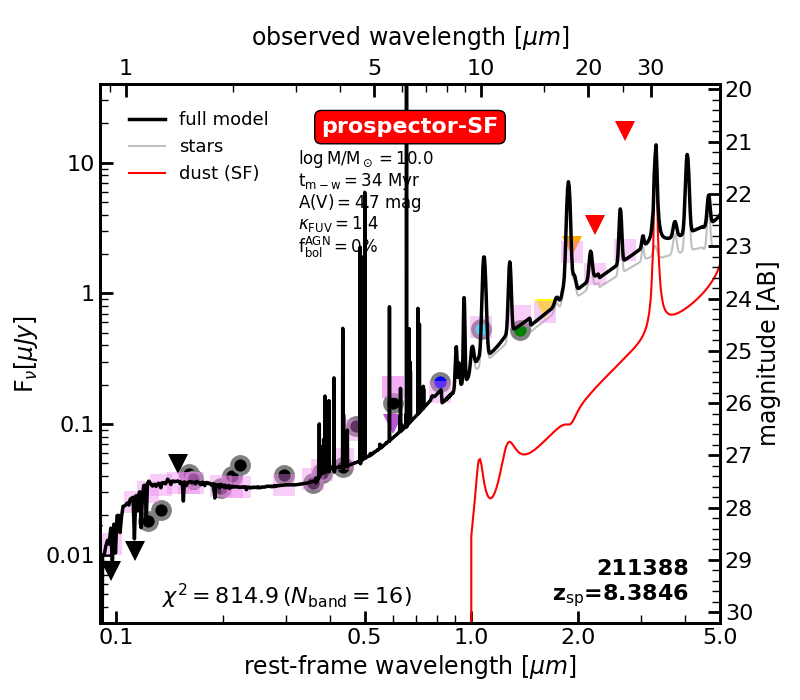}
\includegraphics[clip,trim=0.0cm 0.0cm 0.0cm 0.0cm,width=8cm,angle=0]{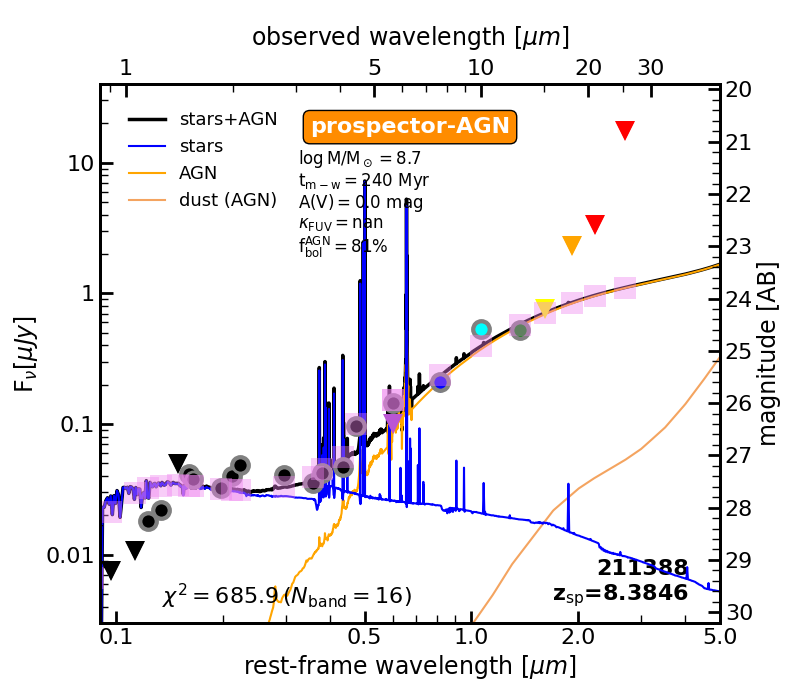}
\includegraphics[clip,trim=0.0cm 1.6cm 0.0cm 1.0cm,width=18cm,angle=0]{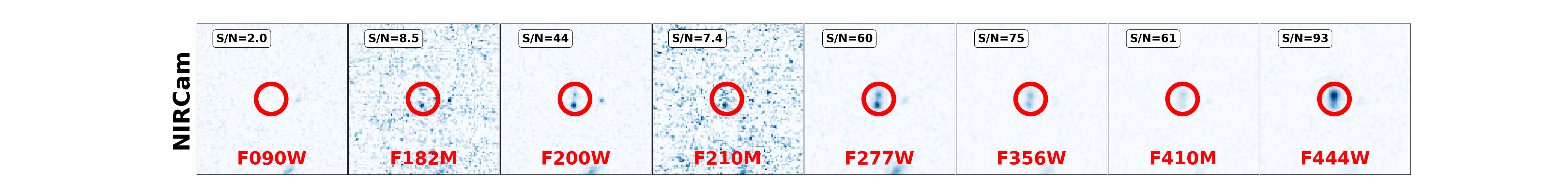}
\includegraphics[clip,trim=0.0cm 1.6cm 0.0cm 1.0cm,width=18cm,angle=0]{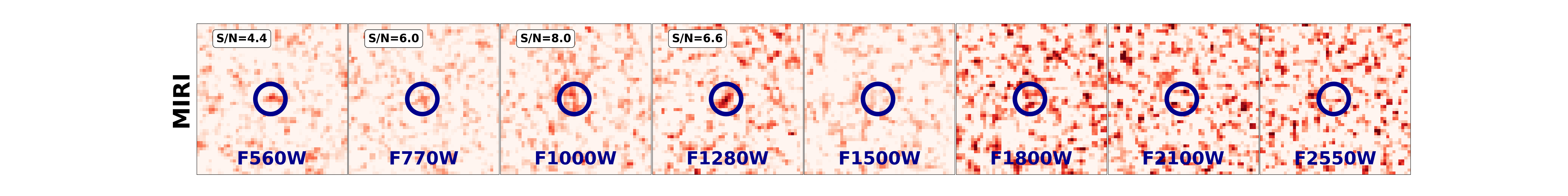}
\includegraphics[clip,trim=0.0cm 1.6cm 0.0cm 1.0cm,width=18cm,angle=0]{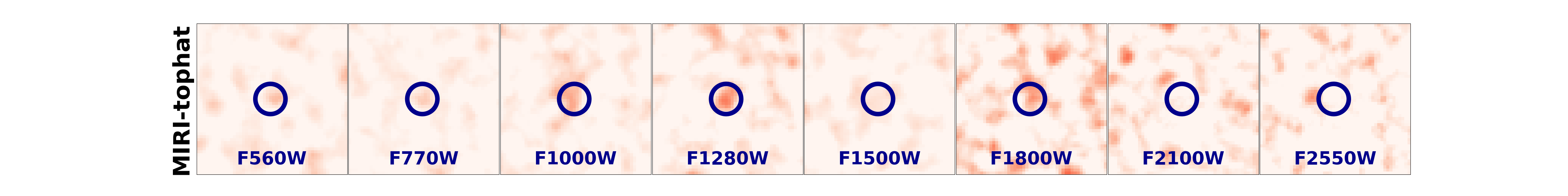}
\caption{Same as Figure~\ref{fig:fits1}, but for source JADES$-$211388, a source detected up to F1280W.}
\label{fig:fits4}
\end{figure*}

\begin{figure*}
\centering
\includegraphics[clip,trim=0.0cm 0.0cm 0.0cm 0.0cm,width=8.cm,angle=0]{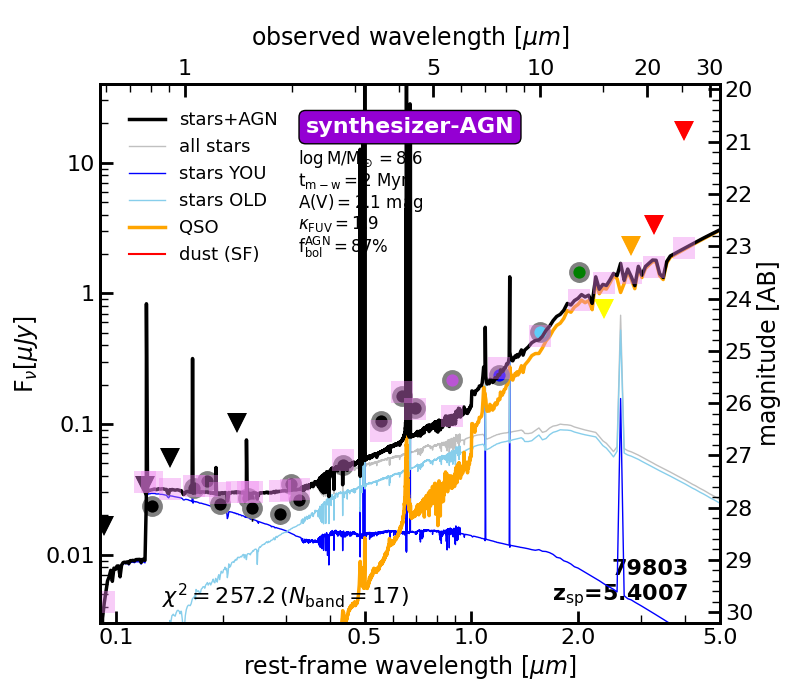}
\includegraphics[clip,trim=0.0cm 0.0cm 0.0cm 0.0cm,width=8.cm,angle=0]{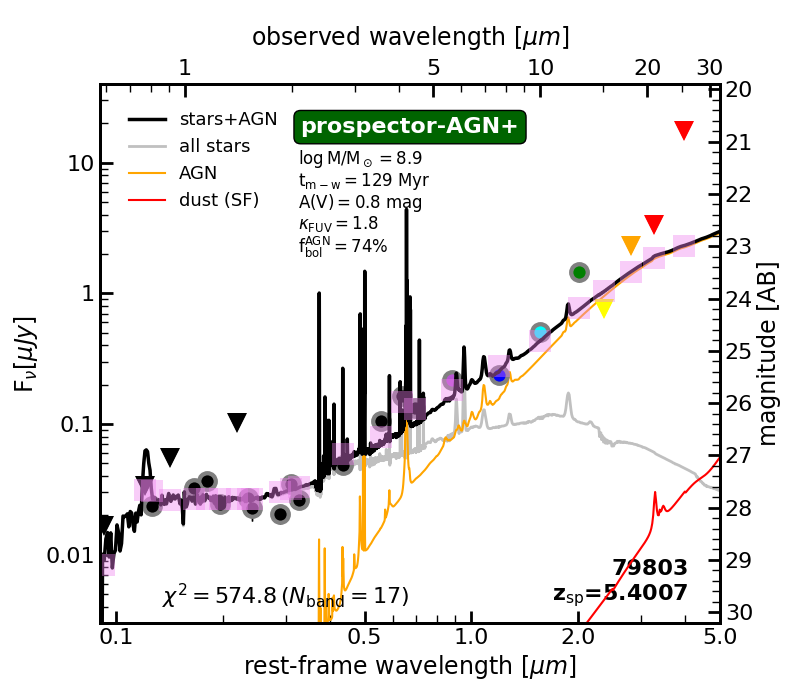}
\includegraphics[clip,trim=0.0cm 0.0cm 0.0cm 0.0cm,width=8.cm,angle=0]{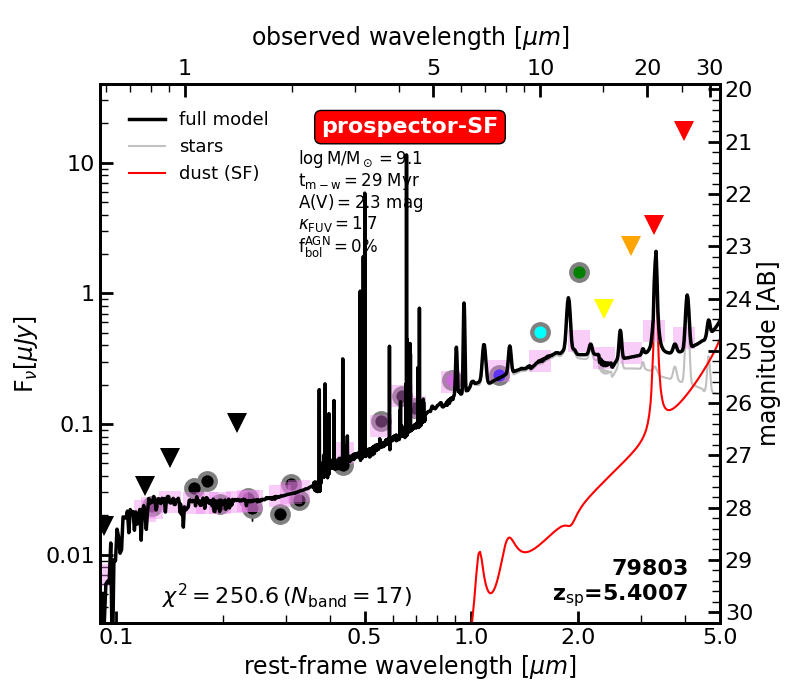}
\includegraphics[clip,trim=0.0cm 0.0cm 0.0cm 0.0cm,width=8cm,angle=0]{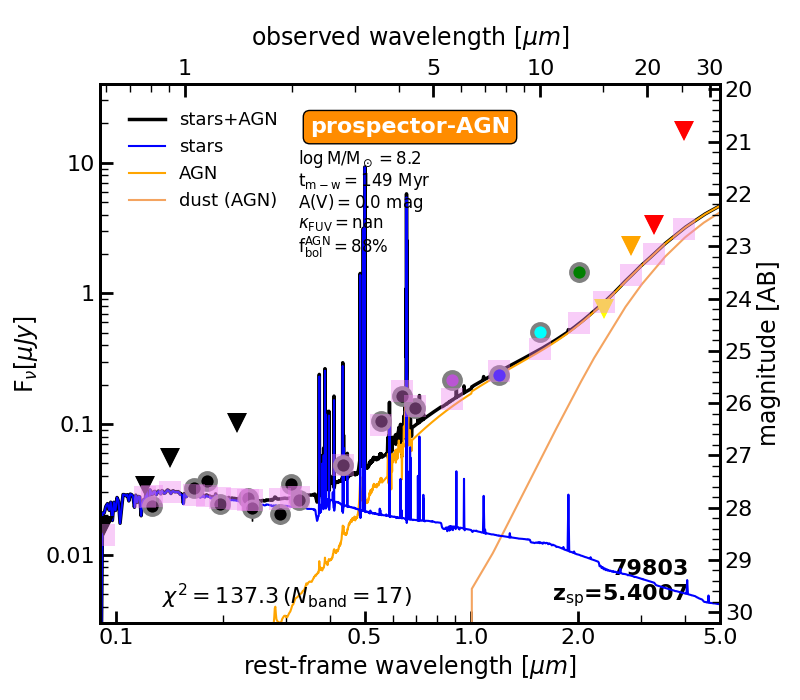}
\includegraphics[clip,trim=0.0cm 1.6cm 0.0cm 1.0cm,width=18cm,angle=0]{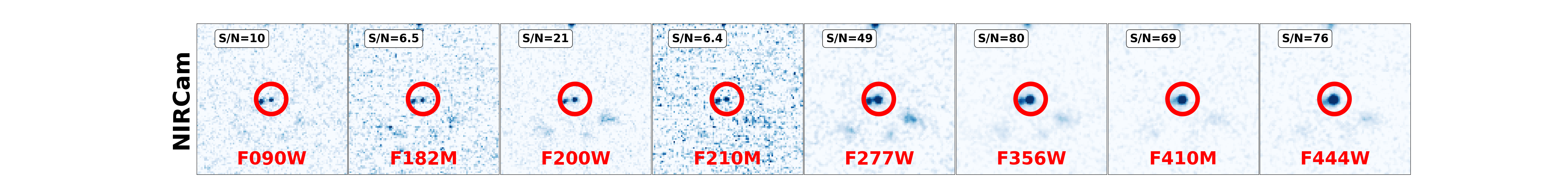}
\includegraphics[clip,trim=0.0cm 1.6cm 0.0cm 1.0cm,width=18cm,angle=0]{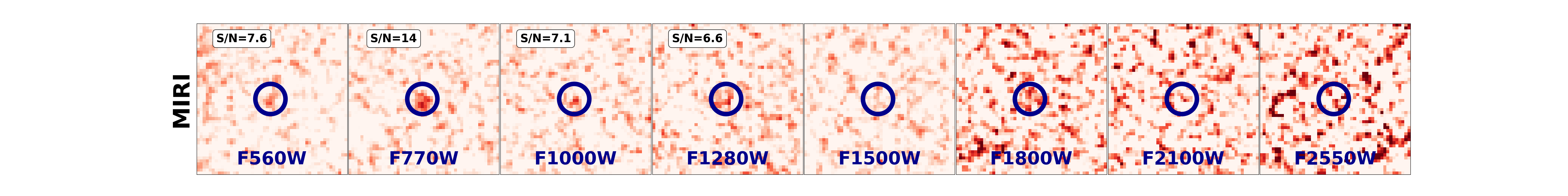}
\includegraphics[clip,trim=0.0cm 1.6cm 0.0cm 1.0cm,width=18cm,angle=0]{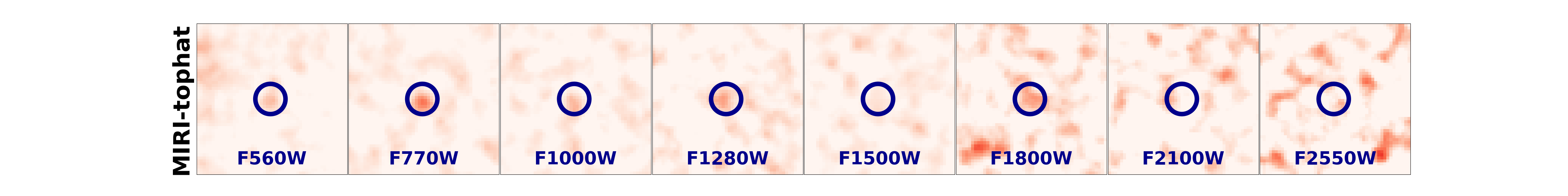}
\caption{Same as Figure~\ref{fig:fits1}, but for source JADES$-$79803, a source detected up to F1280W.}
\label{fig:fits5}
\end{figure*}

\subsubsection{Results for the Golden Five galaxies with {\sc synthesizer-AGN}}

With {\sc synthesizer-AGN}, the rest-frame UV spectral range is fitted with a young stellar population for all the galaxies in the Golden Five sample except for JADES$-$57356, which is fitted with the QSO component. For this galaxy, the fact that no emission lines are observed in the optical ([OIII] or H$\alpha$), but the UV spectral range is flat, favors the contribution of a QSO to the UV emission, outshined by the stars in the optical. Very young stars would contribute strong emission lines in the optical (as in JADES$-$204851), and thus are disfavored by this modeling. The different UV (relatively flat) slopes observed for the Golden Five sample are reproduced with $\sim$1 Myr old (mass-weighted age) stellar population and $\mathrm{A(V)}=0.6$~mag dust attenuation, which  is also able to reproduce the emission lines observed for four of the Golden Five galaxies.

The rest-frame optical range of the Golden Five sample is dominated by slightly more evolved stars in all cases, typically of 10-100~Myr mass-weighted age. They account for most of the stellar mass (typically, more then 90\%) but do not contribute much, in relative terms with respect to the younger stars, to the optical emission lines. This more evolved stellar populations present high attenuations, typically  $\mathrm{A(V)}=2-3$~mag.

If we consider the ratio of the attenuation of the far-UV with respect the optical emission arising from all the stellar populations included in the {\sc synthesizer-AGN} modeling, $\kappa_\mathrm{FUV}$, we typically find flatter slopes, $\sim1.5-2.0$, than what is expected for the \citet{2000ApJ...533..682C} attenuation law ($\kappa_\mathrm{FUV}=2.6$). This effect is, in fact, common to all modeling techniques, they all typically obtain gray attenuation laws.

Finally, the rest-frame near/mid-infrared emission of the Golden Five LRDs is fitted with dust combined with star formation in three Golden Five galaxies, and dust in an AGN torus in the other two. The heating source is indicated by the possible detection of a most probably star-formation related 3~$\mu$m PAH feature in JADES$-$57356, and by the differences in slopes of the SEDs for the other sources, steeper in the case of AGN-dominated fits (JADES$-$211388 and JADES$-$79803).

The {\sc synthesizer-AGN} code presents the smallest $\chi^2$ values for the fits of the SEDs of all the Golden Five galaxies except one.

\subsubsection{Results for the Golden Five galaxies with {\sc prospector-SF}}

By construction, {\sc prospector-SF} only considers stars and dust heated by stars to fit the SEDs. In general, the $\chi^2$ values obtained by this  code and the following one are larger than those obtained with  {\sc synthesizer-AGN} (except for JADES$-$219000) and smaller than {\sc prospector-AGN$+$}.

The {\sc prospector-SF} results imply a $\sim$30~Myr stellar population (mass-weighted age), with large attenuations $\mathrm{A(V)}=3$~mag and a flat attenuation law. These properties translate to larger stellar masses   compared to other codes (a $\sim$0.5~dex difference with respect to {\sc synthesizer-AGN}, for example). We remark that the dust emission models are significantly colder than what is obtained with {\sc synthesizer-AGN} (see Appendix~\ref{app:dust}).

\subsubsection{Results for the Golden Five galaxies with {\sc prospector-AGN}}

The main distinct characteristic of the results achieved with {\sc prospector-AGN} for the Golden Five galaxies is the smaller stellar mass. Given that, by construction, this code only fits the rest-frame UV spectral region with stars, and the optical and infrared with AGN templates (including emission from the accretion disk and the torus), the stellar masses are up to a factor of 100 smaller than those estimated with the other codes. In general, better fits in terms of $\chi^2$ values are obtained with {\sc prospector-AGN} compared to {\sc prospector-SF}, except for JADES$-$57356, the galaxy with possible PAH emission. But better fits are obtained with the mixed models in  {\sc synthesizer-AGN} (favoring stellar emission in the UV and optical, a variety of results in the near-infrared).

\subsubsection{Results for the Golden Five galaxies with {\sc prospector-AGN+}}

The {\sc prospector-AGN$+$} code fits to the SEDs of the Golden Five galaxies are very similar to those outlined for  {\sc synthesizer-AGN}. The rest-frame UV and optical spectral ranges are dominated by stars, except at 0.8-1.0~$\mu$m for two galaxies (JADES$-$211388 and JADES$-$79803) whose infrared emission was fitted with an AGN torus model.    This was also the best solution for {\sc synthesizer-AGN}. The only significant difference with respect to the results achieved with {\sc synthesizer-AGN} is found for the infrared emission of JADES$-$219000, whose slope if we take the MIRI upper limits at face value is too steep for a star formation model, so  {\sc prospector-AGN$+$} (which used upper limits as regular points) prefers a torus template (while the {\sc synthesizer-AGN} star-formation heated dust model lies below the upper limits).  The typical mass-weighted ages for the fits with the {\sc prospector-AGN$+$} code are around 150~Myr, with extinctions around $\mathrm{A(V)}=1.5$~mag with a flatter attenuation law compared to the \citet{2000ApJ...533..682C} recipe.

A detailed analysis of the SED fits for each galaxy in the Golden Five sample is presented in Appendix~\ref{app:allSEDs}.

\subsection{Implications for the nature of the NIR emission of the LRDs}
\label{sec:dust_emission}

\begin{figure*}[htp!]
\centering
\includegraphics[clip,trim=0.0cm 0.0cm 0.0cm 0.0cm,width=8.cm,angle=0]{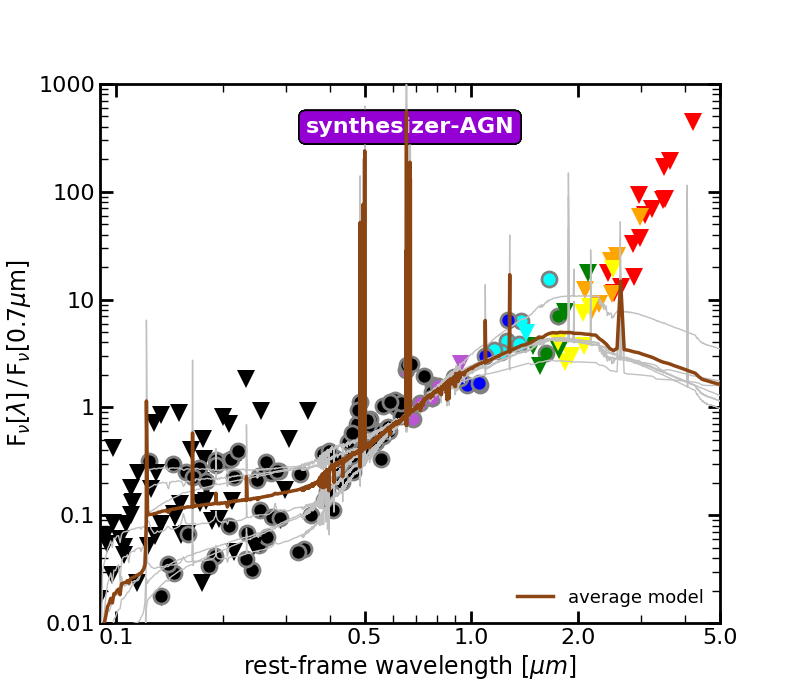}
\includegraphics[clip,trim=0.0cm 0.0cm 0.0cm 0.0cm,width=8.cm,angle=0]{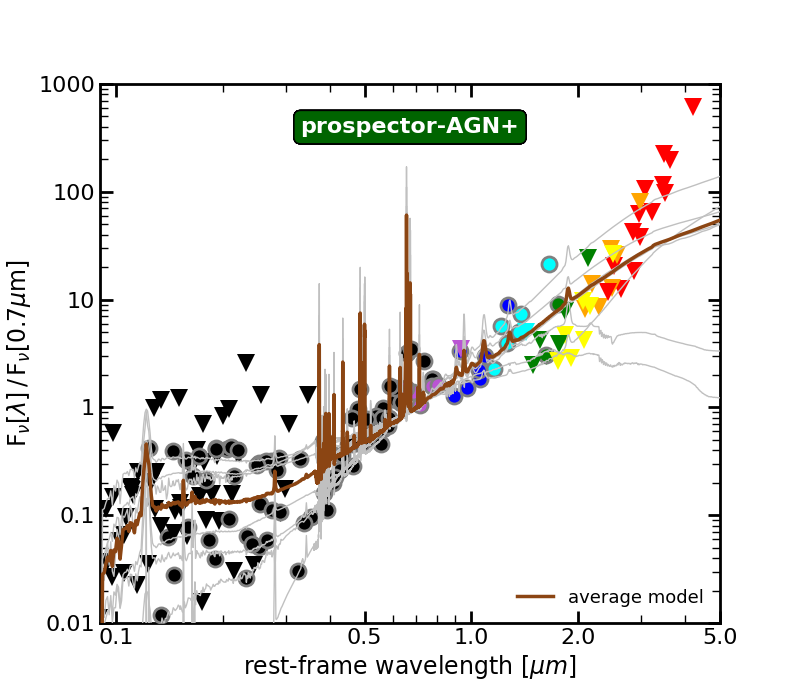}
\includegraphics[clip,trim=0.0cm 0.0cm 0.0cm 0.0cm,width=8.cm,angle=0]{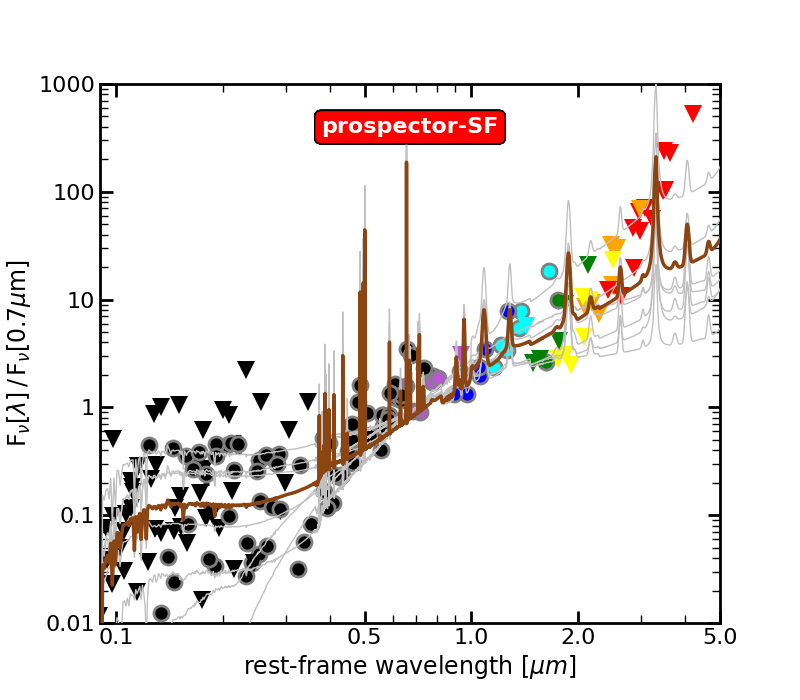}
\includegraphics[clip,trim=0.0cm 0.0cm 0.0cm 0.0cm,width=8cm,angle=0]{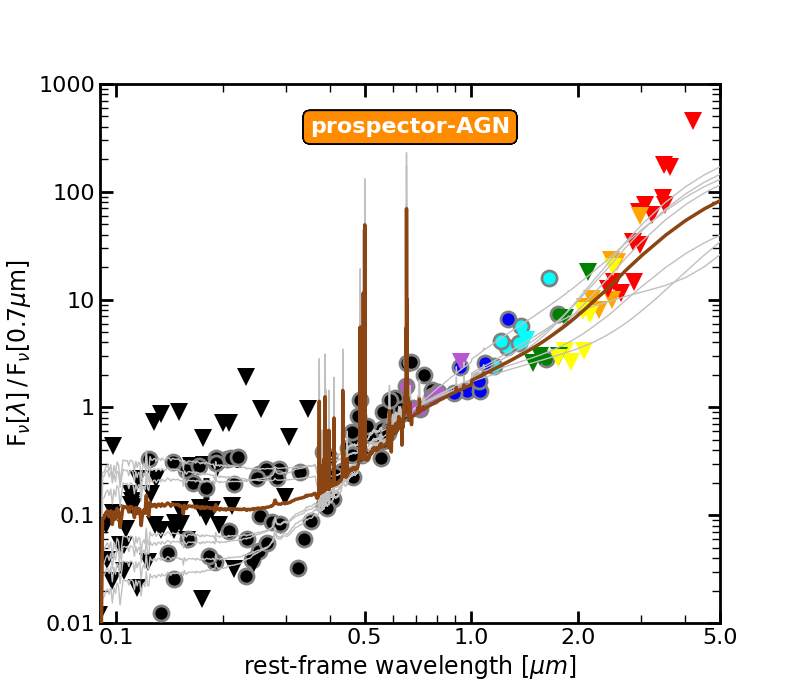}
\caption{SED fitting results for the 7 galaxies detected at F1000W and not beyond (i.e., the plot does not include any of the Golden Five galaxies). SEDs are normalized to 0.7~$\mu$m. In gray, we show the fits to each individual galaxy, and we provide an average in brown.}
\label{fig:fits6}
\end{figure*}

Overall, all four models fit the characteristic bimodal SED of the Golden Five LRDs relatively well. Qualitatively, the UV-optical spectral range is dominated either by 1) stars: two young populations with very different attenuations in {\sc synthesizer-AGN}, or a single population with an extremely gray attenuation in \textsc{prospector-SF} and  \textsc{prospector-AGN$+$}; or 2) emission from an obscured accretion disk combined with a young stellar population that contributes only in the UV. In the next section, we discuss the implied stellar masses and other stellar properties for the Golden Five and all other LRDs that have very similar UV-to-optical SEDs (see, e.g., Figure~\ref{fig:fits6}) probed primarily by the NIRCam bands. However, the Golden Five, and other MIRI-detected subsamples, allow us to probe further into the rest-frame NIR of the LRDs to characterize the origin of the emission in that spectral range.

\begin{figure*}[htp!]
\centering
\includegraphics[clip,width=17cm,angle=0]{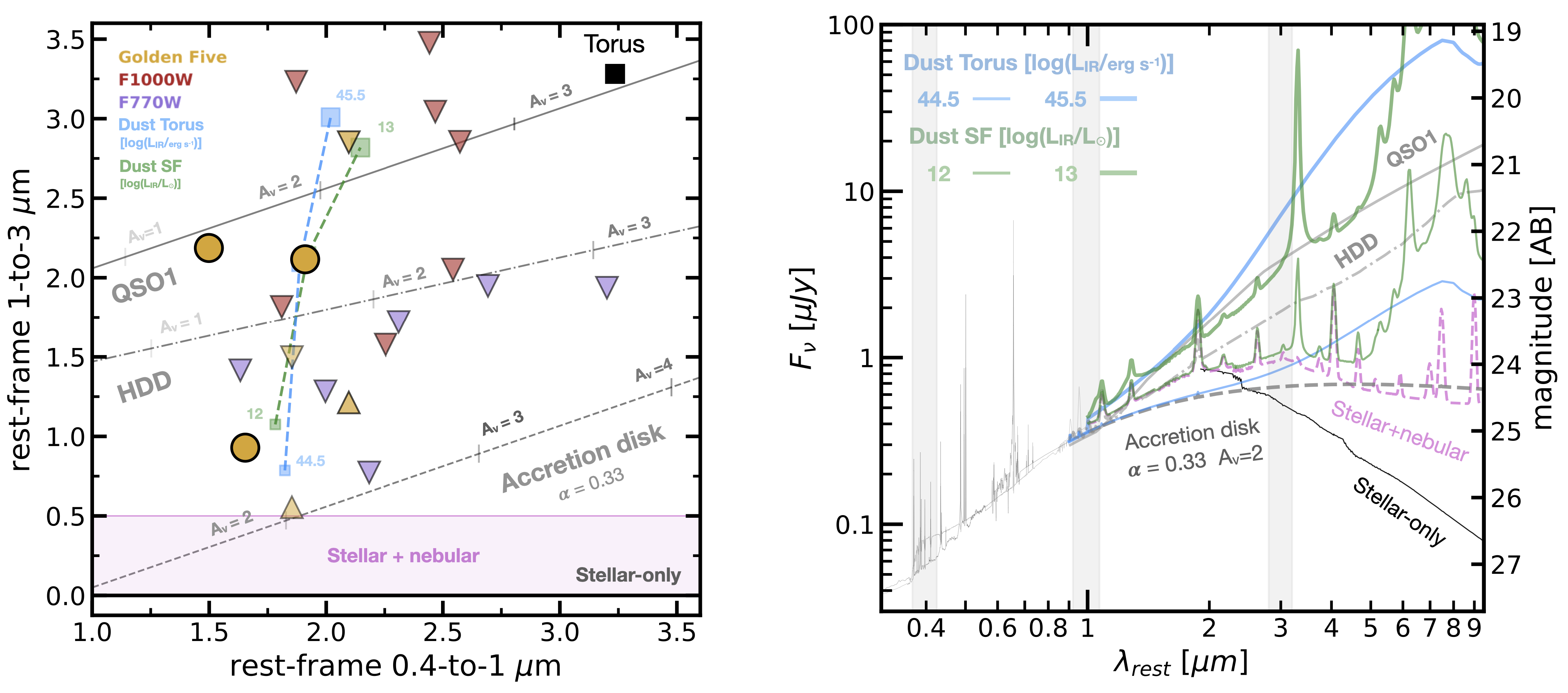}
\caption{Rest-frame optical and NIR colors for the LRDs (left, with measured and upper limit fluxes shown with circles and arrows, respectively) and different templates and best-fit models (right). The 0.4-to-1~$\mu$m color traces the optical slope and thus it is a good proxy for the dust attenuation. The 1-to-3~$\mu$m color tracks the amount of dust emission relative to the stellar or accretion disk emission which dominates the UV-optical SED, probed by the 0.4-to-1.0~$\mu$m color. The LRDs exhibit values in between a stellar-only sequence with 1-to-3~$\mu$$\sim$0 (or up to 0.5~mag with increasing nebular continuum)  indicated by the solid black and magenta dashed lines on the right, and the torus-dominated sequence outlined by the colors of the \citet{2007ApJ...663...81P} QSO1 template with increasing A(V), indicated with a solid grey line on the top-left, converging to the color of a \citet{2007ApJ...663...81P} Torus template (black square). The dashed and dashed-dotted grey lines show a similar sequence for the HDD template of \citet{Lyu2017} and the accretion disk model with slope $\alpha=1/3$ used in \textsc{prospector-AGN}. The green and blue dashed lines and markers illustrate the sequence toward redder 1-to-3~$\mu$m colors with increasing dust emission from star-formation ($\log(L_{\rm IR}/L_{\odot})=12-13$) or the \citet{2008ApJ...685..160N} clumpy torus ($\log(L_{\rm IR}$/erg~s$^{-1}$)=44.0-45.5) relative to the stellar or accretion disk continuum (solid black and dashed grey lines on the right).}
\label{fig:rest_color}
\end{figure*}



Our results  confirm the flattening of the LRD SEDs in the rest near infrared. Previous LRD papers fitting NIRCam-only SEDs with empirical AGN templates (e.g., \citealt{2023ApJ...954L...4K,2023arXiv230514418B,2023arXiv230905714G}) implied that the steep rest-optical slope would continue into the NIR. This trend seemed to be confirmed by the handful of LRDs with MIRI detections up to F770W and F1000W (\citealt{2023ApJ...956...61A}; \citealt{2023arXiv230514418B}). However, as shown in section \ref{sec:sample3} and also \citet{2023arXiv231107483W}, the MIRI data at longer wavelengths indicate that LRDs have a flattening in the SED between 1 and 2~$\mu$m (rest). Interestingly, the SEDs of four out of the Golden Five LRDs at redshifts $z<7$, which have direct detections of the NIR continuum at $\lambda\gtrsim$2~$\mu$m, appear to show upturns with different slopes at longer wavelengths. This suggests that, while it does not dominate, the amount of dust emission from star formation or an AGN can vary substantially from object to object.

Our four codes fit the NIR spectral range with a combination of the dominant source of UV-optical emission (i.e., stars or an accretion disk) plus a variable contribution from dust emission, either from star-formation ({\sc synthesizer-AGN}, and \textsc{prospector-SF}) or from an AGN torus (\textsc{prospector-AGN},
and \textsc{prospector-AGN$+$}  depending on the source). The left panel of Figure~\ref{fig:rest_color} illustrates this variation showing the rest-frame color-color diagram for the LRDs and
some templates and models. The 0.4-to-1~$\mu$m color probes the optical to NIR slope. This color is similar to the F277W-F444W used in the sample selection but it is not affected by emission lines and, thus, is a better proxy for the amount of dust attenuation. The color is also similar to the rest-frame $V-J$ which is a known tracer of large
dust attenuation in the UVJ diagram ($V-J>1.5$~mag for very dusty galaxies, e.g., \citealt{brammer11}, \citealt{wuyts11b}). 

The behavior can be discussed in terms of the color-color behavior in Figure~\ref{fig:rest_color}. The black horizontal line in the figure shows the flat 1-to-3~$\mu$m color of a stellar population with zero contribution from dust emission, which peaks at 1.6~$\mu$m. The magenta region indicates the redder colors up to 0.5~mag relative to the stellar continuum due to increasing amounts of nebular continuum (magenta dashed line in the left panel). The 3 grey lines indicate the color tracks with increasing attenuation (A(V)=1 to 4~mag) for the \citet{2007ApJ...663...81P} QSO1 template (solid), the hot dust deficient (HDD) template of \citet{Lyu2017} used in \textsc{prospector-AGN+} (dashed-dotted), and the accretion disk model with declining slope ($\alpha=1/3$) and zero dust emission used in \textsc{prospector-AGN}. The right panel of Figure~\ref{fig:rest_color} illustrates some of same trends as they affect the SED templates.

Figure~\ref{fig:rest_color} illustrates how the modeling codes populate the color-color diagram {\it between} the flat (color $\sim0$~mag) stellar-only sequence, and the hot-dust dominated sequence of the QSO1 template with increasing contributions from dust emission (i.e., larger IR-luminosities) from
star-formation or an AGN, relative to the stellar or accretion disk continua. The color-color tracks ranging from moderate to high infrared luminosities (indicated in log$(L(L_\odot))$ for the star-forming case and log$(L(ergs~ s^{-1})$ for AGNs) are computed by scaling the dust emission (f$_{\rm dust}$/f$_{\rm total}$$=$5\% to 60\% at 2~$\mu$m) of JADES$-$57356. As expected, the star-forming dust templates exhibit larger luminosities than the torus at similar 1-to-3~$\mu$m colors because their SEDs extend to longer wavelengths with a more prominent peak (see also Figure~\ref{fig:appB1} in the appendix). While a single NIR color is not able to capture all the nuances of the different dust emission templates, Figure~\ref{fig:appB1} shows that the torus and star-forming dust templates in \citet{2007A&A...461..445S} and \citet{2008ApJ...685..160N} have similar slopes for the same normalization (i.e., same f$_{\rm dust}$/f$_{\rm total}$ value). Consequently, the four  modeling codes can all, in principle, reproduce the NIR continuum   of the LRDs.

 The Golden Five galaxies with direct detections beyond rest-frame $\sim$2~$\mu$m exhibit colors that are at least 1~mag redder than a flat, stellar-only SED. In particular, two of them (JADES$-$57356 and JADES$-$79803) have very red colors, [1-to-3~$\mu$m]$\sim$~2~mag, indicative of large dust emissions and IR luminosities. JADES-219000 and JADES-211388 are not well constrained beyond 2~$\mu$m because of their higher redshifts, but upper/lower triangles show the range in possible colors spanned between the \textsc{prospector-SF} and \textsc{synthesizer-AGN} best-fits, which feature different amounts of dust emission within the upper limits of the redder MIRI bands. These limits fall within the overall behavior of all the sources in the figure.  The broad range overall in Figure~\ref{fig:rest_color} 
 highlights the need for deep, long wavelength MIRI data to constrain precisely the amount and heating nature of dust emission in LRDs.

However, we can already see that fitting the colors with a purely 
 AGN-dominated model requires an accretion disk model with declining slope (dashed grey line) and only a small contribution from dust torus emission (thin blue line) to successfully reproduce the bluest 1-to-3~$\mu$m$\lesssim$1.5 colors at the lower limit of the HDD template. This would differ significantly from lower redshift AGN, which tend to have strong emission from their circumnuclear tori. This issue is mitigated by  \textsc{prospector-AGN+}, which  can reproduce those colors with a hybrid of AGN {\it and} stellar emission. 

We now discuss the LRDs detected only at shorter wavelengths. Figures~\ref{fig:fits6} and \ref{fig:fits7} show the stacked SEDs for the LRDs detected up to F1000W and F770W, which reveal that their SEDs are only well constrained up to rest-frame $\sim$1.6~$\mu$m and $\sim$1~$\mu$m, respectively. Consequently, the 1-to-3~$\mu$m colors of their best-fit models span a much larger range from the \textsc{synthesizer-AGN}, stellar-only fits with [1-to-3~$\mu$m]$\sim$~0~mag (e.g., top-left panel of Figure~\ref{fig:fits6}), to the much redder best-fit models of \textsc{prospector-AGN+} and \textsc{prospector-AGN} (bottom-left and right), which sometimes fit the MIRI upper limits with pronounced upturns at $\lambda>$2~$\mu$m. It is worth mentioning the F770W-only sample places more restrictive constraints against very red QSO-like colors than the F1000W-only sample. This is because, as shown in Figure~\ref{fig:colors}, the F1000W-only LRDs have redder F777W-F1000W colors than the upper limits of the F770W-only LRDs and thus lead to a stronger flattening of the SED in the 1-2$\mu$m range. Consequently, the possible upturn after 2~$\mu$m is not nearly as red.
 
In summary, the exact contribution from dust emission to the NIR SEDs of the LRDs is still poorly constrained by the available MIRI data at long wavelengths. Nonetheless, the constraints point to a certain diversity in the dust emission. This emission is clearly larger than stellar-only SEDs but in many cases lower than prior expectations based on QSO templates. Overall, the main conclusion is that most LRDs could harbor some (relatively large) amount of dust emission and the heating source must be intense, although not necessarily a dominant obscured AGN.

\subsection{Physical properties of the LRDs detected in the bluest MIRI bands}

All other LRDs in our sample detected up to F1000W, apart from the Golden Five, are shown jointly in the SED plots provided in Figure~\ref{fig:fits6}. In the same way, the SED fits for all F770W detections not included in any previous SED plot, as well as all sources not detected by MIRI are shown and discussed in Appendix~\ref{app:allSEDs}. Physical properties derived from each code for individual galaxies as well as statistical properties obtained for the whole sample and subsamples are given in Tables~\ref{tab:props} and \ref{tab:stats}.

Overall, the properties of those samples with fewer MIRI points in their SEDs are similar to the Golden Five galaxies; the clearest difference is that some blue and/or very flat SED sources start to enter the selection. These are selected due to a very strong F444W emission probably linked to a high-EW emission line, but the slope of the SED is not very different in the rest of the SW and LW filters, with a quite flat slope (or even blue, as in the case of JADES$-$187025).

Even though the dust emission spectral region is not fully probed for the 10~$\mu$m sample, and to an even  lesser  extent the 7.7~$\mu$m and non-MIRI samples, the upper limits imposed by the MIRI data at longer wavelengths, more specifically, at 12.8 and 15~$\mu$m, indicate a very similar behavior of the spectral range around 2-3~$\mu$m (rest) compared to what we showed for the Golden Five galaxies. Indeed, the SED flattens, indicating that possible dust emission powered by stellar or AGN heating is not dominant and could only start adding significant flux redward of 3~$\mu$m. The properties we infer from the UV-to-NIR SEDs are also similar for the F1000W, F770W, and no-MIRI subsamples compared to those obtained for the Golden Five galaxies. There are, however, some observational trends, which translate to differences in physical properties.

First, the Golden Five galaxies are brighter than the rest of the sources in the full sample (cf. Figure~\ref{fig:selection1}). The median and quartiles for F444W are $25.4_{25.1}^{26.1}$~mag, compared to $26.1_{25.8}^{26.3}$~mag for the F1000W sample, $25.9_{25.7}^{26.1}$~mag for the F770W sample, and
$27.6_{26.8}^{27.8}$ for the sources with no MIRI detection. Concerning colors, the F1000W sample is redder than the Golden Five, and the overall sample. As shown in Figures~\ref{fig:selection1} and \ref{fig:colors}, the F1000W sample is among the reddest in both the NIRCam and MIRI colors: F277W-F444W=2.2~mag, F444W-F770W=1~mag and F770W-F1000W=0.9~mag, versus for the overall sample they are 1.4~mag, 0.7~mag and 0.2~mag respectively.  Comparatively, the largest difference is in F770W-F1000W color difference, where the F1000W sample has similar colors to 2 galaxies among the Golden Five, JADES-79803 and JADES-211388, whose MIR spectral range is fitted with dust tori. We note that for JADES-211388, at $z_\mathrm{sp}=8.3846$, F1000W lies on top of the He~I$+$Pa-$\gamma$ line, which partially explains the red color.

Looking at the rest-frame colors and stacked SEDs in Figures~\ref{fig:rest_color} and \ref{fig:fits6}, we find that the F1000W sample has also redder 0.4-to-1.0~$\mu$m colors than the Golden Five sources, suggesting that they are dustier (see next paragraphs). Interestingly, using a longer baseline color 0.25-to-1.0~$\mu$m (similar to NUV-J), which probes into the relatively flat UV SED of the LRDs, we find even larger differences between the Golden Five galaxies and the F1000W samples, 0.25-to-1.0~$\mu$m$~=~$2.1~mag for the Golden Five vs. 3.5~mag. These colors and the multi-component SED modeling discussed in the previous section indicate that the colors
relative to rest 1~$\mu$m are partially driven by differences in the relative luminosity of the component dominating the rest-UV (young, unobscured stellar population) and the component dominating the rest-optical (older stellar population or obscured accretion disk). As the flat UV component scales up in brightness, it leads to bluer 0.25-to-1.0~$\mu$m and 0.4-to-1.0~$\mu$m colors, but perhaps not because of a change in the intrinsic properties of the component dominating the optical range. It could be that a brighter UV reveals a larger fraction of the starburst emission percolating through the compact dust cloud (possibly linked to a higher burst strength or younger age, apart from dust-star relative geometry), or perhaps it shows a more massive stellar host for the obscured AGN. This interpretation also helps to explain the larger scatter in the UV region of the stacked SEDs relative to the optical region. That is, while all LRDs have distinctive blue-UV and red-optical SEDs, there is a larger diversity in the UV emission for a similar optical-to-NIR slope that might reflect variations in the relative luminosity of two different components.

The overall colors of the LRDs range from 0.25-to-1.0~$\mu$m~$=$~2 to 4.5~mag. Taking 0.25-to-1~$\mu$m~$=$~3~mag as an intermediate value, we find that all of the Golden Five galaxies exhibit bluer colors, versus only 30\% (2/7) of the galaxies in the F1000W and F770W samples. This emphasizes again that the Golden Five galaxies are intrinsically bluer than the other samples in all colors. Interestingly, we also find a trend toward stronger emission lines (larger EWs) with bluer 0.25-to-1.0~$\mu$m colors. For example, this trend is seen among the bluest Golden Five sources (JADES$-$79803 and JADES$-$204851) and in the handful of galaxies with bluer colors in the F1000W sample (JADES$-$210600, JADES$-$214552 and JADES$-$217926) and F770W sample (JADES$-$187025 and JADES$-$197348).

\begin{figure*}[htp!]
\centering
\includegraphics[clip,trim=2.0cm 1.0cm 2.3cm 2.5cm,width=18.cm,angle=0]{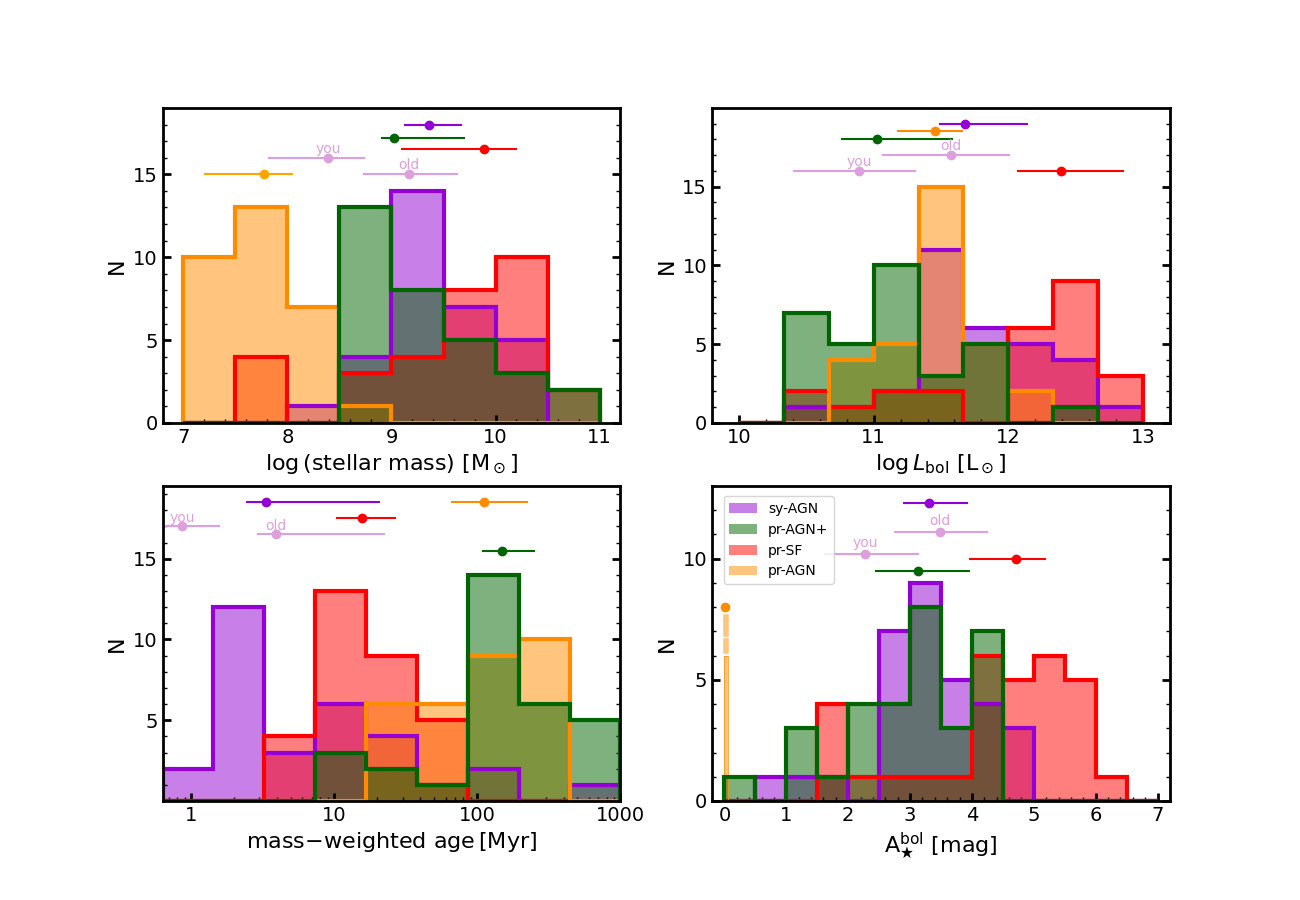}
\caption{Statistical stellar properties of  LRDs, according to the 4 SED-fitting codes described in Section~\ref{sec:modeling}. From top to bottom, left to right, we show stellar masses, bolometric luminosities (obtained by integrating the stellar emission correcting for the effects of dust attenuation), mass-weighted ages, and bolometric stellar luminosity attenuation. Medians and quartiles are shown for each distribution. For the results provided by \textsc{synthesizer-AGN}, we separate statistics for the young and old stellar populations (marked as {\it you} and {\it old}) as well as the integrated values. }
\label{fig:props1}
\end{figure*}

Based on the previous observational differences, our modeling of the SEDs provides some trends also in physical properties. The Golden Five galaxies lie at smaller redshifts than the rest of sources, median values $z=5.5$, compared to $z=6.3$, $z=5.9$, and $z=7.4$ for the F1000W, F770W, and non-MIRI subsamples (check Table~\ref{tab:stats} for more statistical information). So part of the reason for the detections of the Golden Five in many MIRI bands can be linked to redshift.

The stellar masses of the Golden Five galaxies are 0.2-0.4~dex larger than those of the other subsamples, with values around $10^{9.6}\,\mathrm{M}_\odot$ for the former.
Interestingly, the stellar population attenuation is lower for the Golden Five galaxies, $\mathrm{A(V)}=3.0$~mag, compared to F1000W sources ($\mathrm{A(V)}=3.2$~mag), F770W galaxies ($\mathrm{A(V)}=3.3$~mag), and non-MIRI sources ($\mathrm{A(V)}=3.9$~mag).

We conclude that sources detected only in the bluest MIRI filters are not just fainter (less massive) versions of the Golden Five galaxies, lying at higher redshifts. Indeed, there are also other differences in physical properties (deriving from differences in SEDs) which point to larger attenuations. If the fainter MIRI sources present larger attenuations, but still are not detected by MIRI at the longest wavelengths, the interpretation would be that the host dust emission is not enhanced compared to the Golden Five galaxies, which would lead to a dominant role of dust-enshrouded star formation rather than obscured nuclear activity for a significant fraction of these sub-populations.

\subsection{Statistical stellar and AGN properties of LRDs}
\label{sec:propstats}

In this section, we discuss the general properties of our sample of LRDs. Figure~\ref{fig:props1} shows the distributions of stellar mass, bolometric luminosity, mass-weighted age, and bolometric stellar light attenuation. All statistical information is summarized in Table~\ref{tab:stats}.

The typical stellar mass of LRDs is $\log\mathrm{M_\star/M_\odot}=9.4_{9.1}^{9.7}$ (median and quartiles) according to \textsc{synthesizer-AGN}. Considering the full redshift range of our sample of LRDs, $5\lesssim z\lesssim 9$, and the number density of galaxies in the stellar mass range $9.0<\log \mathrm{M}/\mathrm{M}_\odot<10$ detected by CEERS based on the v0.51 catalogs \citep{2023arXiv231104279F}, we calculate that the LRDs account for $14\pm3$\% of the full population of galaxies (subject to uncertainties due to cosmic variance). This translates to a comoving density of LRDs of $10^{-4.0\pm0.1}$~Mpc$^{-3}$, which is quite constant across the $5<z<9$ redshift range, with differences $<0.1$~dex between $5<z<7$ and $7<z<9$. The estimates for both the LRDs and the other galaxies are subject to a number of potential systematic errors, but the estimate indicates that the LRDs represent a significant, but not a dominant, population over this redshift range. Given the ranges of  Universe ages probed by our sample, 200~Myr for  $7<z<9$ and 400~Myr for $5<z<7$, the $\sim10$\% frequency within the global population could be interpreted in terms of a duty cycle around 20--40~Myr, which points to  starburst behavior. 

\begin{figure}[htp!]
\centering
\includegraphics[clip,trim=1.5cm 12.0cm 15.5cm 2.0cm,width=8.5cm,angle=0]{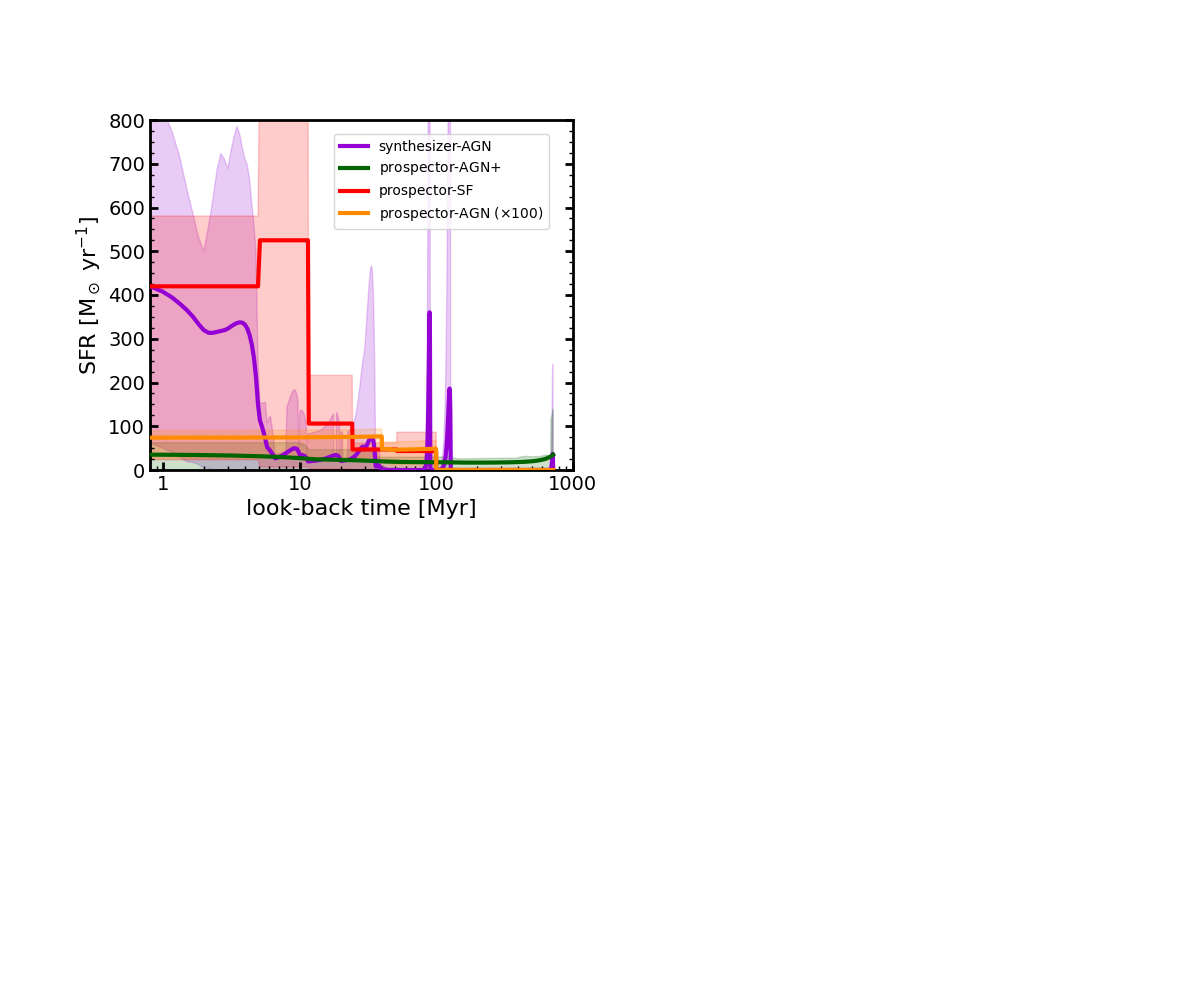}
\caption{Average SFHs of LRDs according to the fitting codes presented in Section~\ref{sec:modeling}. Averages and scatter are shown as lines and shaded regions.}
\label{fig:props2}
\end{figure}

Larger masses are obtained by \textsc{prospector-SF}, $\log\mathrm{M_\star/M_\odot}=9.9_{9.2}^{10.2}$, mainly because the mass-weighted ages are older, between 10 and 100~Myr, with median and quartiles being $t_\mathrm{m-w}=16_{10}^{27}$~Myr. \textsc{synthesizer-AGN} fits the SEDs with significantly younger stellar populations, typically $t_\mathrm{m-w}=3_{2}^{20}$~Myr. The 2 stellar populations considered by \textsc{synthesizer-AGN} are typically younger than 20~Myr. In fact, the average SFH of LRDs obtained by \textsc{synthesizer-AGN} and shown in Figure~\ref{fig:props2} indicates that LRDs are experiencing a very intense episode of star formation extending for nearly 10~Myr and with a very compact size. The burst would be in part heavily dust-enshrouded, with some younger stars having cleared the interstellar medium and being directly observable through much smaller dust optical paths. The young ages are expected for starbursts with strong emission lines (as present in some of our galaxies), with large amounts of dust (confirmed for a significant fraction of the whole sample), and with gas also feeding a SMBH. \textsc{synthesizer-SF} provides a similar average SFH, extended almost at a constant level up to approximately 10~Myr, and decaying afterwards. However, the first age bin considered by \textsc{synthesizer-SF} encompasses the 2 bursts obtained by  \textsc{synthesizer-AGN}, the former adding more mass in ages around 10~Myr (with a large scatter, shown by the shaded region in Figure~\ref{fig:props2}).

Coming back to stellar content, even smaller masses compared to \textsc{synthesizer-AGN} are obtained by \textsc{prospector-AGN}, $\log\mathrm{M_\star/M_\odot}=7.8_{7.2}^{8.0}$, which only considers contributions of stellar light to the UV spectral region. These masses are similar to the values obtained for the youngest population in the \textsc{synthesizer-AGN} modeling. Given that \textsc{prospector-AGN} fits the optical and NIR spectral regions with an AGN, the stellar mass estimates are significantly smaller than what is needed to reproduce the emission at those wavelengths with stars. 

We conclude that the \textsc{prospector-AGN} stellar masses should be considered lower limits, since they assume little contribution of stellar light to the optical spectral range. \textsc{synthesizer-AGN} and \textsc{prospector-AGN$+$} typically obtain fits for which the optical emission is dominated by stars, hence the stellar mass estimates they obtain should be interpreted as more realistic or upper limits. The differences in stellar masses derived with \textsc{synthesizer-AGN} and \textsc{prospector-SF} exemplify the effect of the SFH, which is relatively important (0.4-0.6~dex) for these young galaxies whose mass-to-light ratios can change significantly as the most massive stars disappear.

The bolometric luminosities (including dust-absorbed energy) vary by a factor of 10 between the \textsc{prospector-SF} and \textsc{prospector-AGN} runs, while \textsc{synthesizer-AGN} lie in between, with a typical value of $\log\mathrm{L/L_\odot}_\mathrm{bol}=12.0_{11.6}^{12.4}$. Combining with the stellar masses, we conclude that LRDs present mass-to-light ratios of 1/400, typical of OB stellar associations (a B2 star would have that value, approximately), as expected based on the young ages. We remind the reader that given these very young ages and the starburst nature of LRDs, the {\it a priori} assumption of a universal IMF is quite relevant. The amount of OB stars formed, the quick and efficient formation of metals and dust, and the inferred stellar masses (or even the growth of a SMBH) are all affected by the IMF.

Finally, the bottom-right panel of Figure~\ref{fig:props1} shows the total attenuation of the stellar light in LRDs. All codes are consistent in assigning large dust content to this type of galaxy, with attenuation around 3-4~mag (i.e., 95-99\% of the light being absorbed by dust).

\begin{figure*}[htp!]
\centering
\includegraphics[clip,trim=2.0cm 1.0cm 2.3cm 2.5cm,width=18.cm,angle=0]{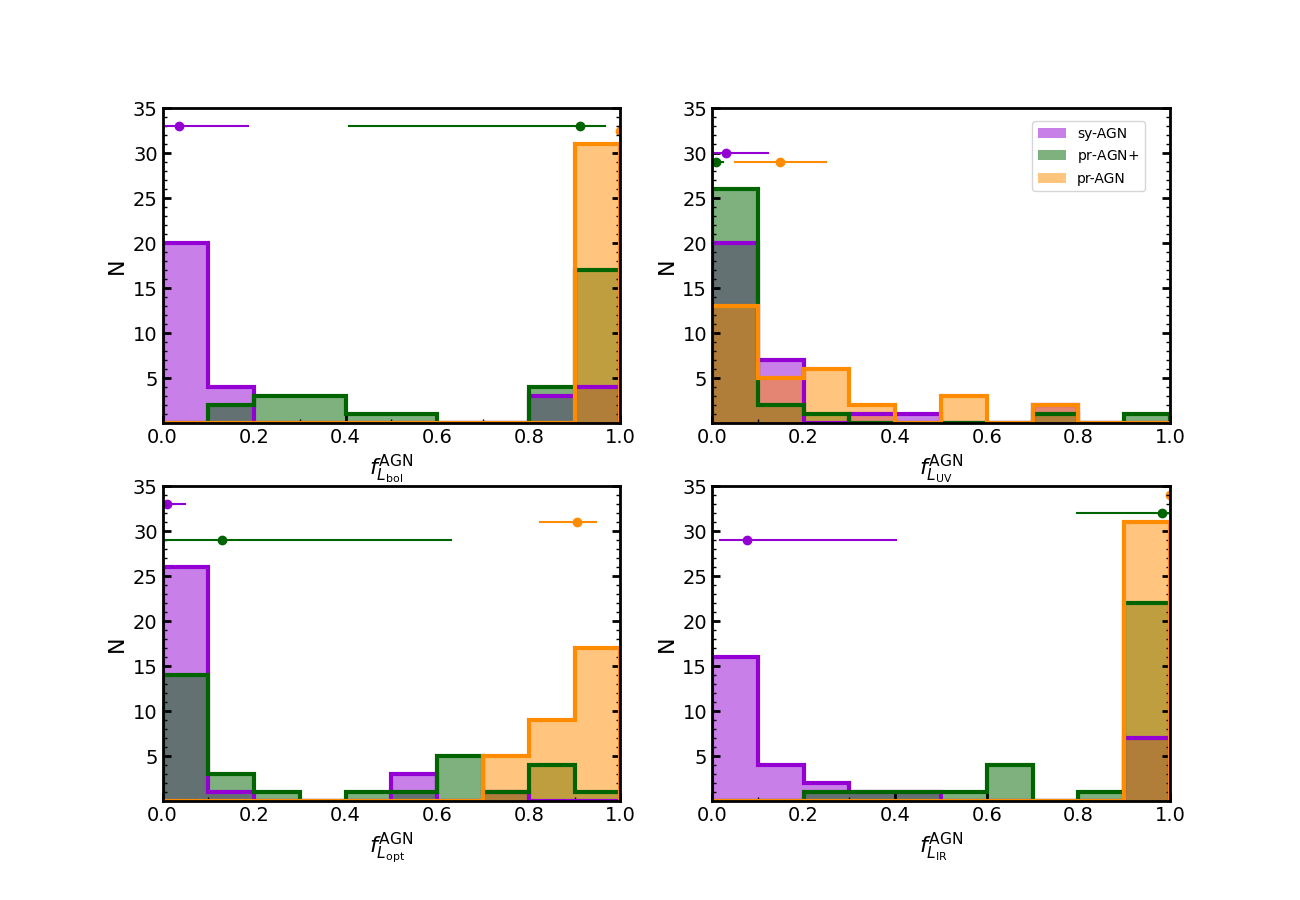}
\caption{Histograms of the fraction of integrated luminosity coming from an AGN, according to the SED fits presented in Section~\ref{sec:modeling}. On the top-left, we show results for the bolometric luminosity. On the top-right, for the UV luminosity, integrated up to 0.4~$\mu$m. On the bottom-left, results for the optical luminosity are provided, with $L_\mathrm{opt}$ defined as the integral between 0.4 and 2~$\mu$m. The bottom-right panel shows the histograms for IR wavelengths longer than 2~$\mu$m. Medians and quartiles for each spectral range and code are displayed at the top of the panels. }
\label{fig:props3}
\end{figure*}

The relative importance of the AGN and stellar components in LRDs is presented in Figure~\ref{fig:props3}. Here we show the fraction of the total luminosity coming from the AGN and integrated in several spectral ranges. We show bolometric, UV (integrated up to 0.4~$\mu$m), optical (from 0.4 to 2~$\mu$m), and IR (from 2~$\mu$m redwards) luminosity ratios for the AGN emission (with the rest coming directly from stars or dust heated by stars). We note that we only include 3 of the 4 codes in this plot, since \textsc{prospector-SF} does not consider any AGN contribution (although the dust models imply intense radiation fields which could be easily identified with an AGN).

For \textsc{synthesizer-AGN}, the bolometric luminosity of most of the sample is dominated by stars, with only $\sim$20\% of sources presenting AGN luminosity fractions larger than $f_{L_\mathrm{bol}}^\mathrm{AGN}>0.5$. In contrast,  \textsc{prospector-AGN$+$} obtains a much larger AGN contribution for most galaxies, with nearly 70\% of the galaxies presenting a bolometric luminosity fraction $>50$\%. This is a direct consequence, however, of this code fitting the MIRI non-detections assuming the 5$\sigma$ upper limit as an actual flux. This means that the \textsc{prospector-AGN$+$} results about the AGN luminosity ratio should be regarded as upper limits. In the same sense, \textsc{synthesizer-AGN} could be regarded as lower limits, since the MIR is not fitted for MIRI non-detections nor for F770W-only sources, since all bands with measured fluxes are well reproduced by stellar models alone (i.e., the dust emission is loosely constrained and it is not fitted). 

More consistent results are found for the UV luminosity fraction identified with an AGN. All codes agree that the UV spectral region of LRDs is best-reproduced by a young stellar population with varying but low dust content and relatively young ages (1-10~Myr). We remark here that \textsc{prospector-AGN} forces the UV to be dominated by unobscured stars, while the other 2 codes leave complete freedom in this spectral range in terms of a possible contribution from a UV-bright AGN or the amount of dust.

A similar behavior is observed in the optical, where both \textsc{synthesizer-AGN} and \textsc{prospector-AGN$+$} find that the emission is dominated by stars in most ($>75$\%) of the sample. The {\it a priori} assumption of \textsc{prospector-AGN} is that the AGN dominates the IR spectral region and results in the optical also being dominated by the nuclear activity for most sources; consequently, as we mentioned, the estimated stellar masses are considerably smaller than  the values obtained by the other codes.

Finally, the bottom-right panel of Figure~\ref{fig:props3} shows the fraction of the IR (dust) luminosity linked to the AGN. \textsc{prospector-AGN$+$}  and \textsc{synthesizer-AGN} show opposite distributions, but two facts must be taken into account to interpret this behavior. First, as mentioned earlier, \textsc{prospector-AGN$+$} fits upper limits as regular flux points. Those upper limits increase with a slope similar to what can be expected for an AGN torus. Second, the dust emission fits provided by \textsc{synthesizer-AGN} imply very intense radiation fields in dense, compact hot spots, whose properties would be indistinguishable between OB associations or an AGN (see discussion on Figure~\ref{fig:fits1} and Appendix~\ref{app:dust}).

\subsection{Spectroscopic properties of LRDs found in the literature}

Apart from JADES$-$204851, which has broad  H$\alpha$ emission at $z=5.4790$ ($\sim$2000~km~s$^{-1}$) \citep{2023arXiv230605448M}, as mentioned in Section~\ref{sec:properties}, there are other sources in our sample with relevant spectroscopic information. JADES$-$197348 was included in the JADES NIRSpec initial data release \citep{2023arXiv230602467B} and identified with a broad-line AGN \citep{2023arXiv230801230M}. Its spectrum shows a $\sim2500$~km~s$^{-1}$ wide component that accounts for two thirds of the H$\alpha$ total flux, while [OIII] shows no such component. Our fits to this source are dominated by a QSO-like spectrum in the optical and NIR in the case of \textsc{synthesizer-AGN} and \textsc{prospector-AGN}. JADES$-$154428 is found to present a broad-line component with FWHM $\sim$1800~km~s$^{-1}$ (Sun et al. 2024, in prep.); our fits include a non-negligible contribution from a QSO-like spectrum, dominating the SED (\textsc{prospector-AGN} and \textsc{prospector-AGN$+$} or accounting for $\sim$50\% of the emission at specific wavelengths (\textsc{synthesizer-AGN}). No other broad H$\alpha$ or H$\beta$ line component has been reported in the remaining 18 galaxies with available spectroscopy, i.e., 17\% of the spectroscopic sample are confirmed AGN hosts. For the rest of the spectroscopic sample, the presence of an AGN cannot be ruled out, since the broad-line region could be hidden due to geometrical effects. We also note that the typical $5\sigma$ depth of FRESCO NIRCam 3.9–-5.0-$\mu$m grism spectroscopy is $\sim 5 \times 10^{-18}$\,erg\,s$^{-1}$\,cm$^{-2}$ for broad emission line (FWHM\,$\sim$\,1000\,km\,s$^{-1}$) from a point source and, therefore, faint broad line emission (possibly from an AGN not dominating the continuum) can remain undetected.

\section{Summary and conclusions}
\label{sec:conclusions}

We characterize the nature of Little Red Dots (LRDs) in the JADES field by analyzing their spectral energy distributions including the mid-infrared fluxes provided by the SMILES program for all MIRI broad-band filters. These data probe the rest-frame near-and mid-infrared where stellar emission and/or obscured AGN emission peak. After removing brown dwarfs, which contaminate our sample at the 15\% level, we arrive at a sample of 31 LRDs, the surface density being 0.9~arcmin$^{-2}$. This translates to a number density $10^{-4.0\pm0.1}$~Mpc$^{-3}$, accounting for 10--15\% of the global population of galaxies with similar redshifts ($z\sim7$) and stellar mass ($\log \mathrm{M_\star/M_\odot}=9.5$). Two thirds are detected in the F560W and F770W filters (all sources brighter than F444W$<$26.5~mag), two fifths in F1000W, one seventh in F1500W, one thirteenth in F1800W, and one source in F2100W, down to 5$\sigma$ limits between 26.1 and 22.6~mag. The MIRI detection fraction is largely dependent on the F444W brightness, but we find an additional trend towards more detections at 10~$\mu$m or beyond with bluer F277W-F444W colors.

We find that the observed MIRI colors of the LRDs, in combination with the reddest NIRCam bands, are bluer than the typical obscured QSO templates, which are dominated by the torus warm/hot dust emission at $\lambda\gtrsim1$~$\mu$m. Indeed, the rest-frame NIR spectral range exhibits a much shallower slope that is consistent with the peak of stellar emission at around $\sim1.6$~$\mu$m.

We modeled the rest-frame ultraviolet to mid-infrared spectral energy distributions with a battery of codes that include AGN and stellar emission templates. The various outputs allow us to identify the best fits to the distinctive short wavelength blue plus long wavelength red colors of LRDs under a range of assumptions. They also let us examine which of our conclusions are the most robust (e.g., are reflected by a number of the modeling techniques). 


In general, stellar-dominated models obtain a better agreement to the near-infrared at 1-2~$\mu$m as well as in the UV at $\lambda<0.4$~$\mu$m. The AGN-dominated models where emission from an obscured accretion disk dominates the optical and NIR and the torus takes over at longer wavelengths $\lambda\gtrsim2$~$\mu$m provide a better agreement than the typical QSO templates, but still worse than the stellar models. Furthermore, this AGN-dominated model also has conceptual problems given that many of the LRDs do not present emission lines,  which should be expected for the direct detection of the accretion disk (and its broad- and narrow-line regions). Consequently, we favor the interpretation that the UV-to-optical spectral range of most LRDs is dominated ($>50$\% luminosity ratio) by stars.

In the rest-frame near-infrared, we find that the LRDs detected in the reddest MIRI bands, beyond rest wavelengths $\lambda\gtrsim2$~$\mu$m, have color differences redder than can be expected for stars alone (even accounting for nebular emission associated with a young starburst), and consistent with some amount of emission from dust heated by star formation or an AGN. The upper limit of the MIRI colors for most LRDs also rules out that the near-infrared emission is strongly AGN-dominated, but the loose constraints in some of them can still accommodate similar amounts of dust emission as in the LRDs detected in the rest-frame mid-infrared.

Given that the rest-frame UV/optical spectral range is dominated by stars, we estimate stellar masses. The modeling of the stellar emission must consider the large attenuations implied by the red NIRCam long-wavelength colors and MIRI fluxes and the need for two distinct (in terms of age) stellar populations, with differential attenuation levels or an extremely gray attenuation law, to account for the blue/flat NIRCam-SW emission as well the presence of emission lines. With this in mind, we obtain typical stellar masses for LRDs around $\log \mathrm{M_\star/M_\odot}=9.4_{9.1}^{9.7}$, significantly smaller than what can be obtained with simple recipes for the Star Formation History (e.g., single exponential burst) and attenuation law (e.g., single Calzetti law, as noted by \citealt{2023arXiv230514418B}). This mass estimate can be biased due to uncertainties in the Star Formation History at the 0.5~dex level (overestimated, if, for example, the stellar emission only contributes significantly to the ultraviolet emission), and can be affected by a possible (but less favored by most of our models) significant contribution of an AGN to the optical and near-infrared spectral range at the 1.5~dex level.

Very young stellar ages (typically around 10~Myr or younger) are supported by high equivalent width lines seen in some LRDs (with spectroscopic and imaging data), and the presence of large amounts of dust, which is a common feature of all models for the full sample, and can be expected for gas-rich, quickly enriched dense starbursts with large amounts of OB stars (affected by the IMF). However, despite the presence of large amounts of dust, LRDs are characterized by a relatively blue emission at wavelengths $<0.4$~$\mu$m, which can only be reproduced if this spectral range is dominated by (1) a QSO; (2) has holes in the interstellar medium surrounding star-forming regions that allow us to see unobscured very young star formation; or (3) have a gray attenuation law typically linked to significant scattering (with implications to the relative geometry of dust and stars and the dust clumpiness). Indeed, the large $\sim$3~mag scatter in the UV-to-NIR colors within the typical blue-red SED of the LRDs might indicate different burst strengths or different fractions of percolating light through the compact dust envelope. The presence of a QSO is hinted by some medium-band colors consistent with MgII emission. Nevertheless, for most of the sources, the best spectral energy distribution fit is obtained with just stars presenting a differentiated and/or gray dust attenuation law. A significant fraction of the total stellar emission of LRDs comes from OB stellar clumps mostly embedded in dense dusty regions with large optical paths, $\mathrm{A(V)}\gtrsim10$~mag, with the integrated stellar emission obscured at the  90-95\% level, which would explain the near- and mid-infrared emission jointly with some contribution from an obscured AGN.

\begin{acknowledgments}
We thank the referee for their constructive comments to our original manuscript. PGP-G acknowledges support from grant PID2022-139567NB-I00 funded by Spanish Ministerio de Ciencia e Innovaci\'on MCIN/AEI/10.13039/501100011033,
FEDER {\it Una manera de hacer
Europa}. This work was supported by NASA grants NNX13AD82G
and 1255094. The work was also  supported by  NIRCam Development
Contract NAS5-02105 from NASA Goddard Space Flight
Center to the University of Arizona. This work is based on
observations made with the NASA/ESA/CSA James Webb
Space Telescope. The data were obtained from the Mikulski
Archive for Space Telescopes at the Space Telescope Science
Institute, which is operated by the Association of Universities
for Research in Astronomy, Inc., under NASA contract NAS
5-03127 for JWST. These observations are associated with
program \#1207. DP acknowledges support by the Huo Family Foundation through a P.C. Ho PhD Studentship. BER acknowledges support from the NIRCam Science Team contract to the University of Arizona, NAS5-02015, and JWST Program 3215. AJB acknowledges funding from the ``FirstGalaxies'' Advanced Grant from the European Research Council (ERC) under the European Union’s Horizon 2020 research and innovation programme (Grant agreement No. 789056).
\end{acknowledgments}

\vspace{5mm}
\facilities{JWST (NIRCam), JWST (MIRI), HST (ACS), HST (WFC3). All the JWST data used in this paper can be found in MAST: \dataset[10.17909/jmxm-1695]{http://dx.doi.org/10.17909/jmxm-1695} and \dataset[ 10.17909/8tdj-8n28]{http://dx.doi.org/10.17909/8tdj-8n28}.}

\software{astropy \citep{2013A&A...558A..33A,2018AJ....156..123A},  
          Cloudy \citep{2013RMxAA..49..137F}, {\sc prospector} \citep{leja17, 2019ApJ...876....3L, 2021ApJS..254...22J}, {\sc synthesizer} \citep{2003MNRAS.338..508P,2008ApJ...675..234P}.
          }

\begin{deluxetable*}{llcccl}
\caption{Sample of LRDs}
\tablehead{\colhead{Galaxy name} & \colhead{JADES DR2 ID} & \colhead{F444W} & \colhead{F277W-F444W} & \colhead{F150W-F200W} & \colhead{MIRI subsample}\\
& & [mag] & [mag] & [mag]}
\startdata
JADES-GS-53.12497$-$27.86835 & 42645  & $26.57\pm0.04$ & $+1.01\pm0.09$ & $+0.26\pm0.19$ & no MIRI/SMILES      \\
JADES-GS-53.11532$-$27.85922 & 57356  & $25.34\pm0.01$ & $+1.11\pm0.02$ & $+0.51\pm0.14$ & Golden Five         \\
JADES-GS-53.09116$-$27.85260 & 68544  & $27.91\pm0.05$ & $+2.47\pm0.49$ & $-0.7\pm1.8$ & no MIRI/SMILES      \\
JADES-GS-53.10831$-$27.85101 & 70714  & $27.72\pm0.08$ & $+1.30\pm0.25$ & $+1.1\pm1.2$ & no MIRI/SMILES      \\
JADES-GS-53.11993$-$27.84640 & 75654  & $26.56\pm0.03$ & $+1.18\pm0.08$ & $+0.72\pm0.23$ & no MIRI/SMILES      \\
JADES-GS-53.14763$-$27.84205 & 79803  & $26.50\pm0.04$ & $+0.97\pm0.09$ & $+0.30\pm0.35$ & Golden Five         \\
JADES-GS-53.12540$-$27.83997 & 81400  & $26.66\pm0.04$ & $+2.16\pm0.31$ & --             & F770W sample        \\
JADES-GS-53.13383$-$27.82825 & 90354  & $27.03\pm0.04$ & $+2.63\pm0.31$ & $-0.8\pm1.1$ & no MIRI/SMILES      \\
JADES-GS-53.15905$-$27.81825 & 99267  & $27.82\pm0.05$ & $+1.11\pm0.13$ & $+0.20\pm0.28$ & no MIRI/SMILES      \\
JADES-GS-53.17603$-$27.81740 & 99915  & $27.62\pm0.04$ & $+1.02\pm0.10$ & $-0.14\pm0.20$ & no MIRI/SMILES      \\
JADES-GS-53.15933$-$27.81175 & 104238 & $26.81\pm0.02$ & $+1.04\pm0.05$ & $-0.07\pm0.22$ & F770W sample        \\
JADES-GS-53.12543$-$27.78744 & 120484 & $25.76\pm0.01$ & $+1.48\pm0.03$ & $+0.19\pm0.42$ & F770W sample        \\
JADES-GS-53.12690$-$27.78615 & 121710 & $25.38\pm0.01$ & $+2.45\pm0.04$ & $+0.13\pm0.23$ & F1000W sample       \\
JADES-GS-53.17280$-$27.78315 & 124327 & $28.07\pm0.10$ & $+1.46\pm0.34$ & $+0.3\pm2.0$ & no MIRI/SMILES      \\
JADES-GS-53.14430$-$27.77986 & 126594 & $27.34\pm0.04$ & $+1.53\pm0.13$ & $+0.13\pm0.52$ & F770W sample        \\
JADES-GS-53.20400$-$27.77210 & 132229 & $26.07\pm0.01$ & $+2.34\pm0.09$ & $-0.28\pm0.24$ & F1000W sample       \\
JADES-GS-53.19077$-$27.76788 & 136872 & $27.71\pm0.05$ & $+0.97\pm0.12$ & $-0.04\pm0.25$ & no MIRI/SMILES      \\
JADES-GS-53.14795$-$27.75993 & 143133 & $27.62\pm0.06$ & $+1.60\pm0.24$ & $+0.52\pm0.60$ & no MIRI/SMILES      \\
JADES-GS-53.15817$-$27.73913 & 154428 & $25.79\pm0.01$ & $+2.07\pm0.06$ & $+0.04\pm0.20$ & F1000W sample       \\
JADES-GS-53.09642$-$27.85309 & 184838 & $26.97\pm0.02$ & $+1.41\pm0.08$ & $+0.46\pm0.28$ & no MIRI/SMILES      \\
JADES-GS-53.10605$-$27.84823 & 187025 & $25.48\pm0.01$ & $+0.95\pm0.06$ & $-0.27\pm0.09$ & F770W sample        \\
JADES-GS-53.14365$-$27.82553 & 194809 & $28.00\pm0.10$ & $+1.03\pm0.22$ & $-0.11\pm0.38$ & no MIRI/SMILES      \\
JADES-GS-53.12654$-$27.81809 & 197348 & $26.05\pm0.02$ & $+1.44\pm0.05$ & $-0.10\pm0.13$ & F770W sample        \\
JADES-GS-53.12142$-$27.79491 & 203749 & $26.00\pm0.01$ & $+3.04\pm0.13$ & $+1.30\pm0.81$ & F1000W sample       \\
JADES-GS-53.13859$-$27.79025 & 204851 & $24.47\pm0.00$ & $+1.78\pm0.01$ & --             & Golden Five         \\
JADES-GS-53.16611$-$27.77204 & 210600 & $26.21\pm0.02$ & $+1.24\pm0.06$ & $-0.18\pm0.11$ & F1000W sample       \\
JADES-GS-53.18354$-$27.77016 & 211388 & $26.55\pm0.02$ & $+1.01\pm0.04$ & $-0.05\pm0.08$ & Golden Five         \\
JADES-GS-53.17922$-$27.75872 & 214552 & $26.45\pm0.01$ & $+1.25\pm0.04$ & $+0.07\pm0.13$ & F1000W sample       \\
JADES-GS-53.18478$-$27.74405 & 217926 & $26.83\pm0.03$ & $+1.22\pm0.08$ & $+0.07\pm0.32$ & F1000W sample       \\
JADES-GS-53.16137$-$27.73767 & 219000 & $25.10\pm0.01$ & $+1.68\pm0.02$ & $+0.09\pm0.08$ & Golden Five         \\
JADES-GS-53.19212$-$27.75251 & 297689 & $26.60\pm0.02$ & $+1.14\pm0.05$ & $-0.04\pm0.19$ & F770W sample        \\
\enddata
\tablecomments{\label{tab:sample}Table with basic information (used for selection) about the sample of galaxies in this paper: IAU-format name, ID based on JADES DR2 catalog \citep{2023arXiv231012340E}, magnitude and colors used in the selection (see Figure~\ref{fig:selection1}), and subsample according to the reddest band counting with a MIRI detection.}
\end{deluxetable*}

\begin{deluxetable*}{lclcccc}
\caption{Physical properties of LRDs (complete version online)}
\tablehead{\colhead{ID} & \colhead{redshift} & \colhead{code} & \colhead{$\log \mathrm{M}/\mathrm{M}_\odot$} & \colhead{$\mathrm{A}_\bigstar^{\lambda}$ [mag]} & \colhead{$f_\lambda^\mathrm{AGN}$} &  \colhead{m-w age [Myr]} \\
 & &  &  & \colhead{$[\mathrm{bol,FUV^{old}_{you},V^{old}_{you}}]$} & \colhead{$[\mathrm{bol,UV,opt,IR}]$} & 
}
\startdata
42645   &  8.5   & sy-AGN      &  9.49 & [3.8,$^{9.7}_{1.5}$,$^{3.8}_{0.6}$] & [0.80,0.73,0.56,0.94]  &   3\\
        &        & pr-AGN$+$   &   8.89 & [1.1,1.4,0.3] & [0.02,0.00,0.00,0.05]  & 147\\
        &        & pr-SF       &  7.92 & [1.5,1.8,0.7] & [0.00,0.00,0.00,0.00]  &   4\\
        &        & pr-AGN      &  7.86 & [0.0,0.0,0.0] & [0.97,0.22,0.94,1.00]  &  20\\
57356   &  5.5   & sy-AGN      & 10.38 & [2.7,$^{7.9}_{2.3}$,$^{3.1}_{0.9}$] & [0.01,0.72,0.15,0.00]  & 122\\
        &        & pr-AGN$+$   &  10.03 & [2.0,3.0,1.7] & [0.00,0.00,0.00,0.00]  & 684\\
        &        & pr-SF       & 10.20 & [5.0,6.6,4.4] & [0.00,0.00,0.00,0.00]  &  46\\
        &        & pr-AGN      &  8.09 & [0.0,0.0,0.0] & [0.98,0.16,0.91,1.00]  & 131\\
68544   &  7.2   & sy-AGN      &  9.21 & [2.6,$^{1.0}_{9.7}$,$^{0.4}_{3.8}$] & [0.02,0.03,0.01,0.03]  &  10\\
        &        & pr-AGN$+$   &   9.07 & [3.8,10.4,2.6] & [0.03,0.02,0.03,0.03]  & 117\\
        &        & pr-SF       &  9.88 & [5.1,13.9,7.6] & [0.00,0.00,0.00,0.00]  &   8\\
        &        & pr-AGN      &  7.00 & [0.0,0.0,0.0] & [1.00,0.07,0.95,1.00]  & 234\\
70714   &  5.8   & sy-AGN      &  8.75 & [4.2,$^{1.0}_{9.7}$,$^{0.4}_{3.8}$] & [0.04,0.03,0.01,0.12]  &   2\\
        &        & pr-AGN$+$   &   8.94 & [2.9,3.7,2.1] & [0.92,0.01,0.63,0.94]  & 186\\
        &        & pr-SF       & 10.28 & [6.3,9.3,7.1] & [0.00,0.00,0.00,0.00]  &  18\\
        &        & pr-AGN      &  7.13 & [0.0,0.0,0.0] & [0.96,0.01,0.86,1.00]  &  25\\
\enddata
\tablecomments{\label{tab:props}Table with physical properties of the sample of galaxies in this paper. Apart from redshift (spectroscopic values given with 4 decimals), we provide results for each of the 4 stellar population synthesis codes plus AGN described in Section~\ref{sec:modeling}: \textsc{synthesizer-AGN}  (sy-AGN), \textsc{prospector-AGN$+$} (pr-AGN$+$), \textsc{prospector-SF} (pr-SF), and \textsc{prospector-AGN} (pr-AGN). We quote stellar masses  (surviving stars), attenuation of the bolometric, far-ultraviolet (FUV, i.e., 150~nm), and visual (V, i.e., 550~nm) emission from stars (including nebular emission, and providing results for the old and young stellar populations for \textsc{synthesizer-AGN} in the FUV and V cases), fraction of the luminosity associated with the AGN for the bolometric, UV, optical, and IR emission (check definition in Section~\ref{sec:propstats}), and mass-weighted ages.}
\end{deluxetable*}

\begin{deluxetable*}{lclcccc}
\caption{Statistical properties of LRDs}
\tablehead{\colhead{Sample (N$_\mathrm{gal}$)} & \colhead{redshift} & \colhead{code} & \colhead{$\log \mathrm{M}/\mathrm{M}_\odot$} & \colhead{$\mathrm{A}_\bigstar^{\lambda}$ [mag]} & \colhead{$f_\lambda^\mathrm{AGN}$} &  \colhead{m-w age [Myr]} \\
 & &  &  & \colhead{$[\mathrm{bol,FUV^{old}_{you},V^{old}_{you}}]$} & \colhead{$[\mathrm{bol,UV,opt,IR}]$} & 
}
\startdata
Full sample (31) & $6.9_{5.9}^{7.7}$ & sy-AGN      & $ 9.4_{ 9.1}^{ 9.7}$ & [$3.3_{2.9}^{3.9}$,$^{7.6_{5.9}^{9.8}}_{2.3_{1.5}^{4.3}}$,$^{3.0_{2.3}^{3.8}}_{0.9_{0.6}^{1.7}}$] & [$0.02_{0.00}^{0.10}$,$0.03_{0.00}^{0.12}$,$0.01_{0.00}^{0.05}$,$0.04_{0.01}^{0.21}$]  & $  3_{  2}^{ 20}$\\ 
 &  & pr-AGN$+$   & $ 9.0_{ 8.9}^{ 9.7}$ & [$2.5_{1.5}^{3.0}$,$3.0_{2.0}^{4.3}$,${1.8_{1.0}^{2.3}}$] & [$0.74_{0.01}^{0.89}$,$0.01_{0.00}^{0.02}$,$0.13_{0.00}^{0.63}$,$0.82_{0.02}^{0.94}$]  & $150_{110}^{249}$\\
 &  & pr-SF   & $ 9.9_{ 9.2}^{10.2}$ & [$4.7_{4.0}^{5.2}$,$6.6_{4.9}^{9.1}$,${4.0_{2.8}^{5.1}}$] & [$0.00_{0.00}^{0.00}$,$0.00_{0.00}^{0.00}$,$0.00_{0.00}^{0.00}$,$0.00_{0.00}^{0.00}$]  & $ 16_{ 10}^{ 27}$\\
 &  & pr-AGN   & $ 7.8_{ 7.2}^{ 8.0}$ & [$0.0_{0.0}^{0.0}$,$0.0_{0.0}^{0.0}$,${0.0_{0.0}^{0.0}}$] & [$0.98_{0.95}^{0.99}$,$0.15_{0.05}^{0.25}$,$0.91_{0.82}^{0.95}$,$1.00_{1.00}^{1.00}$]  & $111_{ 67}^{222}$\\
Golden Five (5) & $5.5_{5.5}^{6.8}$ & sy-AGN      & $ 9.6_{ 9.1}^{10.4}$ & [$3.0_{2.7}^{3.1}$,$^{6.1_{6.1}^{6.6}}_{1.5_{1.5}^{2.3}}$,$^{2.4_{2.4}^{2.6}}_{0.6_{0.6}^{0.9}}$] & [$0.01_{0.01}^{0.87}$,$0.12_{0.01}^{0.12}$,$0.15_{0.06}^{0.51}$,$0.00_{0.00}^{0.99}$]  & $  3_{  2}^{122}$\\ 
 &  & pr-AGN$+$   & $ 9.4_{ 9.0}^{10.0}$ & [$1.6_{1.2}^{2.0}$,$2.9_{1.5}^{3.0}$,${1.4_{0.8}^{1.7}}$] & [$0.74_{0.00}^{0.89}$,$0.00_{0.00}^{0.01}$,$0.06_{0.00}^{0.46}$,$0.85_{0.00}^{0.95}$]  & $129_{101}^{453}$\\
 &  & pr-SF   & $10.0_{ 9.9}^{10.2}$ & [$4.3_{3.9}^{5.0}$,$5.5_{5.0}^{6.4}$,${3.3_{2.8}^{4.4}}$] & [$0.00_{0.00}^{0.00}$,$0.00_{0.00}^{0.00}$,$0.00_{0.00}^{0.00}$,$0.00_{0.00}^{0.00}$]  & $ 34_{ 29}^{ 46}$\\
 &  & pr-AGN   & $ 8.2_{ 8.1}^{ 8.2}$ & [$0.0_{0.0}^{0.0}$,$0.0_{0.0}^{0.0}$,${0.0_{0.0}^{0.0}}$] & [$0.88_{0.87}^{0.94}$,$0.09_{0.05}^{0.16}$,$0.82_{0.77}^{0.83}$,$1.00_{1.00}^{1.00}$]  & $149_{131}^{240}$\\
F1000W (7) & $6.3_{5.5}^{6.7}$ & sy-AGN      & $ 9.2_{ 9.2}^{ 9.5}$ & [$3.2_{3.0}^{3.6}$,$^{7.9_{6.5}^{8.9}}_{2.3_{1.8}^{4.3}}$,$^{3.1_{2.6}^{3.5}}_{0.9_{0.7}^{1.7}}$] & [$0.01_{0.00}^{0.04}$,$0.02_{0.00}^{0.12}$,$0.00_{0.00}^{0.01}$,$0.04_{0.01}^{0.09}$]  & $  9_{  3}^{ 24}$\\ 
 &  & pr-AGN$+$   & $ 9.1_{ 9.0}^{ 9.9}$ & [$2.3_{1.5}^{2.6}$,$3.7_{2.4}^{4.0}$,${2.0_{1.3}^{2.2}}$] & [$0.80_{0.37}^{0.93}$,$0.03_{0.01}^{0.18}$,$0.62_{0.07}^{0.74}$,$0.82_{0.41}^{0.95}$]  & $150_{115}^{393}$\\
 &  & pr-SF   & $ 9.7_{ 9.6}^{10.1}$ & [$4.9_{4.4}^{5.3}$,$7.7_{6.5}^{8.3}$,${4.0_{3.4}^{5.0}}$] & [$0.00_{0.00}^{0.00}$,$0.00_{0.00}^{0.00}$,$0.00_{0.00}^{0.00}$,$0.00_{0.00}^{0.00}$]  & $ 13_{ 11}^{ 16}$\\
 &  & pr-AGN   & $ 7.9_{ 7.8}^{ 8.0}$ & [$0.0_{0.0}^{0.0}$,$0.0_{0.0}^{0.0}$,${0.0_{0.0}^{0.0}}$] & [$0.97_{0.96}^{0.98}$,$0.15_{0.08}^{0.22}$,$0.87_{0.82}^{0.92}$,$1.00_{1.00}^{1.00}$]  & $112_{ 89}^{174}$\\
F770W (7) & $5.9_{5.5}^{7.4}$ & sy-AGN      & $ 9.4_{ 9.2}^{ 9.6}$ & [$3.3_{3.1}^{4.0}$,$^{9.9_{7.5}^{10.1}}_{2.8_{1.7}^{4.0}}$,$^{3.9_{3.0}^{4.0}}_{1.1_{0.7}^{1.5}}$] & [$0.03_{0.01}^{0.09}$,$0.03_{0.02}^{0.10}$,$0.01_{0.00}^{0.02}$,$0.05_{0.03}^{0.20}$]  & $ 13_{  2}^{ 16}$\\ 
 &  & pr-AGN$+$   & $ 9.6_{ 8.9}^{ 9.8}$ & [$2.7_{1.6}^{3.2}$,$3.7_{1.9}^{8.1}$,${2.1_{1.1}^{2.7}}$] & [$0.01_{0.01}^{0.43}$,$0.00_{0.00}^{0.01}$,$0.00_{0.00}^{0.33}$,$0.02_{0.01}^{0.45}$]  & $229_{116}^{252}$\\
 &  & pr-SF   & $10.0_{ 9.3}^{10.2}$ & [$4.5_{3.6}^{4.9}$,$6.9_{4.5}^{9.7}$,${3.7_{2.8}^{4.4}}$] & [$0.00_{0.00}^{0.00}$,$0.00_{0.00}^{0.00}$,$0.00_{0.00}^{0.00}$,$0.00_{0.00}^{0.00}$]  & $ 27_{ 20}^{ 33}$\\
 &  & pr-AGN   & $ 7.7_{ 7.4}^{ 8.1}$ & [$0.0_{0.0}^{0.0}$,$0.0_{0.0}^{0.0}$,${0.0_{0.0}^{0.0}}$] & [$0.96_{0.90}^{0.99}$,$0.24_{0.07}^{0.54}$,$0.94_{0.80}^{0.96}$,$1.00_{1.00}^{1.00}$]  & $212_{ 86}^{239}$\\
no-MIRI (12) & $7.4_{7.0}^{8.3}$ & sy-AGN      & $ 9.4_{ 9.0}^{ 9.7}$ & [$3.9_{3.0}^{4.2}$,$^{7.5_{1.4}^{9.8}}_{3.1_{2.0}^{9.7}}$,$^{3.0_{0.6}^{3.8}}_{1.2_{0.8}^{3.8}}$] & [$0.02_{0.00}^{0.09}$,$0.03_{0.00}^{0.11}$,$0.01_{0.00}^{0.03}$,$0.04_{0.01}^{0.20}$]  & $  3_{  3}^{  8}$\\ 
 &  & pr-AGN$+$   & $ 8.9_{ 8.9}^{ 9.1}$ & [$2.8_{2.1}^{3.1}$,$3.3_{2.5}^{5.7}$,${1.9_{1.3}^{2.6}}$] & [$0.82_{0.03}^{0.89}$,$0.01_{0.00}^{0.02}$,$0.18_{0.02}^{0.66}$,$0.85_{0.03}^{0.92}$]  & $136_{117}^{165}$\\
 &  & pr-SF   & $ 9.6_{ 8.0}^{10.2}$ & [$4.7_{1.7}^{5.7}$,$7.3_{2.0}^{9.3}$,${4.7_{0.8}^{6.0}}$] & [$0.00_{0.00}^{0.00}$,$0.00_{0.00}^{0.00}$,$0.00_{0.00}^{0.00}$,$0.00_{0.00}^{0.00}$]  & $  9_{  7}^{ 16}$\\
 &  & pr-AGN   & $ 7.3_{ 7.1}^{ 7.7}$ & [$0.0_{0.0}^{0.0}$,$0.0_{0.0}^{0.0}$,${0.0_{0.0}^{0.0}}$] & [$0.99_{0.99}^{1.00}$,$0.16_{0.05}^{0.23}$,$0.94_{0.90}^{0.95}$,$1.00_{1.00}^{1.00}$]  & $ 71_{ 33}^{119}$\\
\enddata
\tablecomments{\label{tab:stats}Table with statistical physical properties of the sample of galaxies in this paper. We provide medians and quartiles for the physical properties mentioned in Table~\ref{tab:props} for the whole sample and the subsamples defined based on the detection in different MIRI bands.}
\end{deluxetable*}

\appendix
\section{Bespoke procedures for the reduction of MIRI data: the SMILES case}
\label{app:mirireduction}

The MIRI data used in this paper were reduced with the Rainbow JWST pipeline developed within the European Consortium MIRI GTO Team to deal with MIRI, NIRCam, and NIRISS imaging data. The pipeline relies on the {\it jwst} official pipeline and adds some offline steps to improve the results, mainly dealing with the varying, full-of-structure background observed in MIRI data, especially at the shortest wavelengths. 

The Rainbow pipeline starts with a default execution of the 3 stages of the  {\it jwst} official pipeline, which provides a first full mosaic, now also dealing (to some extent, but not completely) with cosmic ray showers (or snow balls for NIRCam). This mosaic is used to detect sources with {\it sextractor} in order to produce a mask for further refinements of the calibration. We use a relatively shallow detection limit since the mosaic presents intense background gradients and structure, indicating that a deep detection would be dominated by background, not real sources. Apart from the {\it sextractor} detection, we performed a 5-pixel dilation of the segmentation maps to account for the faint outskirts of extended objects, also including the emission from PSF spikes and the cruciform feature of short wavelength MIRI data \citep{2021PASP..133a4504G}.

After masking sources from the previous task, stage 2 data (i.e., cal.fits files) are median-filtered in rows and columns and a smooth 4-order surface is subtracted. The new calibrated data products are again mosaicked with the official pipeline. This new mosaic presents a much more well-behaved background and allows a more aggressive source detection. The new mask is used in a final step of the Rainbow pipeline which implements a super-background strategy to obtain the best results. 

The super-background strategy consists in homogenizing the background of a given stage 2 image using all the other images taken in the same program. For programs extending over several epochs and with enough data, PRIMER for example, we use the closest data in time. We first subtract the median of the background (after masking sources detected in the previous step) for each image. Then, for each frame to reduce, we build a stacked median background image with the rest of images. If not enough data are available for a given pixel, i.e., when very few images are available and the dithering is small compared to the size of (some) objects in the field, we replace the pixel value by a random nearby background pixel chosen from the 100 closest non-masked pixels. The stacked median super-background image is subtracted from the frame we are considering, and the filtering in rows and columns is performed, as well as the subtraction of a smooth surface.

The astrometry of the new background-homogenized cal.fits files is calibrated with the {\it tweakreg} external routine provided by the CEERS collaboration \citep{2023ApJ...946L..12B}, using an external catalog constructed with IRAF's {\it center} task in centroid mode (which was checked to perform better than photutils), using in the SMILES case the JADES catalog as the  WCS reference. Then stage 3 of the pipeline is executed, switching off the {\it tweakreg} step and setting the pixel scale to 60 milliarcsec in the case of MIRI data. The WCS of the final mosaic is checked again against the reference WCS catalog and a final background subtraction is performed with {\it sextractor}.

\begin{figure}[htp!]
\centering
\epsscale{1.15}
\plotone{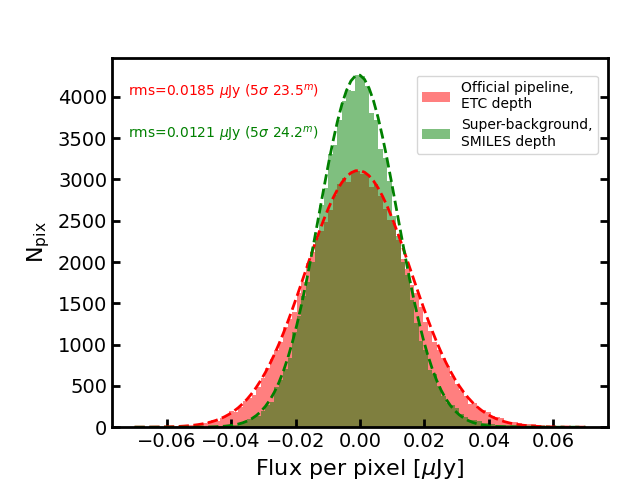}
\caption{\label{fig:appA1}Histogram of pixel values (transformed to $\mu$Jy) of the F1500W SMILES mosaic, after subtracting the median background. In red, we show the results for the mosaic produced by the official pipeline. In green, the results obtained with the bespoke version of the pipeline, embedded in the Rainbow database, and implementing a super-background strategy. The 2 histograms are fitted to a gaussians, whose dispersion is translated to $5\sigma$ depths for a point-like source, calculations based on measurements in an circular of radius equal to the FWHM of the PSF, and corrected for the limited size of the aperture using the calibration available in pmap 1138.}
\end{figure}

The procedure is evaluated in Figure~\ref{fig:appA1}, where we compare the histograms of pixel signal for the final mosaic of the F1500W filter reduced just with the {\it jwst} pipeline (after subtracting the median value of the full image) and reduced with our super-background method. The histograms are fitted to a gaussian, showing that our procedure is able to reduce the noise by a factor of $\sim1.5$. This translates to 0.5~mag deeper detection limits in the final mosaic for this band compared to what can be achieved with the official pipeline alone, which roughly agrees with the expectations provided by the ETC version v3.0.

\startlongtable
\begin{deluxetable*}{lccccccccc}
\tablehead{\colhead{Filter} & \colhead{NGROUP} & \colhead{NINT} & \colhead{NDITHER} & \colhead{texp} & \colhead{background (ETC)} & \colhead{FWHM} & \colhead{5$\sigma$ depth (ETC)} & \colhead{ap.corr}\\
 &  &  &  & s & MJy/sr &  & AB mag & AB mag }
\startdata
F560W  &  59 & 1 & 4 &  720 & 1.3 (1.0) & 0.21\arcsec & 26.0 (25.4) & 0.73 \\
F770W  &  78 & 1 & 4 &  923 & 4.5 (5.6) & 0.27\arcsec & 25.5 (24.9) & 0.64 \\
F1000W &  58 & 1 & 4 &  665 &  13  (15) & 0.33\arcsec & 24.8 (24.2) & 0.53 \\
F1280W &  68 & 1 & 4 &  755 &  25  (31) & 0.42\arcsec & 24.3 (23.6) & 0.50 \\
F1500W & 101 & 1 & 4 & 1121 &  43  (49) & 0.49\arcsec & 24.2 (23.5) & 0.49 \\
F1800W &  68 & 1 & 4 &  755 &  93  (97) & 0.59\arcsec & 23.0 (22.6) & 0.46 \\
F2100W &  32 & 6 & 4 & 2186 & 233 (200) & 0.67\arcsec & 22.6 (22.6) & 0.49 \\
F2550W &  18 & 4 & 4 &  832 & 876 (755) & 0.80\arcsec & 20.8 (20.8) & 0.48\\
\enddata
\tablecomments{\label{table:mirismiles}(1) MIRI filter. (2)-(4) Observational strategy in the SMILES survey: number of groups, number of integrations, number of dithering positions. (5) Exposure time per pixel (seconds). (6) Average background measured in the data. In parenthesis, the ETC v3.0 predictions for Dec 7, 2022 are given. Units (MJy/sr). (7) FWHM of the PSF in each filter (arcsec). (8) Depth of the data for a point-like source measure in a circular aperture with radius equal to the FWHM of the PSF, aperture corrected. In parenthesis, the ETC v3.0 predictions for Dec 7, 2022 are given. Units are AB mag. (9) Aperture correction applied to previous column results (AB mag), based on files released in pmap-.}
\end{deluxetable*}

Table~\ref{table:mirismiles} provides the observation strategy, total exposure times per pixel, average background levels, and depths of the SMILES data reduced with the Rainbow JWST pipeline, compared with ETC predictions.

\section{Dust emission modeling}
\label{app:dust}

\begin{figure*}[htp!]
\centering
\epsscale{1.0}
\plotone{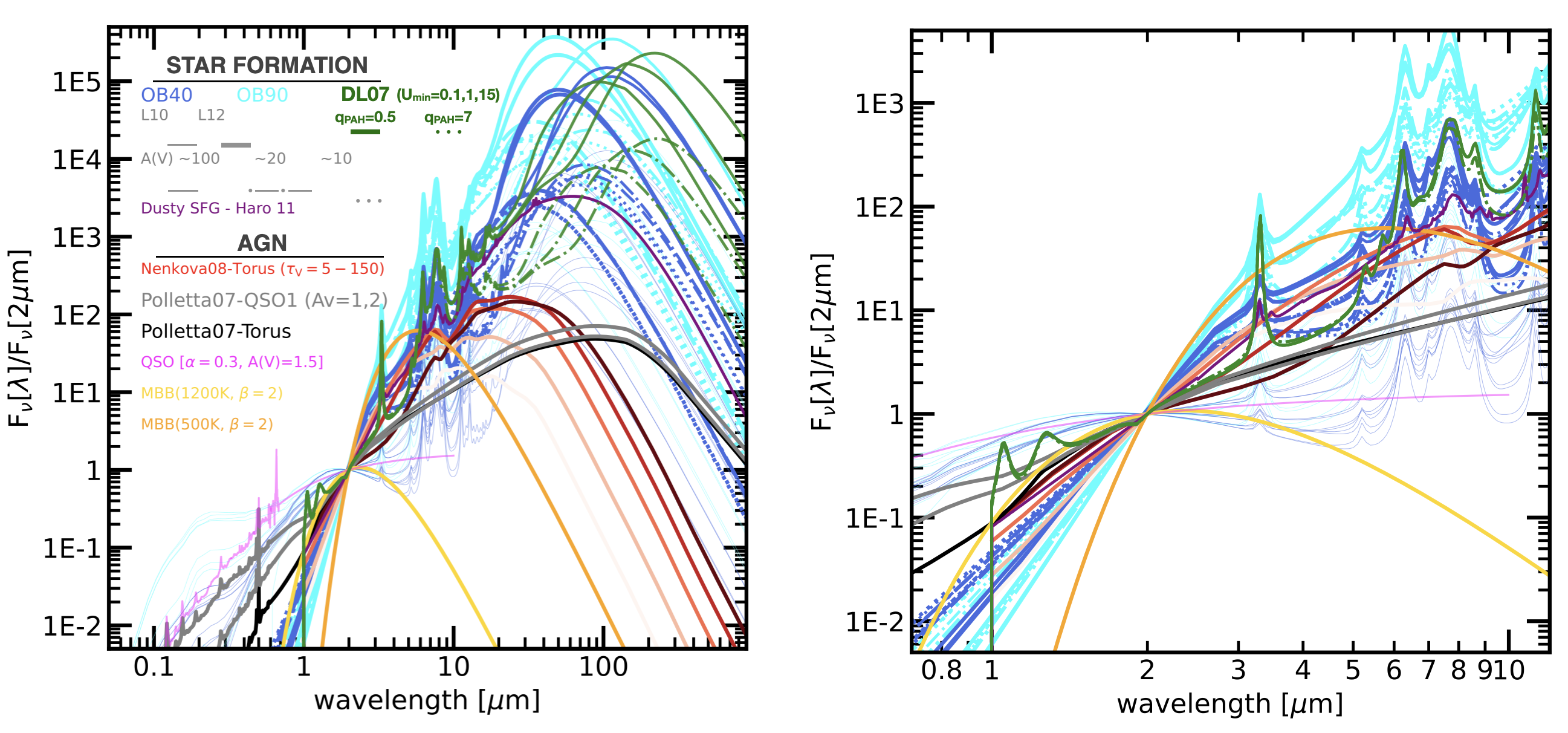}
\caption{\label{fig:appB1}Dust emission models used in the analysis of LRDs presented in this paper. Five different models from \citet{2008ApJ...685..160N} used by \textsc{prospector-AGN} are shown in red colors, showcasing the shift to shorter wavelengths (from $\sim$30 to $\sim$10~$\mu$m) of the dust emission peak arising from an AGN torus with different dust optical depths, namely, $\tau_V$ equal to 5,10, 20, 40, and 150. Gray lines show the dust torus and QSO templates (the latter, extincted by 1 and 2~mag following a \citealt{2000ApJ...533..682C} law) in \citet{2007ApJ...663...81P}, the former being used by \textsc{synthesizer-AGN}. Thick lines show emission from dust, thin lines show the full models, which include stellar emission. Two black-body models for warm dust (500-1200~K) are depicted. The dust emission models for nuclear starbursts, also used by \textsc{synthesizer-AGN} \citep{2007A&A...461..445S}, are shown in blue and cyan colors, with representative values of the different parameters (total luminosity, luminosity arising from OB stars, hot spot hydrogen density, and total attenuation in the $V$-band). The star-forming model for Haro~11 used by \textsc{prospector-AGN$+$} \citep{Lyu2016,DeRossi2018} is shown in purple, and the models from \citet{2007ApJ...657..810D} used by \textsc{prospector-SF} in green.}
\end{figure*}

In this Appendix, we describe in detail the dust emission models used in this paper. Different recipes and emission origins are considered by the four codes described in Section~\ref{sec:modeling}. The main characteristics of our dust emission templates are summarized in Figure~\ref{fig:appB1}. 

The \textsc{synthesizer-AGN} code uses the radiative transfer models of starburst nuclei and (ultra)luminous infrared galaxies (LIRG) presented in \citet{2007A&A...461..445S}. These models assume an intense star formation event (on top a more evolved stellar population) where a fraction of the most massive (OB) stars are embedded in compact dusty clouds (hot spots in their terminology) that dominate the mid-infrared emission. The models are parametrized in terms of the total radiated luminosity (ranging from sub-LIRG to hyper-LIRG values), the size of the star-forming region, the total amount of dust described by the total $V$-band attenuation, the fraction of the total luminosity linked to the OB stars in the hot spots, and the gas/dust density of the hot spot clouds. 

In Figure~\ref{fig:appB1}, we show models for a $10^{10}$~L$_\odot$ and $10^{12}$~L$_\odot$ compact (350~pc in size) region, attenuations $\mathrm{A(V)}\sim10,20,100$~mag, OB luminosity ratios 40 and 90\%, and hydrogen number densities $10^2$ and $10^4$ cm$^{-3}$. The templates include a fixed stellar population, which is removed for our modeling of LRDs. Models with and without stars are shown in Figure~\ref{fig:appB1}. 

We remark that these radiative transfer models nicely recover the blue$+$red nature of the UV-to-NIR emission of LRDs. The change in slope of the stellar UV and optical emission is governed by the OB luminosity ratio, i.e., cyan models (OB90) change in slope at longer wavelengths, around 0.4~$\mu$m rest-frame,  similarly to LRDs. Their dust emission peaks at shorter wavelengths, implying a significant amount of warm dust. The attenuation of the FUV emission in these compact starburst models is, however, steeper than what is observed for our galaxies. All \citet{2007A&A...461..445S} models, which assume Milky Way type dust, present more or less prominent PAH bands. The typical mid- to far-IR flux density ratio is nearly 1000, which means that some of our galaxies, which present MIRI short wavelength fluxes around 1~$\mu$Jy, are also constrained by the non-detections in {\it Herschel} bands (with 5$\sigma$ limits around 2-3~mJy at 100 and 160~$\mu$m and 10~mJy at 250-500~$\mu$m).

The MIR slopes of the \citet{2007A&A...461..445S} models are very similar to the slopes of AGN torus emission presented in \citet{2008ApJ...685..160N}, which are used in our \textsc{prospector-AGN} fits. \citet{2008ApJ...685..160N} parametrizes the emission from clumpy dust tori in terms of the number of clouds intercepting the visual, which depend on other parameters such as the torus thickness (outer to inner radii), the density and angular distribution of clumps and the viewing angle. The subset of templates included in FSPS and then Prospector adopts typical assumptions for the majority of these parameters (see e.g., \citealt{leja18}) leaving e only the scaling (overall luminosity) and the optical depth of an individual dust clump at 5500\AA ($\tau_\mathrm{V}$ from 5 to 150) as free parameters. For the prospector-AGN fits we remove any contribution from the accretion disk in the dust tori templates and we instead model the disk separately following a combination of empirical QSO template plus a power-law f$_{\nu}$=$\nu^{\alpha}$ (equivalent to $\nu$f$_{\nu}$=$\nu^{\alpha+1}$) with variable slope $\alpha=-0.5$ to 0.5, as discussed in \citealt{2016MNRAS.463.2064H} or more recently in \citealt{2023arXiv230714414B}, attenuated by a Calzetti law. For reference, the accretion disk model with A(V)=1.5 is shown in Figure~\ref{fig:appB1} in magenta. The main difference between the \citet{2008ApJ...685..160N} models and the ones in \citet{2007A&A...461..445S} dominated by star formation is the absence of PAH emission (especially relevant for our observations, the one at 3~$\mu$m) and the silicate absorption present in some dust models. Overall, a dust torus (and an AGN template) is almost featureless. 

For reference, Figure~\ref{fig:appB1} also shows the torus (also used in our \textsc{synthesizer} and \textsc{prospector-AGN$+$} fits) and QSO template presented in \citet{2007ApJ...663...81P}, the latter additionally attenuated with a \citet{2000ApJ...533..682C} law assuming $\mathrm{A(V)}=2$~mag. These templates present a similar slope as the \citet{2008ApJ...685..160N} torus models, adding the contribution from the accretion disk that dominates at wavelengths shorter than $\sim1$~$\mu$m and a tail of colder dust peaks at around 100~$\mu$m, as the star-forming models. 

The AGN SED models used in \textsc{prospector-AGN$+$} are largely based on empirical observations with the AGN emission strength calibrated against various observations. For example, the AGN hot dust emission predicted from the SED template is confirmed with NIR image decomposition of HST observations of low-$z$ quasars \citep{Lyu2017}. The AGN mid- to far-IR SED shape is checked against MIR spectral decomposition and PAH strengths \citep[e.g.,][]{Lyu2017b}. Compared to radiation transfer models, these empirical SED models can provide more realistic descriptions of the observations across a very wide range of AGN luminosity and redshift with fewer free parameters and less model degeneracy, which make them particularly preferred for AGN identifications (see review by \citealt{Lyu2022b}). The galaxy dust emission template used in \textsc{prospector-AGN$+$} is based on an empirical SED template of Haro~11, which has been also tested against real observations of very high-$z$ galaxies \citep{Lyu2016, DeRossi2018}. In contrast, the other galaxy dust emission models used in other works typically do not capture some key features of high-$z$ galaxy ISM, such as low-metallicity and the possibly different dust compositions, as argued in \citet{DeRossi2018}.

We have normalized all models at 2~$\mu$m. If the emission of LRDs were dominated by the dust torus at these wavelengths, considering the average redshift of our sample $<z>=6.5$, i.e., 15~$\mu$m observed,  the slope of the torus would translate to the emission at 1~$\mu$m rest-frame, 8~$\mu$m observed, being $\sim$10 times fainter. The F770W-F1500W colors of LRDS are much smaller, indicating that the 1~$\mu$m emission cannot be dominated by the torus even if the 15~$\mu$m flux is linked to an obscured AGN.

\section{SED fits for the full LRD sample in the SMILES/JADES field}
\label{app:allSEDs}

In this Appendix, we present a detailed discussion of the SEDs of the Golden Five galaxies (shown in Figures~\ref{fig:fits1} to \ref{fig:fits5}) and the rest of galaxies in our LRD sample.

\subsection{Galaxies detected up to F1800W: JADES-57356 and JADES-204851}

Figures~\ref{fig:fits1} and \ref{fig:fits2} showed the fits for JADES$-$57356 and JADES$-$204851, the 2 LRDs in our sample detected up to (at least) 18~$\mu$m. 

JADES$-$57356 is a canonical LRD, presenting a change in slope in its SED at around 2~$\mu$m. No prominent emission lines are detected in any medium- and broad-band filter, although some excess is seen in the filters that would cover the Ly-$\alpha$ and MgII  emissions for $z=5.5$\footnote{The photometric redshift probability distribution function for this source presents 2 peaks, one at $z\sim3.5$ and one at $z\sim5.5$. Significantly better $\chi^2$ values are obtained by \textsc{synthesizer-AGN} and \textsc{prospector-AGN} for our finally chosen redshift $z\sim5.5$.}. To reproduce the characteristic, bimodal SED of the LRDs, 2 of the 4 codes use two distinct components. 


The SED fit with \textsc{synthesizer-AGN} uses 2 stellar populations (with independent attenuations), one with a young (60~Myr) and mildly unobscured ($\mathrm{A(V)}\sim1$~mag) starburst, which also contributes to the faint emission lines, and an older (250~Myr) stellar population with much larger reddening, $\mathrm{A(V)}\sim4$~mag, that dominates the stellar mass content (98\% of the total).


\textsc{Prospector-AGN} uses a similarly young (130~Myr) stellar component for the UV emission and AGN emission from a dust-obscured accretion disk for the optical. \textsc{Prospector-SF}, on the other hand, uses only stars to reproduce the SED up to the optical and redder wavelengths, as also obtains \textsc{Prospector-AGN$+$}. However, \textsc{Prospector-SF} requires an unusual, extremely gray attenuation law (n$\sim$0.4). Indeed, \citet{2023arXiv230514418B} noted that  a typical Calzetti law (n$=$0) and a single attenuation parameter would not reproduce the SED of the LRDs. In fact, as can be extracted from the data in Table~\ref{tab:props} and the parameters written in the plots, the ratio between the far-UV (at $\sim$150~nm, FUV) and optical ($\sim550$~nm) attenuations is $\sim$1.5  for \textsc{Prospector-SF} and $\sim$1.5 for \textsc{Prospector-AGN$+$}, smaller than the $\sim2.6$ implied by the Calzetti law. We note that  the \textsc{synthesizer-AGN} fits assume such as law, but the combination of the independent attenuations for the old and young stellar populations results in a FUV-to-optical attenuation ratio  close to the values given above ($\kappa_\mathrm{FUV}=1.8$).

We also remark that \textsc{synthesizer-AGN} does use a  Calzetti law but assumes independent extinctions for the 2 stellar populations in the host, and also for the AGN, which allows a good fit to the data for JADES$-$57356. \textsc{Prospector-AGN$+$} does not require a two-component fit but it allows for hybrid AGN+galaxy models, which can sometimes fit the SED with 2 distinct components, each dominating a different spectral range, but can also fit the whole SED with only one of those components. This is the case for JADES$-$57356 which, as noted above, exhibits a similar best-fit model to the prospector-SF model for the UV-to-NIR region. 

Overall, all 4 models provide good fits to the UV-to-NIR SED, with some differences. The three stellar-dominated models suggest a relatively large stellar mass, $\mathrm{M}_\star=10^{10.4}\,\mathrm{M}_\odot$ for \textsc{synthesizer-AGN}. In contrast, in \textsc{Prospector-AGN} the emission from the obscured accretion disk ($\mathrm{A(V)}=2.8$~mag) dominates the SED up to $\lambda=1~\mu$m (and the AGN at even bluer wavelengths), which leads to a much smaller stellar mass for the galaxy host that is only visible in the rest-UV.

While all the models obtain similarly good UV-to-NIR fits, they differ substantially in the MIR emission, where the SED shows a clear flattening beyond 1~$\mu$m (F770W-F1000W$\sim$0 mag) followed by an upturn around 2$\mu$m that steepens quickly towards the redder bands. This SED shape is well reproduced by the stellar-continuum-dominated models that combine a stellar peak around 1.6$\mu$m with dust emission presenting a prominent 3~$\mu$m PAH line in the MIRI F2100W band. For \textsc{synthesizer-AGN}, the overall best-fitting code ($\chi^2$ values given in the plots), the dust emission model is quite extreme: the best template corresponds to a ULIRG with $\mathrm{A(V)}=10$~mag, 350~pc star-forming region size, 60\% luminosity ratio of OB stars in hot spots compared to total luminosity, and dust density in hot spots $10^2$~cm$^{-3}$. This dust emission model peaks at rest-frame $\sim30$~$\mu$m (see Figure~\ref{fig:appB1} in Appendix~\ref{app:dust}), quite a blue wavelength compared to the more typical $\sim70-100$~$\mu$m for (U)LIRGs \citep[e.g.,][]{2009ApJ...692..556R}, revealing the important role of warm/hot dust in LRDs, and also implying a relatively low emission at (sub-)mm wavelengths, where the flux of LRDs is faint \citep{2023arXiv230607320L,2023arXiv231107483W}.



This \citet{2007A&A...461..445S} model points to the existence of very hot dust bathed by a very intense radiation field, coming from dust-buried OB stars (the total attenuation of stellar light in the models ranges from 2 to 4.5~mag), as indicated by the \textsc{synthesizer-AGN} and \textsc{prospector-AGN$+$} fits. Such an intense and compact heating source also matches well what can be expected from an AGN, which in principle could contribute to some extent to the MIR emission. Indeed, \textsc{synthesizer-AGN} and \textsc{prospector-AGN$+$} do show some contribution, although faint and thus uncertain, from an AGN.

However, \textsc{prospector-AGN}, the only code that is forced to be AGN-dominated in this spectral range, provides a worse fit to the MIRI bands because the transition from disk-dominated to torus dominated emission around $\lambda\sim2\,\mu$m leads to redder MIRI colors due to the steeper slope of the dust torus relative to the dust associated with star formation  in the other models (note the bad fits to the F560W and F1000W fluxes). To some degree, \textsc{prospector-SF} also has problems fitting the rest-NIR region because the nebular emission partially hides the flattening of the stellar continuum and provides a worse fit to those MIRI bands.

In summary, for this source the fit to the dust emission presents a smaller $\chi^2$ when including the, most probably star-formation related, 3~$\mu$m PAH. The relatively large stellar mass and old mass-weighted age (50--700~Myr) could support the AGN nature of the dust emission (for several reasons, e.g., the stars are relatively old and that could intuitively imply that there are not so obscured, or the evolved state of the galaxy gives time for the SMBH to grow), but the fits are significantly worse.

JADES$-$204851, whose SED and postage stamps are shown in Figure~\ref{fig:fits2}, is a source with spectroscopic redshift $z_\mathrm{sp}=5.4790$ discussed previously in \citet[][GOODS-S-13971 in that paper]{2023arXiv230605448M}, where they reported on the presence of a broad 2200~km~s$^{-1}$ H$\alpha$ component (S/N$\sim$5 and with some artifacts in the spectrum --certainly, this is not one of the clearest BLR sources in \citealt{2023arXiv230605448M}--) arising from an AGN with a $10^{7.5}$~M$_\odot$ SMBH. This AGN would account for half of the H$\alpha$ emission, according to the spectroscopic analysis. 


In our {\sc synthesizer-AGN} fits for JADES$-$204851 (top left panel of Figure~\ref{fig:fits2}), the AGN emits around 10\% of the total emission at 2~$\mu$m, also contributing to emission lines such as H$\alpha$, [OIII], and MgII. The MIR emission would also have a significant contribution from star-formation powered dust emission, with a $\sim10$\% possible contribution from a QSO-like emission. For this galaxy, the best fitting dust emission model corresponds to one 350~pc ULIRG star-forming region with highly embedded stars ($\mathrm{A(V)}=72$~mag), with a 40\% ratio of OB stars in hot spots compared to the total luminosity, and dust density in hot spots $10^4$~cm$^{-3}$. The optical-to-NIR is fitted with a combination of two stellar populations. One of them is a very young (1~Myr) unobscured ($\mathrm{A(V)}=0.5$~mag)  starburst, which takes care of the strong emission seen in spectroscopy and detected in the NIRCam imaging and the blue continuum. The other population is slightly older (10~Myr) affected by a large reddening ($\mathrm{A(V)}\sim2.5$~mag) and dominating (94\%) the total stellar mass ($\mathrm{M}_\star=10^{9.6}\,\mathrm{M}_\odot$).


The {\sc prospector-AGN$+$} results are very similar to those obtained with {\sc synthesizer-AGN} in the rest-frame optical and near-infrared: the SED is dominated by stars and dust heated by stars. However, {\sc prospector-AGN$+$} reproduces the UV part of the SED with stars alone, while some contribution from a (unobscured) QSO is obtained with {\sc synthesizer-AGN}, mainly due to the possible presence of a MgII line at observed wavelength $\sim$2~$\mu$m. The contribution is small, and thus uncertain, but the spectroscopy hints that there is a broad-line component (at 5$\sigma$ confidence level).
This difference between both codes might also be caused by the different configurations of the stellar extinction laws. As pointed out by \citet{Kriek2013}, the standard Calzetti law (used by {\sc synthesizer-AGN}) typically provides poor fits at UV wavelengths for high-$z$ galaxies and thus the AGN component is likely selected from the model to fit the SED in {\sc synthesizer-AGN}. Meanwhile, {\sc prospector-AGN$+$} used the updated galaxy extinction law introduced 
by \citet{Kriek2013}, which is supposed to be more realistic. On the other hand, the independent treatment of the attenuation for old and young stars in {\sc synthesizer-AGN} can overcome the problems with a single extinction parameter. 

The {\sc prospector-SF} also provides a qualitatively good fit with a single stellar component and a modest A(V)$=$1.2~mag, but very gray attenuation, $n=0.4$ (translating to $\kappa_\mathrm{FUV}\sim1.5$, same value obtained by {\sc synthesizer-AGN}). Interestingly, the fit to the MIRI bands is worse due to the presence of strong emission lines in the model which would imply a larger flux in F1500W than is observed. The dust emission in this model contributes less than 10\% of the total emission at 2~$\mu$m and has virtually no impact in the best-fit SED.  

The {\sc prospector-AGN} fit also provides a good overall fit of the NIRCam and MIRI photometry, using stars for the UV and an AGN for the optical/IR (as in the LRD characterized in \citealt{2023arXiv231203065K}). In this galaxy, which exhibits significantly bluer MIRI colors than JADES$-$57356, the obscured emission from the disk (A(V)$=2.3$~mag) dominates the optical and NIR emission up to $\lambda\sim2\mu$m. The emission from the torus is mostly unconstrained but it would require low IR luminosities or a very large opacity ($\tau_V\gtrsim$100). As before, the implied stellar mass of the host is the smallest of all the models, $\mathrm{M}_\star=10^{8.1}\,\mathrm{M}_\odot$.

In any case, the different morphology seen in F814W compared to the LW NIRCam bands, with the emission in the former  being dominated by a knot located to the NW of the very concentrated (and very red) emission seen in the latter, points to star formation dominating the UV spectral range. This morphological difference between the rest-frame UV and NIR is also seen in other LRDs, for example, JADES$-$211388 or JADES$-$79803, discussed later.

\subsection{Galaxy detected up to F1500W: JADES-219000}


Figure~\ref{fig:fits3} discussed the results for JADES$-$219000, the third LRD in our sample detected up to 15~$\mu$m. This source is presented in Sun et al. (2024, in prep.), with a spectroscopic redshift of $z=6.8119$. Similarly to JADES$-$57356, the best-fit with {\sc synthesizer-AGN} indicates a UV-to-NIR SED dominated by stars with a relatively high mass ($\mathrm{M}_\star=10^{10.4}\,\mathrm{M}_\odot$), and dust emission revealing a very intense radiation field. The dust emission model corresponds to an intense starburst with 90\% luminosity arising from OB stars embedded in a $\mathrm{A(V)}=72$~mag compact (350~pc), dense ($10^4$~cm$^{-3}$) and clumpy dust cloud. The very hot dust present in this object could also be heated by an AGN, but the typical torus emission used in \textsc{prospector-AGN$+$} fails to fall within the F1800W and F2100W upper limits. This, in part, explains the larger $\chi^2$ value compared to the other codes, jointly with discrepancies in the FUV which could be interpreted as attenuation law effects.

The {\sc prospector-SF} fit is also relatively good with similar best-fit values as the other galaxies, i.e., a young stellar population, with a small and very gray attenuation (A(V)$=$1.6~mag, n$=$0.33, $\kappa_\mathrm{FUV}\sim2$). The dust emission does not contribute significantly to the MIRI fluxes ($<1\%$ at $\lambda=2\mu$m), the nebular emission dominates (but the F1000W is not well-fitted). The {\sc prospector-AGN} fit also agrees well with the data. As before, the best-fit model also implies a low stellar mass for the host and a moderate attenuation for the disk (A(V)=1.6~mag), but it favors a more luminous torus with large opacity, $\tau_V=100$. This means that the torus emission starts to contribute significantly at $\lambda\sim1\,\mu$m, but it peaks at longer wavelengths ($\lambda=30\,\mu$m) than the low opacity tori (see Figure~\ref{fig:appB1} of the appendix), and therefore the model does not over-estimate the F1800W and F2100W upper limits.

\subsection{Galaxies detected up to F1280W: JADES-211388 and JADES-79803}

Figure~\ref{fig:fits4} showed the SED fits for JADES$-$211388, a spectroscopically confirmed LRD at $z=8.3846$. This is one of the highest redshift LRDs and, consequently, the MIRI detections probe shorter rest-frame wavelengths than in previous galaxies. 

The {\sc synthesizer-AGN} best-fit indicates that the SED is dominated by young (3~Myr), slightly extincted ($\mathrm{A(V)}\sim1$~mag) stars in the blue, with some contribution from slightly older (5~Myr) and more obscured ($\mathrm{A(V)}=2.5$~mag) stars in the red, for a total stellar mass of $\mathrm{M}_\star=10^{9.1}\,\mathrm{M}_\odot$. Remarkably, the NIR emission is dominated by a heavily obscured AGN as opposed to the star-formation powered dust emission of previous galaxies.

The obscured AGN template is very similar to the best-fit model with \textsc{prospector-AGN$+$} and resembles the AGN-dominated fits in the NIR of \citet{2023arXiv230514418B} or \citet{2023arXiv230607320L}, but with significant contribution from stars up to rest-frame wavelength $\sim2$~$\mu$m, where the AGN would contribute with 50\% of the total flux),  but with a lower luminosity that is still consistent with the upper limits. 

The {\sc prospector-SF} best-fit parameters are almost identical to the previous galaxies with a young stellar population obscured with a mild but gray dust attenuation (A(V)$=$1.7~mag, $n=0.39$, $\kappa_\mathrm{FUV}\sim1.4$, still failing to reproduce the FUV) and a relatively flat NIR emission dominated by the nebular continuum in the $\lambda\sim1-2~\mu$m range.

{\sc prospector-AGN} also provides a good fit to the optical-to-NIR SED but it has some issues fitting the sharp Lyman break indicated by the blue NIRCam bands (we remind the reader that the attenuation is fixed to 0~mag in these models). The best-fit properties are again consistent with a young, low-mass galaxy host $\mathrm{M}_\star=10^{8.7}\,\mathrm{M}_\odot$ and an obscured AGN, $\mathrm{A(V)}=3.4$~mag. The lack of MIRI constraints at long wavelengths leads to best-fit models dominated only by the accretion disk emission even up to $\lambda=$4~$\mu$m. However, a more significant contribution from the torus emission at $\lambda\gtrsim2\,\mu$m would still be consistent with the upper limits, as shown in the {\sc synthesizer-AGN} and \textsc{prospector-AGN$+$} fits.

Finally, Figure~\ref{fig:fits5} shows the SED fits for JADES$-$79803, another spectroscopically confirmed LRD, this time at $z=5.4007$. The SED exhibits a flattening around $\lambda\sim1\mu$m, probed by F560W and F770W,  followed by a steep rise in F1000W and F1280W and then another flattening indicated by the upper limits in the reddest bands. The best-fit parameters with {\sc prospector-SF} are similar to JADES$-$211388, but the stellar mass is a bit smaller, $\mathrm{M}_\star=10^{8.6}\,\mathrm{M}_\odot$, with all stars being younger than 20~Myr and presenting different attenuations between 0.5 and 2~mag. The near-to-mid IR emission is reproduced again by an obscured AGN, but in this case, the F1280W detection provides stronger evidence of an upturn. 

The \textsc{prospector-AGN$+$} also reproduces the SED with a similarly young stellar population and an obscured AGN dominating the near-to-mid IR emission. 

The {\sc prospector-SF} best-fit continues on the same trends as before, but it provides a worse fit to the reddest MIRI bands whose steep rise can not be easily reproduced with emission from star-formation heated dust. 

{\sc prospector-AGN} provides a good overall fit, the best among all codes for this source, with the same trends in previous galaxies (low-mass host and obscured disk dominating the optical emission). In this case, the additional MIRI constraints motivate a fit with a larger contribution of the torus emission to the IR emission starting around $\lambda\sim$2~$\mu$m, but consistent with the upper limits.

\subsection{Galaxies detected up to F770W and with no MIRI detections}

\begin{figure*}[htp!]
\centering
\includegraphics[clip,trim=0.0cm 0.0cm 0.0cm 0.0cm,width=8.cm,angle=0]{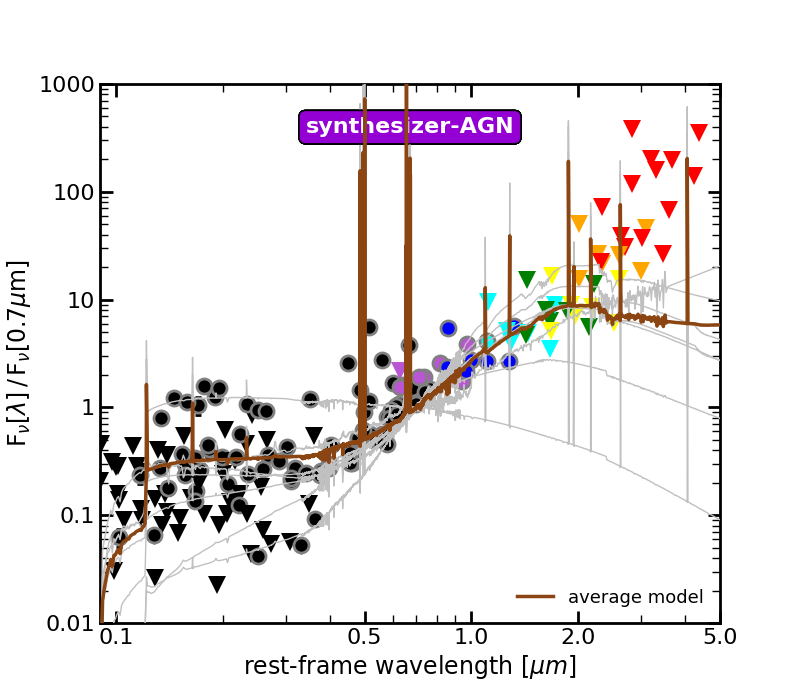}
\includegraphics[clip,trim=0.0cm 0.0cm 0.0cm 0.0cm,width=8.cm,angle=0]{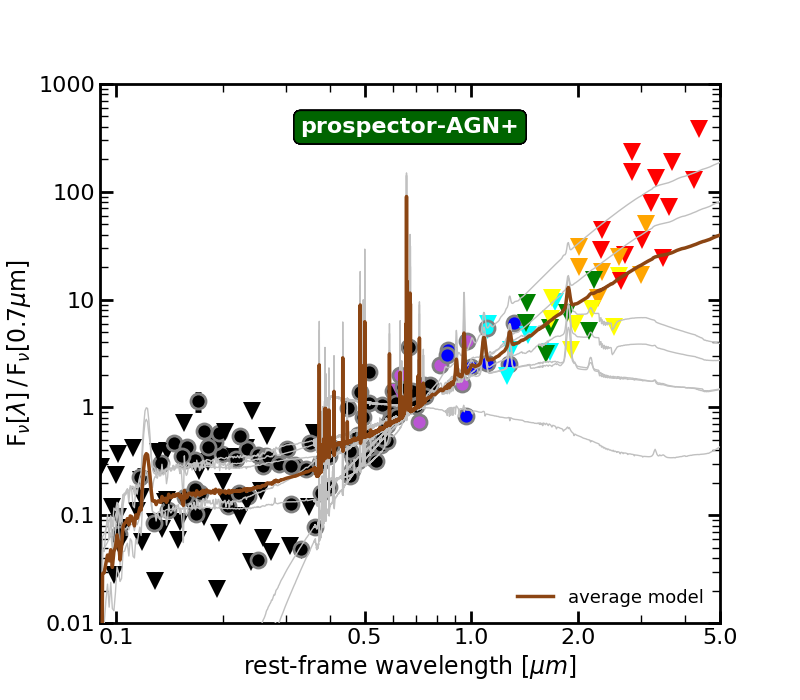}
\includegraphics[clip,trim=0.0cm 0.0cm 0.0cm 0.0cm,width=8.cm,angle=0]{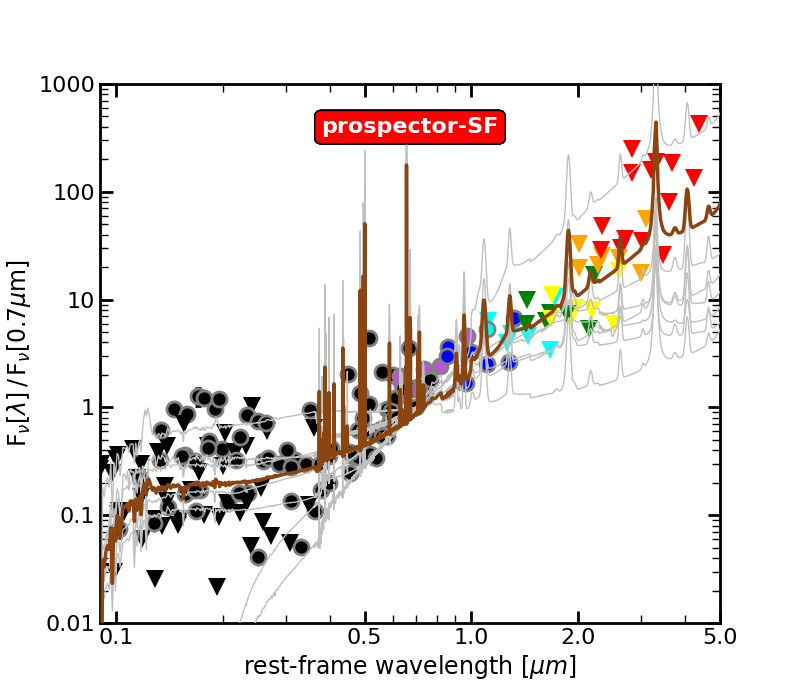}
\includegraphics[clip,trim=0.0cm 0.0cm 0.0cm 0.0cm,width=8cm,angle=0]{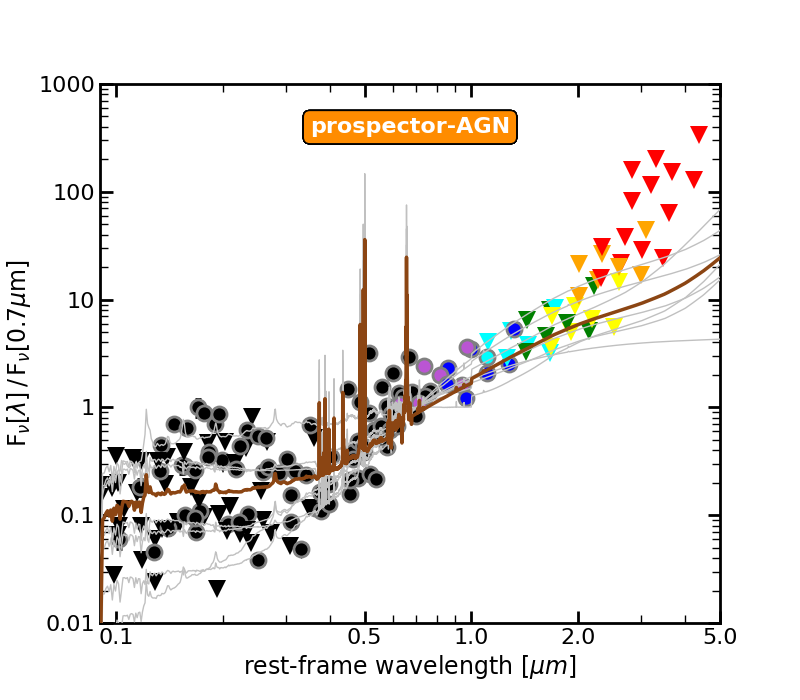}
\caption{SED fitting results for the 7 galaxies detected at F770W and not beyond (i.e., the plot does not include any of the Golden Five or F1000W galaxies discussed in the main text). SEDs are normalized to 0.7~$\mu$m. In gray, we show the fits to each individual galaxy, and we provide an average in brown.}
\label{fig:fits7}
\end{figure*}

\begin{figure*}[htp!]
\centering
\includegraphics[clip,trim=0.0cm 0.0cm 0.0cm 0.0cm,width=8.cm,angle=0]{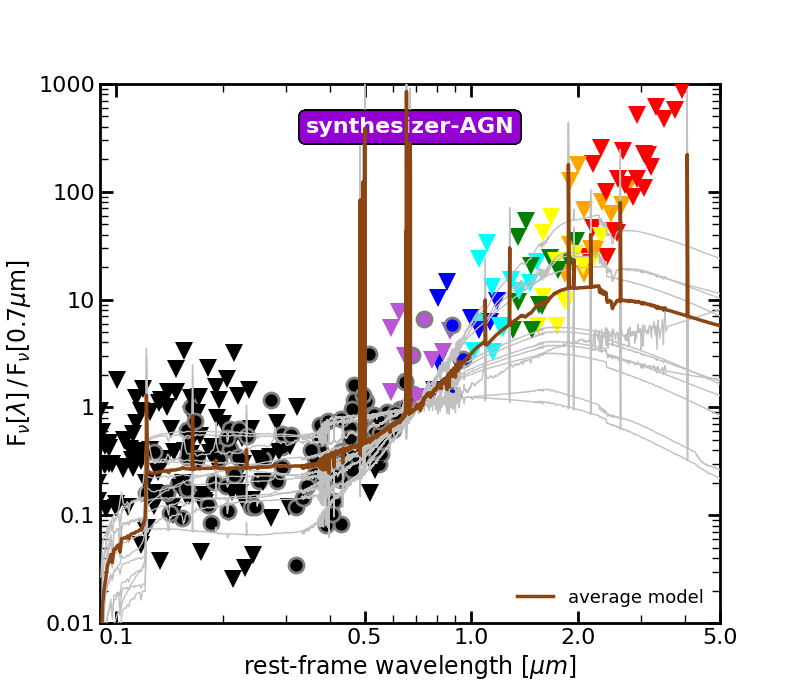}
\includegraphics[clip,trim=0.0cm 0.0cm 0.0cm 0.0cm,width=8.cm,angle=0]{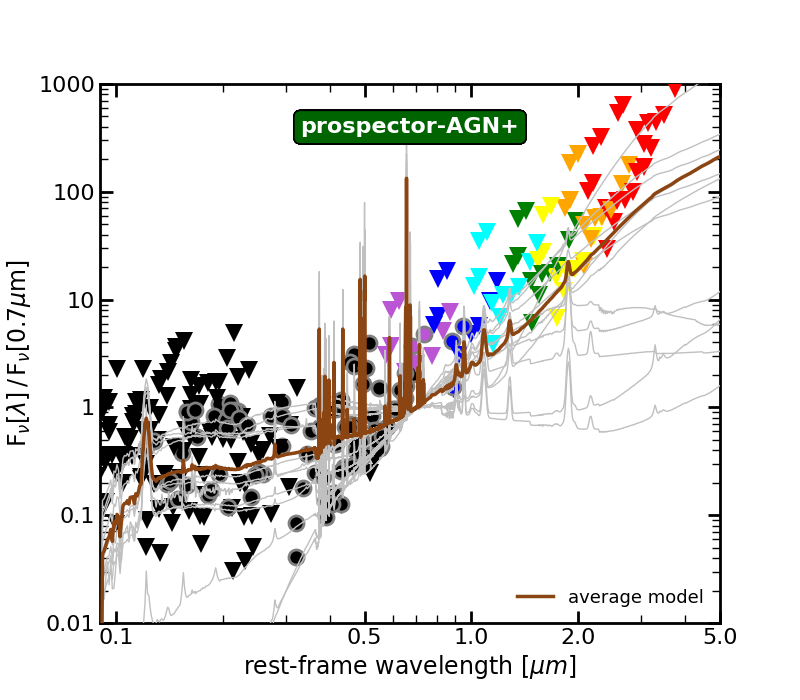}
\includegraphics[clip,trim=0.0cm 0.0cm 0.0cm 0.0cm,width=8.cm,angle=0]{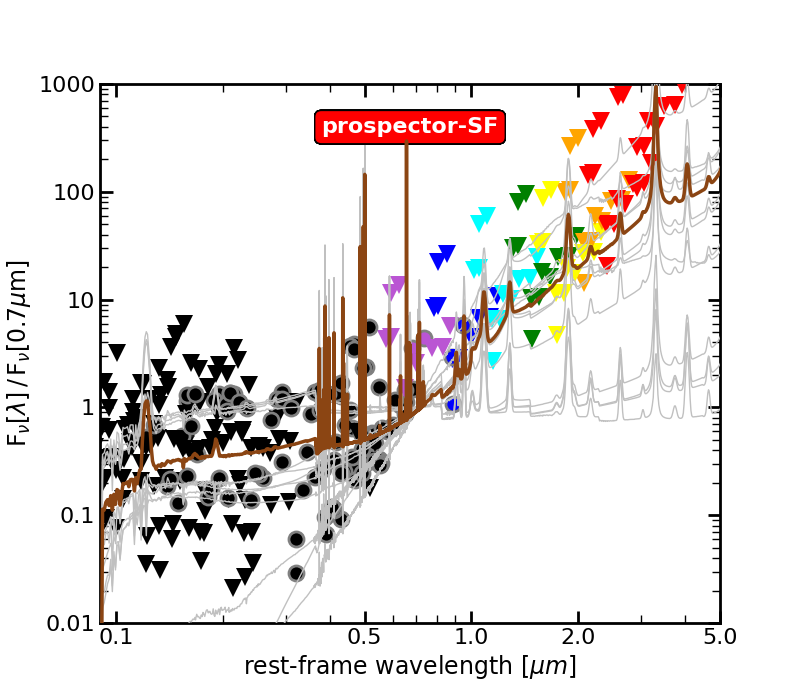}
\includegraphics[clip,trim=0.0cm 0.0cm 0.0cm 0.0cm,width=8cm,angle=0]{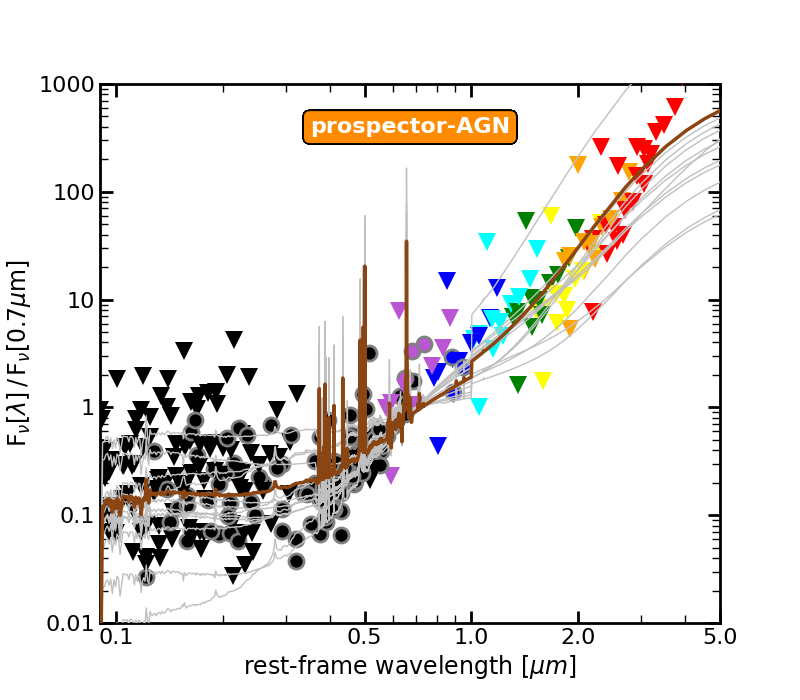}
\caption{SED fitting results for the 12 galaxies not detected in any MIRI band. Same format as previous figure.}
\label{fig:fits8}
\end{figure*}

Figures \ref{fig:fits7} and \ref{fig:fits8} show the SED fits for all F770W-detected sources not presented in the main text and the source with no detection in any MIRI band, respectively. Results for the 4 fitting codes described in Section~\ref{sec:modeling} are provided. 

The F770W sample shown in Figure~\ref{fig:fits7} presents  more heterogeneous SEDs than the Golden Five or F1000W samples discussed in Section~\ref{sec:modeling}. The global slope of the SED is more constant, with the difference between the SW and LW bands being less marked than for the other LRDs. In fact, this sample includes a  very blue galaxy (187025, $z_\mathrm{sp}=6.9076$), which only entered the sample because of the enhanced F444W flux due to an emission line. The SEDs, including MIRI upper limits (remarkably, for F1000W and F1280W), are very flat in the 1-2~$\mu$m range, completely compatible with being dominated by stars without the need of much more contribution from other emitting components. Compared to other subsamples, the F770W sources are more extincted (nearly 3~mag on average for the stars dominating the mass) and present older mass-weighted ages (nearly 10~Myr, compared to the 3-4~Myr for galaxies detected at longer MIRI wavelengths).

Figure~\ref{fig:fits8} shows a similar result for MIRI undetected sources, the available data (limited to NIRCam) is well fitted by stars only. The upper limits for MIRI would be consistent with hot dust emission from an AGN torus, but the current data is inconclusive.

Considering the full sample of 31 sources, the median/quartiles $\chi^2$ for \textsc{synthesizer-AGN} are  $342_{207}^{800}$, \textsc{prospector-AGN$+$} obtains larger relative values compared to the former by a factor of $2.3_{1.3}^{4.8}$, as is the case for \textsc{prospector-SF} with $1.2_{0.6}^{2.5}$ times larger $\chi^2$ values,  and \textsc{prospector-AGN}, $1.3_{0.6}^{3.7}$.

\bibliography{lrds_smiles}{}
\bibliographystyle{aasjournal}

\end{document}